\RequirePackage{fix-cm}
\documentclass[smallextended]{svjour3wide}
\smartqed

\usepackage{amsmath,amssymb,amsfonts,bm}
\usepackage{mathrsfs,amsbsy,latexsym}
\usepackage{graphicx}
\usepackage{url}
\usepackage[hidelinks]{hyperref}

\usepackage[dvipsnames]{xcolor}

\newcommand{\pder}[2]{\frac{\partial #1}{\partial  #2}}
\newcommand{\pderf}[3]{\left(\frac{\partial #1}{\partial  #2}\right)_{#3}}
\newcommand{\pdertf}[3]{\left( \frac{\partial^2 #1}{{\partial  #2}^2}\right)_{#3}}
\newcommand{\pderc}[3]{\frac{\partial^2 #1}{\partial  #2 \partial  #3}}
\newcommand{\pdert}[2]{\frac{\partial^2 #1}{{\partial  #2}^2}}
\newcommand{\der}[2]{\frac{d #1}{d  #2}}

\newcommand{\nm}{\nonumber\\}

\DeclareMathOperator*{\argmin}{arg\,min}

\newcommand{\ep}{\varepsilon}

\newcommand{\kB}{k_\mathrm{B}}

\newcommand{\eq}{\mathrm{eq}}

\newcommand{\ex}{\mathrm{ex}}

\newcommand{\vdW}{\mathrm{vdW}}
\newcommand{\eff}{\mathrm{eff}}
\newcommand{\res}{\mathrm{res}}
\newcommand{\mech}{\mathrm{mech}}
\newcommand{\sat}{\mathrm{sat}}
\newcommand{\bias}{\mathrm{bias}}

\newcommand{\homo}{\mathrm{homo}}
\newcommand{\hetero}{\mathrm{hetero}}
\newcommand{\saddle}{\mathrm{saddle}}

\newcommand{\topP}{p(L)}
\newcommand{\botP}{p(0)}
\newcommand{\topT}{T(L)}
\newcommand{\botT}{T(0)}

\newcommand{\subC}{\mathrm{c}}

\newcommand{\subL}{\mathrm{L}}
\newcommand{\subG}{\mathrm{G}}

\newcommand{\subl}{\mathrm{\ell}}
\newcommand{\subu}{\mathrm{u}}

\newcommand{\Tc}{T_\mathrm{c}}

\newcommand{\xm}{x_\mathrm{m}}
\newcommand{\xint}{x_\theta}

\newcommand{\Ps}{p_\mathrm{s}}

\newcommand{\pint}{p_\theta}

\newcommand{\bT}{\tilde{T}}
\newcommand{\bP}{\bar{p}}
\newcommand{\bmu}{\tilde\mu}

\newcommand{\calN}{\mathcal{N}}
\newcommand{\calV}{\mathcal{V}}
\newcommand{\calF}{\mathcal{F}}

\makeatletter
\@addtoreset{equation}{section}
\makeatother

\journalname{Journal of Statistical Physics}

\begin{document}
\title{Global thermodynamics for heat-conducting fluids under weak gravity}

\titlerunning{Global thermodynamics under heat conduction and weak gravity}

\author{Naoko Nakagawa \and Shin-ichi Sasa}
\authorrunning{N. Nakagawa and S.-i. Sasa}

\institute{
N. Nakagawa \at
Department of Physics, Ibaraki University, Mito 310-8512, Japan
\email{naoko.nakagawa.phys@vc.ibaraki.ac.jp}
\and
S.-i. Sasa \at
Department of Physics, Kyoto University, Kyoto 606-8502, Japan
\email{sasa.shinichi.6n@kyoto-u.ac.jp}
}

\date{\today}

\maketitle

\begin{abstract}
We study liquid-gas coexistence under gravity and heat conduction from the viewpoint of global thermodynamics. We construct a variational free-energy function for the fixed-global-temperature description and decompose it into two parts. The first has the same configurational form as the equilibrium weak-gravity free energy with gravity replaced by the effective gravity, and it determines the first-order configurational transition between the two separated liquid-gas arrangements. The second is a residual excess-latent-heat contribution that vanishes without heat conduction. Although it does not decide which separated liquid-gas arrangement is thermodynamically favored, this residual part is needed to derive the fundamental relation in the laboratory variables and to recover thermodynamic observables such as the spatially averaged pressure. The same residual contribution reshapes the barrier geometry, ridge/valley structure, and interfacial anomalies of the fixed-global-temperature free-energy landscape. Numerical examples based on the van der Waals model illustrate the resulting landscape structure, and estimates of experimental scales suggest a setup for detecting the effective-gravity inversion.
\end{abstract}


\setcounter{tocdepth}{3}
\tableofcontents

\section{Introduction}
Phase coexistence under sustained driving is a basic setting in which nonequilibrium thermodynamics is expected to make quantitative predictions.
In fluids, heat transport, phase change, and external forcing are coupled in boiling, evaporation, two-phase convection, and gravity-confined liquid--gas systems \cite{Thome04,Ahlers93,Zhong09,Cristea10,Wagner23,DavisGupta25}.
Thermal driving can alter phase conversion and heat transport in two-phase convection \cite{Zhong09,WeissAhlers13,Urban13}; under gravity, it can change which phase arrangement is realized \cite{Cristea10,Wagner23,DavisGupta25}.

Liquid--gas coexistence under gravity and heat conduction is a direct state-ordering problem.
Gravity favors the denser phase at lower positions, whereas heat conduction can bias the liquid toward the colder side.
A conventional route would be to determine steady profiles from hydrodynamic balance equations, heat conduction, interfacial conditions, and local equations of state.
This route is natural, but it requires nontrivial interfacial input.

The nontrivial part of this conventional route is already visible at the liquid--gas interface.
Heat-conducting or evaporating interfaces can exhibit thermal resistance and temperature jumps, and the thermodynamic quantities assigned to the interface need not be reducible to a naive local-saturation condition alone \cite{Muscatello17,FangWard99}.
One may therefore try to refine the hydrodynamic description and the associated interfacial boundary conditions.
In this paper we take a different, complementary route.
Rather than constructing a microscopic or hydrodynamic theory of interfacial transport, we ask whether competing liquid--gas arrangements can be compared by a global thermodynamic balance.
The question is not only how to construct candidate steady profiles, but how to rank coarse-grained configurations when gravity and heat flow favor different arrangements.
Such a comparison is naturally expressed in terms of a quantity that treats the competing configurations as states of the whole system, even when candidate local profiles can be constructed.

The framework used here is global thermodynamics for phase coexistence: it describes steady states by global variables and compares them through a variational free-energy function.
For heat-conducting coexistence, it predicted the deviation of the interface temperature from the equilibrium transition temperature and the possible stabilization of a supercooled gas near the interface \cite{NS17,NS19}.
The formulation was later sharpened through a maximum-entropy principle \cite{NS22}, tested numerically in a Hamiltonian Potts model \cite{KNS23}, and connected to fluctuating-hydrodynamic theory in a diffusive density-field model \cite{SN25}.
For fluids under gravity, the equilibrium reference problem was formulated from equilibrium statistical mechanics as a variational free-energy problem in the global variables \cite{NSgravity25}, and the competition between gravity and heat flow leads to the global-thermodynamic question of whether the macroscopic ordering can be described by an effective-gravity-like bias \cite{NSgeff26}.
A related numerical study, separate from that formulation, observed heat-induced liquid hovering in liquid-gas coexistence under gravity \cite{YNS24}.

The global-thermodynamic results summarized above point to a simple possibility.
Heat conduction may act, at the level of the macroscopic liquid--gas arrangement, as a correction to the gravitational bias.
If so, the nonequilibrium problem would contain an effective-gravity-like ordering principle, as briefly reported in Ref.~\cite{NSgeff26}.
It remains to determine, however, whether such a reduction can be derived from a global thermodynamic structure, and whether it can describe more than the ordering of separated configurations.
In particular, even if local nonequilibrium signatures are invisible in the macroscopic ordering rule, the global thermodynamic structure must still account for them through its thermodynamic relations.

Adopting this global-thermodynamic perspective also fixes the status of the interface in the present formulation.
Nonequilibrium liquid--gas interfaces themselves have rich thermodynamic and transport structures, as studied in interfacial nonequilibrium thermodynamics \cite{Schweizer16,Struchtrup23}.
A related issue also appears in stochastic order-parameter descriptions of heat-conducting coexistence, where thermodynamic quantities are assigned to the interfacial region through a variational principle \cite{SNIN21}.
Broader reviews and comparisons of evaporation and condensation models further emphasize that the interfacial transport problem contains several possible levels of description \cite{PersadWard16,Rauter22}.
Here we do not model such interfacial transport processes.
Instead, the interface is represented as a sharp boundary between two bulk regions.
Within this sharp-interface description, the question is how the global variational construction encodes the nonequilibrium mismatch of local fields across that boundary.

The task of this paper is therefore to formulate this global balance explicitly.
We construct a variational free-energy function and determine whether a part of it can be written in terms of an effective gravity $g_{\eff}$ that selects the lower configuration.
We then ask what remains beyond that ordering rule: whether a residual nonequilibrium contribution is required, how it enters the thermodynamic relation, and how it accounts for the local nonequilibrium signatures described above.
This distinction is important because a stationary free-energy value and its derivatives do not carry the same information.
A contribution that does not decide the ordering of separated heterogeneous states may still be needed to recover the physical coefficients in the thermodynamic relation.

The paper is organized as follows.
We first introduce the setup and formulate the variational free-energy function.
We then decompose it into the effective-gravity part $\calF_{g,\eff}$ and the residual part $\calF_{g,\res}$, and show that the effective gravity $g_\eff$ determines the ordering of the two separated liquid--gas configurations.
We next derive the fundamental relation and evaluate the global chemical potential from local fields.
We then examine the hierarchy and stability of the stationary solutions, and describe how the fixed-global-temperature free-energy landscape reflects both the effective-gravity ordering and the residual nonequilibrium contribution.
We finally discuss experimental scales for testing the effective-gravity inversion and outline open problems suggested by the landscape description.

\section{Setup}

We consider a fluid of $N$ particles of equal mass $m$ in a rectangular container $V=AL$, where $L$ is the height and $A$ is the cross-sectional area of the container. Throughout this paper, we fix the vertical coordinate to $x\in[0,L]$, so that the bottom and top of the container are located at $x=0$ and $x=L$, respectively. Gravity is represented by the force component $-mg$ along the $x$-axis for each particle; $g>0$ denotes ordinary downward gravity and $g<0$ denotes the sign-reversed case.
In this paper, we assume that the local states are uniform in the horizontal plane; that is, all local quantities are functions only of $x$. We define $\rho(x)$ as the local number density satisfying $N=A\int_{0}^{L}dx \, \rho(x)$,
and $T(x)$ is the local temperature. The local specific volume is $v(x)\equiv 1/\rho(x)$.
The free energy per particle $f(T,v)$ is the single-phase thermodynamic function used to characterize local single-phase states and, when needed below, their metastable continuations; see Ref.~\cite{NS19} for this use of single-phase continuations. The entropy per particle, pressure, and chemical potential are given by
\begin{align}
s(T,v)&=-\left.\pderf{f}{T}{v}\right|_{(T,v)}, \\
p(T,v)&=-\left.\pderf{f}{v}{T}\right|_{(T,v)},\\
\mu(T,v)&=f(T,v)+p(T,v)v,
\end{align}
respectively.

\subsection{Global thermodynamic quantities}

We define the center of mass as
\begin{align}
X = \frac{A}{N}\int_{0}^{L} dx~x \rho(x),
\label{e:X-def}
\end{align}
and the midpoint of the system as
\begin{align}
\xm = \frac{L}{2}.
\end{align}
In order to extend the conventional thermodynamic framework to the system under gravity, the reference of the gravitational potential should be the midpoint $\xm$ as developed in \cite{NSgravity25}.
For equilibrium systems, consistency with the partition function fixes this reference point and the gravitational contribution \cite{NSgravity25}.
We then define the global free energy including gravity by evaluating this single-phase free-energy function along the local fields as
\begin{align}
F_g&=A\int_{0}^{L}\rho(x)[ f(T(x), v(x))+mg(x-\xm)] dx,\nm
&=A\int_{0}^{L}\rho(x) f(T(x), v(x)) dx +Nmg(X-\xm).
\end{align}
Correspondingly, we define the local chemical potential including gravity as
\begin{align}
\mu_g(x)\equiv \mu(x)+mg(x-\xm).
\end{align}

Due to gravity and heat flow, the pressure, chemical potential, and temperature become functions of $x$, whereas in equilibrium without gravity or heat flow they are spatially uniform.
To provide a global thermodynamic description, we define a global pressure as the spatial average of the local pressure,
\begin{align}
&\bP \equiv \frac{1}{L}\int_{0}^{L} dx~p(x),
\label{e:bP-def}
\end{align}
and a global chemical potential containing the gravitational potential as a particle-number-weighted average,
\begin{align}
\bmu_g \equiv \frac{A}{N}\int_{0}^{L} \mu_g(x) \rho(x)dx.
\label{e:bmug-def}
\end{align}

The global temperature $\bT$ is defined as the density-weighted average of the local temperature,
\begin{align}
\bT= \frac{A}{N}\int_{0}^{L} T(x)\rho(x) dx. \label{e:def-bT}
\end{align}
The imposed temperature difference $\Xi$ is defined by
\begin{align}
\Xi=T(L)-T(0).
\end{align}
Introducing the dimensionless parameter
\begin{align}
\ep= \max\left(\frac{|\Xi|}{\bT}, \frac{|mgL|}{\kB \bT}\right)\ll 1,
\label{e:ep-neq}
\end{align}
we focus below on the linear response regime in $\ep$, where both the gravitational and thermal perturbations are small.

\subsection{Setup for phase separation}

In this paper, we divide the system into two regions at the formal dividing position $x=l$, while restricting each region to a single phase, either liquid or gas.

Hereafter, the indices $\subl$ and $\subu$ denote the quantities in the lower and upper regions, respectively.
The volume and the number of particles in the lower region $\subl$ are denoted as
$V^\subl$ and $N^\subl$, where $V^\subl=Al$.
For the upper region $\subu$, $V^\subu=V-V^\subl$ and $N^\subu=N-N^\subl$.

\begin{figure}[tb]
\centering
\includegraphics[width=0.72\linewidth]{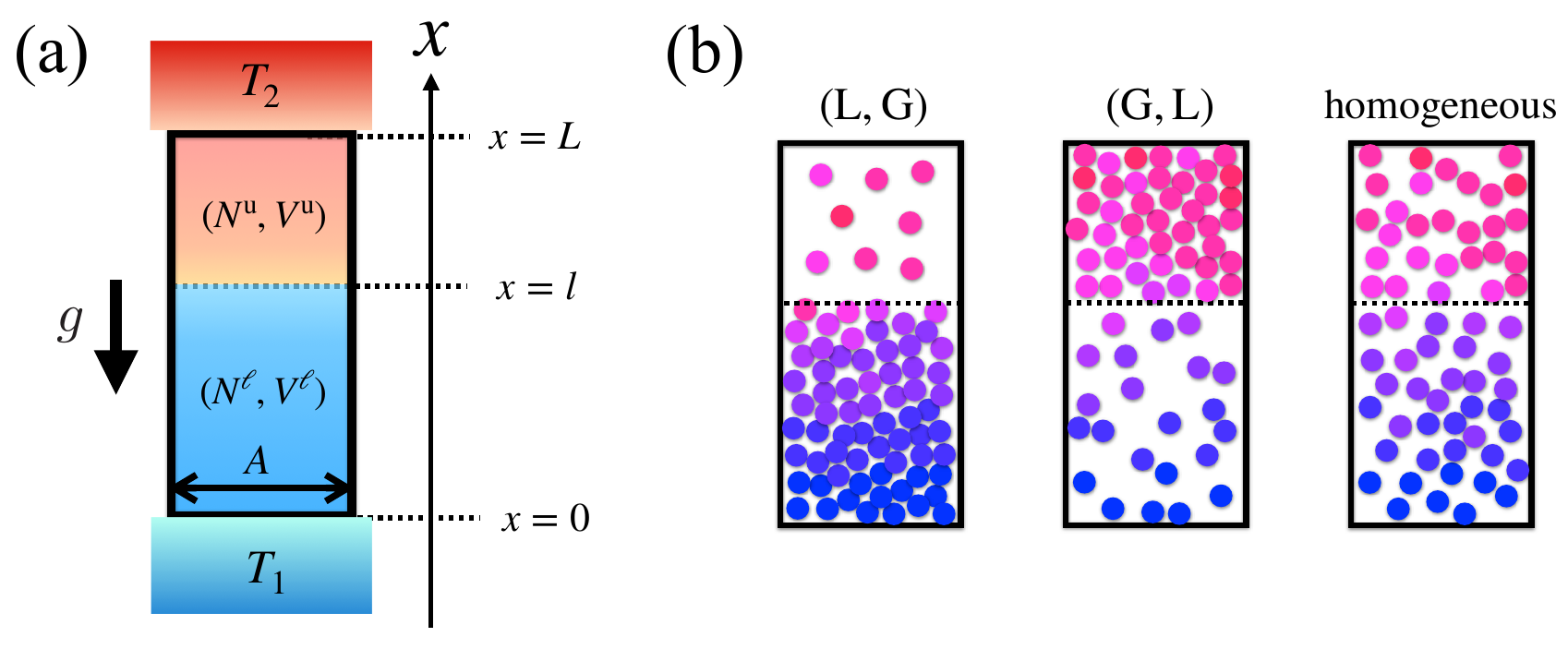}
\caption{(a) System setup under gravity and heat flow. The lower and upper coarse-grained regions are separated by the formal dividing position $x=l$ and are characterized by $(N^\subl,V^\subl)$ and $(N^\subu,V^\subu)$. In a heterogeneous state, the actual liquid-gas interface is at $x=x_\theta$. (b) The three configuration classes: $(\subL,\subG)$, $(\subG,\subL)$, and homogeneous. In the homogeneous panel, the dashed line indicates only the formal partition, not a macroscopic interface.}
\label{fig:setup}
\end{figure}

We impose that each of the lower and upper regions consists of a single phase.
This restriction is expressed by the conditions
\begin{align}
X^\subl = \xm^\subl + O(\ep), \qquad
X^\subu = \xm^\subu + O(\ep),
\label{e:Xl-Xu}
\end{align}
where $X^\subl$ and $X^\subu$ are the centers of mass of regions $\subl$ and $\subu$, respectively, and $\xm^\subl = l/2$, $\xm^\subu = (L+l)/2$ are the midpoints of regions $\subl$ and $\subu$, respectively.
With these constraints, the lower and upper regions can therefore be labeled as
\begin{align}
(\subl,\subu)=(\subL,\subG), ~(\subG, \subL), ~(\subL,\subL), ~(\subG,\subG),
\label{e:config}
\end{align}
where $\subL$ and $\subG$ represent the liquid and gas phases.
The first two represent liquid-gas coexistence, whereas the last two are usually denoted collectively as single-phase states.
We write $(\subL,\subL)$ or $(\subG,\subG)$ explicitly only when the liquid-like and gas-like homogeneous realizations need to be distinguished.
The geometry, notation, and the three coarse-grained configuration classes are summarized in Fig.~\ref{fig:setup}.

\subsection{Thermodynamic quantities in the lower and upper regions}
\label{s:GT-lu}

In weak gravity and weak heat conduction, profiles are approximated as piecewise linear functions.
Within the linear response regime, the profiles in each single-phase region are approximated by linear functions.
Thus, the global thermodynamic quantities in each region are represented by the mean values to leading order in $\ep$.

The global temperatures are
\begin{align}
&\bT^\subl=\frac{A}{N^\subl}\int_{0}^{l} T(x)\rho(x) dx=\frac{T_1+\theta}{2}+O(\ep^2), \\
&\bT^\subu=\frac{A}{N^\subu}\int_{l}^{L} T(x)\rho(x) dx=\frac{T_2+\theta}{2}+O(\ep^2).
\end{align}
Here, $T_1 = T(0)$, $T_2 = T(L)$, and $\theta = T(l)$ is the local temperature at the formal dividing position.
Similarly, the averaged pressures in the lower and upper regions are
\begin{align}
&\bP^\subl = \frac{1}{l} \int_{0}^{l} p(x) dx=\frac{\botP+p_-}{2}+O(\ep^2), \label{e:P_l_def} \\
&\bP^\subu = \frac{1}{L-l} \int_{l}^{L} p(x) dx=\frac{\topP+p_+}{2}+O(\ep^2). \label{e:P_u_def}
\end{align}
Here $p_+$ and $p_-$ are the pressures immediately above and below the formal dividing position at $x=l$.
The average chemical potentials incorporating the gravitational effect are
\begin{align}
\bmu_g^\subl &= \frac{A}{ N^\subl} \int_{0}^{l} [\mu(x)+mg(x-\xm^\subl)]\rho(x) dx
= \frac{A}{ N^\subl} \int_{0}^{l} \mu(x)\rho(x) dx+mg(X^\subl-\xm^\subl) \nm
&=\frac{\mu^\subl(0)+\mu^\subl_-}{2}+O(\ep^2), \\
\bmu_g^\subu &= \frac{A}{N^\subu} \int_{l}^{L} [\mu(x)+mg(x-\xm^\subu)]\rho(x) dx
= \frac{A}{N^\subu} \int_{l}^{L} \mu(x)\rho(x) dx+mg(X^\subu-\xm^\subu)\nm
&=\frac{\mu^\subu(L)+\mu^\subu_+}{2}+O(\ep^2), \label{e:mu_lu_def}
\end{align}
where $\mu^\subl(x)$ and $\mu^\subu(x)$ are the equilibrium chemical potentials in the phases $\subl$ and $\subu$.
The quantities $\mu^\subl_-$ and $\mu^\subu_+$ are the chemical potentials immediately below and above the formal dividing position at $x=l$.
The density-weighted mean of $\mu(x)$ in each single-phase region is represented by the endpoint average up to $O(\ep^2)$.
These global quantities for the lower and upper regions satisfy
\begin{align}
&\bP^\subl=p(\bT^\subl, V^\subl, N^\subl)+O(\ep^2), \qquad
\bP^\subu=p(\bT^\subu, V^\subu, N^\subu)+O(\ep^2),\label{e:p-lu-def}\\
&\bmu_g^\subl=\mu^\subl(\bT^\subl, \bP^\subl)+O(\ep^2),\qquad
\bmu_g^\subu=\mu^\subu(\bT^\subu, \bP^\subu)+O(\ep^2),\label{e:mu-lu-def}
\end{align}
where $p(T,V,N)$, $\mu^\subl(T,p)$ and $\mu^\subu(T,p)$ are the equilibrium thermodynamic functions.

A similar argument can be applied to extensive quantities.
The entropies of the lower and upper regions are
\begin{align}
S^\subl=A\int_{0}^{l}\rho(x)s(T(x), v(x)) dx, \qquad S^\subu= A\int_{l}^{L}\rho(x)s(T(x), v(x))dx,
\end{align}
and the corresponding free energies are
\begin{align}
F^\subl = A\int_{0}^{l}\rho(x)f(T(x), v(x)) dx,
\qquad
F^\subu= A\int_{l}^{L}\rho(x)f(T(x), v(x))dx.
\end{align}
Using $X^\subl=\xm^\subl+O(\ep)$ and $X^\subu=\xm^\subu+O(\ep)$, we have 
\begin{align}
F_g^\subl=F^\subl+O(\ep^2), \qquad F_g^\subu=F^\subu+O(\ep^2).
\end{align}
Here and below, for an extensive quantity, $O(\ep^k)$ denotes an extensive correction whose relative order is $\ep^k$.
Using
\begin{align}
&S^\subl=S(\bT^\subl, V^\subl, N^\subl), \qquad
S^\subu=S(\bT^\subu, V^\subu, N^\subu),\label{e:S-lu-def}\\
&F^\subl=F(\bT^\subl, V^\subl, N^\subl), \qquad
F^\subu=F(\bT^\subu, V^\subu, N^\subu),\label{e:F-lu-def}
\end{align}
we obtain the following fundamental thermodynamic relations,
\begin{align}
&dF^\subl = -S^\subl d\bT^\subl - \bP^\subl dV^\subl + \bmu_g^\subl dN^\subl, \label{e:fund_rel_l} \\
&dF^\subu = -S^\subu d\bT^\subu - \bP^\subu dV^\subu + \bmu_g^\subu dN^\subu. \label{e:fund_rel_u}
\end{align}
These relations hold for each region, with the understanding that the thermodynamic quantities represent spatially averaged values within each region, even in the presence of gravity and a temperature gradient (within the linear response regime).

\section{Variational principle for the configuration of liquid and gas in coexistence}

\subsection{\texorpdfstring{Equilibrium state determined by the lever rule at $g=0$ and $\Xi=0$}{Equilibrium state determined by the lever rule at g=0 and Xi=0}}

We start by reviewing the thermodynamics of the system in the absence of gravity and temperature gradient ($g = 0$ and $\Xi = 0$). This provides a reference for the arguments in the nonequilibrium case.
Let $p=p(T,v)$ be the equation of state, with $v=V/N$.
A state satisfying $p=p(T,v)$ is assumed to be in a single phase, either liquid or gas, including metastable states.
By integrating $p=p(T,v)$ with respect to $v$, the free energy per particle $f(T,v)$ is obtained.
$f(T,v)$ shows a constant slope as a function of $v$ when the liquid and gas coexist.
The slope corresponds to $-\Ps(T)$.
This satisfies $\Ps(T)=p(T,v_\subC^\subL(T))$ and $\Ps(T)=p(T,v_\subC^\subG(T))$,
where $v_\subC^\subL(T)$ and $v_\subC^\subG(T)$ denote the specific volumes of liquid and gas at saturation.
The single-phase state is not globally stable in the region $v_\subC^\subL(T) < v < v_\subC^\subG(T)$; outside the spinodal region it may remain metastable.

The number of particles and volumes,
$(N_0^\subL, V_0^\subL)$ and $(N_0^\subG, V_0^\subG)$, for the liquid and gas are determined using the lever rule.
Specifically, the condition
\begin{align}
&V_0^\subL = v_\subC^\subL(T)N_0^\subL, \quad
V_0^\subG = v_\subC^\subG(T)N_0^\subG, \quad
N_0^\subL + N_0^\subG = N, \quad
V_0^\subL + V_0^\subG = V,
\end{align}
yields
\begin{align}
&\frac{N_0^\subL}{N} = \frac{v_\subC^\subG(T) - \bar v}{v_\subC^\subG(T) - v_\subC^\subL(T)}, \quad
\frac{V_0^\subL}{V} = \frac{v_\subC^\subL(T)}{\bar v} \frac{v_\subC^\subG(T) - \bar v}{v_\subC^\subG(T) - v_\subC^\subL(T)}, \label{e:NV0-L}\\
&\frac{N_0^\subG}{N} = \frac{\bar v - v_\subC^\subL(T)}{v_\subC^\subG(T) - v_\subC^\subL(T)}, \quad
\frac{V_0^\subG}{V} = \frac{v_\subC^\subG(T)}{\bar{v}} \frac{\bar v - v_\subC^\subL(T)}{v_\subC^\subG(T) - v_\subC^\subL(T)}. \label{e:NV0-G}
\end{align}
Thus, $(N_0^\subL/N, V_0^\subL/V)$ and $(N_0^\subG/N, V_0^\subG/V)$ are uniquely determined for a given $(T, \bar{v})$.

\subsection{Variational principle under weak gravity in equilibrium}
\label{s:var-eq}

We review here the variational principle for phase separation enforced by weak gravity in equilibrium.
This principle was established from equilibrium statistical mechanics in Ref.~\cite{NSgravity25}.
Consider constant-temperature systems, $T(x) = T$ everywhere.
To determine the equilibrium liquid-gas separation,
the variational function for fixed $(T, V,N,mgL)$ was derived in Ref.~\cite{NSgravity25} from a canonical ensemble under gravity
in terms of the two regions $\subl$ and $\subu$, each assigned a phase.
Here and below, $(\calN^\subl,\calV^\subl)$ denote variational coordinates, whereas $(N^\subl,V^\subl)$ denote the stationary values determined by the variational principle.
Suppose that $(\calN^\subl,\calV^\subl)$ are the number and volume distributed to the region $\subl$.
We write $\calN^\subu=N-\calN^\subl$ and $\calV^\subu=V-\calV^\subl$.
The variational function is written as a function of variational variables $(\calN^\subl,\calV^\subl)$
\begin{align}
\calF_g(\calN^\subl,\calV^\subl;T,V,N,mgL)
&=F^\subl(T,\calV^\subl,\calN^\subl)
+F^\subu(T,V-\calV^\subl,N-\calN^\subl) 
-\frac{mgL}{2}N
\left(\frac{\calN^\subl}{N}-\frac{\calV^\subl}{V}\right),
\label{e:Fg-var-eq}
\end{align}
with fixed variables $(T,V,N,mgL)$,
where $F^\subl$ and $F^\subu$ are the free energies of the respective regions defined in Sec. \ref{s:GT-lu}.
In the local-density representation used in Ref.~\cite{NSgravity25}, the two-region variational function \eqref{e:Fg-var-eq} corresponds to the spatial integral
\begin{align}
\calF_g=A\int_{0}^{L}\rho(x)[ f(T, v(x))+mg(x-\xm)] dx.
\label{e:Fg-var-eq-local}
\end{align}
Note that the reference point of the gravitational potential is set to $\xm$ in \eqref{e:Fg-var-eq-local}, which was concluded as the unique reference point to provide the global thermodynamic description under gravity \cite{NSgravity25}.
The last term in \eqref{e:Fg-var-eq} is obtained from the gravitational potential term
by using the relation
\begin{align}
\frac{X-\xm}{L}=-\frac{1}{2}\left(\frac{\calN^\subl}{N}-\frac{\calV^\subl}{V}\right)+O(\ep),
\label{e:X-xm}
\end{align}
derived from $NX=\calN^\subl X^\subl+\calN^\subu X^\subu$ and $V\xm=\calV^\subl \xm^\subl+\calV^\subu \xm^\subu$ with $X^\subl=\xm^\subl+O(\ep)$ and $X^\subu=\xm^\subu+O(\ep)$.

The thermodynamic free energy is the minimum value,
\begin{align}
F_g(T,V,N,mgL)=\min_{\mathcal{N}^\subl ,\mathcal{V}^\subl}\calF_g(\calN^\subl,\calV^\subl;T,V,N,mgL).
\end{align}
The stationary values $(N^\subl, V^\subl)$ are determined by the minimization condition
\begin{align}
(N^\subl, V^\subl)=\argmin_{\mathcal{N}^\subl ,\mathcal{V}^\subl}\calF_g(\calN^\subl,\calV^\subl;T,V,N,mgL).
\label{e:ss-min-eq}
\end{align}
$F_g$ satisfies the fundamental relation of global thermodynamics,
\begin{align}
&dF_{\mathit{g}} = -SdT - \bar{p} dV + \bmu_g dN - N\frac{\xm - X}{L}d(m\mathit{g}L).
\label{e:Fg-relation-main}
\end{align}
Thus, imposing gravity, state variables are extended from $(T,V,N)$ to $(T,V,N,mgL)$ to include gravitational effects.

According to the free energy minimum, the required condition for the equilibrium state is given by the variational equations
\begin{align}
\pderf{\calF_g}{\calV^\subl}{*} = 0, \qquad \pderf{\calF_g}{\calN^\subl}{*} = 0. \label{e:var-eq}
\end{align}
Here, the asterisk $(*)$ denotes the derivative evaluated at the equilibrium solution $(N^\subl, V^\subl)$ in fixed $(T,V,N,mgL)$.
Equation \eqref{e:var-eq}, together with \eqref{e:Fg-var-eq} and the fundamental relations \eqref{e:fund_rel_l} and \eqref{e:fund_rel_u}, leads to
\begin{align}
\bP^\subl-\bP^\subu= \frac{mg N}{2A}, \qquad
\bmu_g^\subl-\bmu_g^\subu= \frac{mgL}{2}. \label{e:var-eq-1}
\end{align}
The factor $1/2$ in the first relation reflects that $\bP^\subl$ and $\bP^\subu$ are
the mean pressures of the two regions.  Indeed, inserting
\eqref{e:P_l_def} and \eqref{e:P_u_def} into the first relation in
\eqref{e:var-eq-1}, and using the force balance over the whole system,
$\botP-\topP=Nmg/A$, gives the equality $p_-=p_+$ of the endpoint
pressures at the formal dividing position.  We denote this common endpoint pressure by $p_\theta$:
\begin{align}
p_\theta=p_+=p_-.
\end{align}
The force balance in each region then gives $\botP-p_\theta=mgN^\subl/A$
and $p_\theta-\topP=mgN^\subu/A$.  Substituting these endpoint pressures into
\eqref{e:P_l_def} and \eqref{e:P_u_def}, we obtain
\begin{align}
\bP^\subl=p_\theta+\frac{mgN^\subl}{2A},\qquad
\bP^\subu=p_\theta-\frac{mgN^\subu}{2A}.
\label{e:p-lu}
\end{align}

Using \eqref{e:mu-lu-def}, the second relation in \eqref{e:var-eq-1} is written as
\begin{align}
\mu^\subl(T,\bP^\subl)-\mu^\subu(T,\bP^\subu)=\frac{mgL}{2}.
\end{align}
Applying \eqref{e:p-lu} and expanding around $p_\theta$,
\begin{align}
&\mu^\subl(T,\bP^\subl)=\mu^\subl(T,p_\theta)+\frac{V^\subl}{N^\subl}\frac{mgN^\subl}{2A}+O(\ep^2),\\
&\mu^\subu(T,\bP^\subu)=\mu^\subu(T,p_\theta)-\frac{V^\subu}{N^\subu}\frac{mgN^\subu}{2A}+O(\ep^2).
\end{align}
Recalling that $V^\subl=A l$ and $V^\subu=A(L-l)$, the second relation of \eqref{e:var-eq-1} gives
the equality of the two endpoint chemical potentials,
\begin{align}
\mu^\subl(T,p_\theta)=\mu^\subu(T,p_\theta).
\end{align}
When the lower and upper regions are in different phases, $(\subl,\subu)=(\subL,\subG)$ or $(\subG,\subL)$, this equality gives
\begin{align}
p_\theta=\Ps(T),
\end{align}
that is, the common endpoint pressure equals the saturation pressure.
For the same-phase configurations $(\subL,\subL)$ and $(\subG,\subG)$, the same equation is not a coexistence condition and no saturation pressure is assigned.

According to the equilibrium weak-gravity formulation of Ref.~\cite{NSgravity25}, the minimization principle also determines the stable configuration under gravity,
i.e.,
the equilibrium configuration is specified as $(\subl,\subu)=(\subL,\subG)$ for $\mathit{g}>0$
and $(\subl,\subu)=(\subG,\subL)$ for $\mathit{g}<0$.
Substituting the lever-rule expressions for $(N_0^\subl,V_0^\subl)$
in \eqref{e:NV0-L} and \eqref{e:NV0-G}, chosen according to the configuration,
into the geometric identity \eqref{e:X-xm}, we obtain
\begin{align}
\frac{\xm-X}{L}
= \frac{\mathit{g}}{2|\mathit{g}|} \frac{(v^\subG_\subC(T)-\bar{v})(\bar{v}-v^\subL_\subC(T))}{\bar{v}(v^\subG_\subC(T)-v^\subL_\subC(T))}+O(\ep^2),
\label{e:xm-X-g}
\end{align}
where $v_\subC^\subL(T)$ and $v_\subC^\subG(T)$ denote the specific volumes of liquid and gas at saturation, and $\bar{v}=V/N$.
The sign of $\mathit{g}$ determines the sign of $\xm-X$. The jump of $X$ at $\mathit{g}=0$ is a signature of a first-order transition.
This can be understood as a swap of the primary and secondary minima in the free-energy landscape of $\calF_g$ in the space of $(\calN^\subl,\calV^\subl)$ at $\mathit{g}=0$.
The well-known observation that heavier phases are located below lighter phases
can now be interpreted thermodynamically as a phase transition between these two configurations.

\subsection{Variational function in heat conduction under weak gravity}
\label{s:var-func-neq}

We aim to determine the nonequilibrium steady states of the configurations in \eqref{e:config}, with particular focus on liquid-gas coexistence subject to weak heat flow and gravity.
We employ a variational principle based on the minimization of a suitable variational free energy.
This principle extends, in the local-thermodynamic representation, two established variational structures:
the equilibrium weak-gravity principle reviewed in Sec.~\ref{s:var-eq} and the fixed-volume variational principle for heat-conducting coexistence without gravity \cite{NS19,NS22}.
Accordingly, for fixed $(\bT, V,N,mgL,\Xi)$, we adopt the variational function
\begin{align}
\calF_g(\calN^\subl,\calV^\subl;\bT,V,N,mgL,\Xi)
&=F^\subl(\bT^\subl,\calV^\subl,\calN^\subl)
+F^\subu(\bT^\subu,V-\calV^\subl,N-\calN^\subl) 
-\frac{mgL}{2}N
\left(\frac{\calN^\subl}{N}-\frac{\calV^\subl}{V}\right).
\label{e:Fg-var-neq}
\end{align}
Similarly to the equilibrium case, \eqref{e:Fg-var-neq} can be connected to the spatial integral of the free energy density under gravity
\begin{align}
\calF_g=A\int_{0}^{L}\rho(x)[ f(T(x), v(x))+mg(x-\xm)] dx.
\label{e:Fg-var-neq-local}
\end{align}
Thus, \eqref{e:Fg-var-neq} and \eqref{e:Fg-var-neq-local} are the heat-conducting counterparts of the two-region variational function \eqref{e:Fg-var-eq} and its local-density representation \eqref{e:Fg-var-eq-local}:
the uniform temperature $T$ is replaced by the local temperature profile and the regional global temperatures, while the gravitational term and the reference point $\xm$ are kept as in Ref.~\cite{NSgravity25}.
At $mg=0$, \eqref{e:Fg-var-neq} reduces to the free-energy variational function for global thermodynamics proposed in Ref.~\cite{NS19};
this fixed-$\bT$ free-energy minimization was shown in Ref.~\cite{NS22} to be thermodynamically equivalent, in the linear-response regime, to the extended maximum entropy principle.
With gravity, we regard \eqref{e:Fg-var-neq} as the corresponding local-thermodynamic/global-thermodynamic postulate:
the local free-energy density is the equilibrium one evaluated at local fields, and the gravitational contribution is the weak-gravity contribution with the reference point $\xm$.
We use it only within the assumptions made throughout this paper, namely weak heat flow and weak gravity, a sharp interface, horizontal uniformity, and local thermodynamic relations in each single-phase region.
The variables $(\bT,\Xi)$ are the thermodynamic coordinates of this fixed-global-temperature description;
in a boundary-temperature experiment, they are obtained from the steady temperature profile corresponding to the imposed boundary temperatures.

The variational function \eqref{e:Fg-var-neq} is not explicitly expressed by the global temperature $\bT$ because it is deduced from the spatial integral of $\rho(x)f(T(x),v(x))$ in \eqref{e:Fg-var-neq-local} before imposing a specific temperature profile.
The definition \eqref{e:def-bT} of the global temperature and the continuity of the temperature profile lead to
\begin{align}
&\bT=\frac{\calN^\subl}{N}\bT^\subl + \frac{\calN^\subu}{N}\bT^\subu, \\
&\bT^\subu - \bT^\subl = \frac{\Xi}{2},
\end{align}
which yields
\begin{align}
\bT^\subl=\bT-\frac{\Xi}{2}\frac{\calN^\subu}{N}, \quad
\bT^\subu=\bT+\frac{\Xi}{2}\frac{\calN^\subl}{N}.
\label{e:bT-Tlu}
\end{align}
Applying \eqref{e:bT-Tlu}, we recognize $\calF_g$ in \eqref{e:Fg-var-neq} as a function of $(\calN^\subl,\calV^\subl;\bT,V,N,mgL,\Xi)$.
Substituting \eqref{e:bT-Tlu} into \eqref{e:Fg-var-neq} and
expanding $F^\subl$ and $F^\subu$ around $\bT$ to first order in $\ep$, we obtain
\begin{align}
\calF_g(\calN^\subl,\calV^\subl;\bT,V,N,mgL,\Xi)
&= F^\subl(\bT,\calV^\subl,\calN^\subl)
\!+\!F^\subu(\bT,V-\calV^\subl,N-\calN^\subl) \nm
&\quad
+\frac{mgL}{2}N
\left(\frac{\calV^\subl}{V}-\frac{\calN^\subl}{N}\right)
+\frac{\Xi}{2}\frac{\calN^\subl \calN^\subu}{N}
\left(\frac{\mathcal{S}^\subl}{\calN^\subl}
-\frac{\mathcal{S}^\subu}{\calN^\subu}\right),
\label{e:Fg-var-neq-re1}
\end{align}
with an error of $O(\ep^2)$,
where $\mathcal{S}^\subl=S(\bT,\calV^\subl,\calN^\subl)$ and $\mathcal{S}^\subu=S(\bT,V-\calV^\subl,N-\calN^\subl)$.
Equation~\eqref{e:Fg-var-neq-re1} clarifies that $\calF_g$ is a function of $\calN^\subl$ and $\calV^\subl$ with other parameters $(\bT,V,N,mgL,\Xi)$.

\subsection{Variational principle in heat conduction under weak gravity}
\label{s:var-neq}

The steady state corresponds to the free-energy minimum state,
\begin{align}
(N^\subl, V^\subl)=\argmin_{\mathcal{N}^\subl ,\mathcal{V}^\subl}\calF_g(\calN^\subl,\calV^\subl;\bT,V,N,mgL,\Xi).
\label{e:ss-min}
\end{align}
Then, the required condition for the steady state is given by the variational equations
\begin{align}
\pderf{\calF_g}{\calV^\subl}{*} &= 0, \label{e:varV-neq} \\
\pderf{\calF_g}{\calN^\subl}{*} &= 0. \label{e:varN-neq}
\end{align}
Here, the asterisk $(*)$ denotes the derivative evaluated at the steady state solution $(N^\subl, V^\subl)$ in fixed $(\bT,V,N,mgL,\Xi)$.

We evaluate \eqref{e:varV-neq} and \eqref{e:varN-neq} using the functional form \eqref{e:Fg-var-neq}, rather than \eqref{e:Fg-var-neq-re1},
because \eqref{e:Fg-var-neq} is directly connected to the thermodynamic quantities in the respective regions such as
 $\bP^\subl=p^\subl(\bT^\subl,V^\subl,N^\subl)$, $\bP^\subu=p^\subu(\bT^\subu,V^\subu,N^\subu)$,
 $\bmu_g^\subl=\mu^\subl(\bT^\subl,\bP^\subl)$, and $\bmu_g^\subu=\mu^\subu(\bT^\subu,\bP^\subu)$.
Substituting \eqref{e:Fg-var-neq} into \eqref{e:varV-neq} and \eqref{e:varN-neq}, and applying the fundamental relations  \eqref{e:fund_rel_l} and \eqref{e:fund_rel_u}, we have
\begin{align}
&\bP^\subl-\bP^\subu= \frac{mg N}{2A}, \label{e:varV-neq-1}\\
&\bmu_g^\subl-\bmu_g^\subu-(S^\subl+S^\subu)\frac{\Xi}{2N}= \frac{mgL}{2}, \label{e:varN-neq-1}
\end{align}
where $S^\subl=S(\bT^\subl, \calV^\subl,\calN^\subl)$ and
$S^\subu=S(\bT^\subu, \calV^\subu,\calN^\subu)$ appear
because $\bT^\subl$ and $\bT^\subu$ are functions of $\mathcal{N}^\subl$ but not of $\mathcal{V}^\subl$.

Note that \eqref{e:varV-neq-1} is the same as the equilibrium relation \eqref{e:var-eq-1}.
This implies the equality of endpoint pressures at the formal dividing position,
\begin{align}
p_\theta=p_+=p_-.
\end{align}
The relation \eqref{e:varN-neq-1} appears to differ from the equilibrium relation in \eqref{e:var-eq-1}.
To relate \eqref{e:varN-neq-1} to the chemical potential balance, we consider a first-order expansion
$\mu(T+\Delta T, p+\Delta p)=\mu(T,p)-s \Delta T+v\Delta p$ using the thermodynamic relations $\pderf{\mu}{T}{p}=-s$ and $\pderf{\mu}{p}{T}=v$, where $s$ and $v$ are the entropy and volume per particle.
Using this expansion, we obtain
\begin{align}
&
\mu^\subl(\bT^\subl,\bP^\subl)
=\mu^\subl\left(\bT^\subl+\frac{\Xi}{2}\frac{N^\subl}{N},p_\theta\right)+\frac{\Xi}{2N}S^\subl+\frac{mg l}{2}+O(\ep^2),
\label{e:mu-expand-l}\\
&
\mu^\subu(\bT^\subu,\bP^\subu)
=\mu^\subu\left(\bT^\subu-\frac{\Xi}{2}\frac{N^\subu}{N},p_\theta\right)-\frac{\Xi}{2N}S^\subu-\frac{mg(L-l)}{2}+O(\ep^2),
\label{e:mu-expand-u}
\end{align}
where we applied \eqref{e:p-lu}, $s=S/N$, $\bar v=V/N$, $\bP^\subl-\pint=N^\subl mgL/(2V)$, and $\bP^\subu-\pint=-N^\subu mgL/(2V)$.
Substituting \eqref{e:mu-expand-l} and \eqref{e:mu-expand-u} into \eqref{e:varN-neq-1}, we obtain
the following balance relation
\begin{align}
\mu^\subl\left(\bT^\subl+\frac{\Xi}{2}\frac{N^\subl}{N},p_\theta\right)
=\mu^\subu\left(\bT^\subu-\frac{\Xi}{2}\frac{N^\subu}{N},p_\theta\right).
\label{e:varV-neq-2}
\end{align}
Setting
\begin{align}
T_{\mathrm{s}} \equiv \bT + \frac{\Xi}{2}\frac{N^\subl - N^\subu}{N},
\label{e:Ts}
\end{align}
and applying the expressions of $\bT^\subl$ and $\bT^\subu$ in \eqref{e:bT-Tlu},
\eqref{e:varV-neq-2} is transformed into
\begin{align}
\mu^\subl(T_{\mathrm{s}},p_\theta)=\mu^\subu(T_{\mathrm{s}},p_\theta).
\label{e:mu-continuous-neq}
\end{align}
Equation~\eqref{e:mu-continuous-neq} should be interpreted according to the configuration.
For the liquid-gas configurations $(\subl,\subu)=(\subL,\subG)$ and $(\subG,\subL)$, it gives the coexistence condition at the auxiliary temperature $T_{\mathrm{s}}$,
\begin{align}
T_{\mathrm{s}}=\Tc(p_\theta).
\end{align}
The relation between $T_{\mathrm{s}}$ and the local temperature $\theta=T(l)$ at the formal dividing position will be made explicit in \eqref{e:theta-J}, where the liquid-gas stationary state is considered.
For the same-phase configurations $(\subL,\subL)$ and $(\subG,\subG)$, the same equation is an identity between the same thermodynamic functions and is not a coexistence condition.
Thus no saturation temperature is assigned in the same-phase configurations.

\section{Decomposition into effective-gravity and residual contributions}
\label{s:variational-function-revisited}

Using
\begin{align}
\frac{\calN^\subl \calN^\subu}{N} \left(\frac{\calV^\subl}{\calN^\subl} - \frac{\calV^\subu}{\calN^\subu}\right)
 &= V \left(\frac{\calV^\subl}{V} - \frac{\calN^\subl}{N}\right),
 \label{e:NlNu-relation}
\end{align}
which follows from $\calV^\subu=V-\calV^\subl$ and $\calN^\subu=N-\calN^\subl$,
the variational free energy \eqref{e:Fg-var-neq-re1} is transformed as
\begin{align}
&\calF_g(\calN^\subl,\calV^\subl;\bT,V,N,mgL,\Xi)
= F^\subl(\bT,\calV^\subl,\calN^\subl)
\!+\!F^\subu(\bT,V-\calV^\subl,N-\calN^\subl) \nm
&\qquad +\frac{mgL}{2}N
\left(\frac{\calV^\subl}{V}-\frac{\calN^\subl}{N}\right)
+\frac{\Xi}{2}V
\left(\frac{\calV^\subl}{V}-\frac{\calN^\subl}{N}\right)
\frac{\mathcal{S}^\subl/\calN^\subl-\mathcal{S}^\subu/\calN^\subu}
{\calV^\subl/\calN^\subl-\calV^\subu/\calN^\subu}
\label{e:Fg-var-neq-re1-2}
\end{align}

Regardless of either $(\subl,\subu)=(\subL,\subG)$ or $(\subG,\subL)$, the Clausius--Clapeyron relation is written as
\begin{align}
\der{\Ps}{T}
=
\frac{{S_0^\subl}/{N_0^\subl}-{S_0^\subu}/{N_0^\subu}}{
{V_0^\subl}/{N_0^\subl}-{V_0^\subu}/{N_0^\subu}},
\label{e:CC-relation}
\end{align}
where $(N_0^\subl,V_0^\subl)$ is given by either \eqref{e:NV0-L} or \eqref{e:NV0-G}
according to $(\subl,\subu)=(\subL,\subG)$ or $(\subG,\subL)$.
$S_0^\subl=S(\bT,V_0^\subl,N_0^\subl)$, and $S_0^\subu=S(\bT,V-V_0^\subl,N-N_0^\subl)$.
We separate the last term in \eqref{e:Fg-var-neq-re1-2} by adding and subtracting the Clausius--Clapeyron contribution:
\begin{align}
\frac{\Xi}{2}V\left(\frac{\calV^\subl}{V}-\frac{\calN^\subl}{N}\right)
\left(
\frac{{\mathcal{S}^\subl}/{\calN^\subl}-{\mathcal{S}^\subu}/{\calN^\subu}}
{{\calV^\subl}/{\calN^\subl}-{\calV^\subu}/{\calN^\subu}}-\der{\Ps}{T}
\right)
+\frac{\Xi}{2}V\left(\frac{\calV^\subl}{V}-\frac{\calN^\subl}{N}\right)\der{\Ps}{T}.
\end{align}
Applying \eqref{e:NlNu-relation} to the first term, this is further transformed as
\begin{align}
\frac{\Xi}{2}
\frac{\calN^\subl \calN^\subu}{N} \left[
\left(
\frac{\mathcal{S}^\subl}{\calN^\subl}-\frac{\mathcal{S}^\subu}{\calN^\subu}
\right)
-\der{\Ps}{T}\left(\frac{\calV^\subl}{\calN^\subl} - \frac{\calV^\subu}{\calN^\subu}\right)\right]
+\frac{\Xi}{2}V\left(\frac{\calV^\subl}{V}-\frac{\calN^\subl}{N}\right)\der{\Ps}{T}.
\label{e:qex-pre}
\end{align}
Defining an excess latent heat as
\begin{align}
\hat{q}^\ex(\bT,\calV^\subl,\calN^\subl)
\equiv
\bT
\left(\frac{\mathcal{S}^\subu}{\calN^\subu}-\frac{\mathcal{S}^\subl}{\calN^\subl}\right)
-
\bT \der{\Ps}{T}
\left(\frac{{\calV^\subu}}{{\calN^\subu}}-\frac{{\calV^\subl}}{{\calN^\subl}}
\right),
\label{e:qex-def}
\end{align}
and replacing the last term of \eqref{e:Fg-var-neq-re1-2} with \eqref{e:qex-pre}, \eqref{e:Fg-var-neq-re1-2} is written as
\begin{align}
&\calF_g(\calN^\subl,\calV^\subl;\bT,V,N,mgL,\Xi)
= F^\subl(\bT,\calV^\subl,\calN^\subl)
\!+\!F^\subu(\bT,V-\calV^\subl,N-\calN^\subl) \nm
&\quad
+\frac{mgL}{2}N
\left(\frac{\calV^\subl}{V}-\frac{\calN^\subl}{N}\right)
+\frac{\Xi}{2}V\der{\Ps}{T}
\left(\frac{\calV^\subl}{V}-\frac{\calN^\subl}{N}\right)
-\frac{\Xi}{\bT}\frac{\calN^\subl \calN^\subu}{2N}
\hat{q}^\ex(\bT,\calV^\subl,\calN^\subl).
\label{e:Fg-var-neq-re1-3}
\end{align}

We now define an effective gravity that incorporates the effect of the temperature gradient:
\begin{align}
mg_{\eff} \equiv mg + \bar{v}\frac{d\Ps}{d\bT}\frac{\Xi}{L}.
\label{e:g-eff}
\end{align}
Using $g_\eff$, the variational free energy \eqref{e:Fg-var-neq-re1-3} can be written as
\begin{align}
&\calF_g(\calN^\subl,\calV^\subl;~\bT,V,N,mgL,\Xi)=
\calF_{g,\eff}(\calN^\subl,\calV^\subl;~\bT,V,N,mg_\eff L) 
+\calF_{g,\res}(\calN^\subl,\calV^\subl;~\bT,V,N,\Xi),
\label{e:Fg-var-neq-re2}
\end{align}
where the two contributions are defined by
\begin{align}
&\calF_{g,\eff}(\calN^\subl,\calV^\subl;~\bT,V,N,mg_\eff L)
\equiv F^\subl(\bT,\calV^\subl,\calN^\subl)
+F^\subu(\bT,V-\calV^\subl,N-\calN^\subl)
+\frac{Nmg_\eff L}{2}\left(\frac{\calV^\subl}{V}-\frac{\calN^\subl}{N}\right),
\label{e:Fg-eff-def}
\\
&\calF_{g,\res}(\calN^\subl,\calV^\subl;~\bT,V,N,\Xi)
\equiv -\frac{\Xi}{\bT}\frac{\calN^\subl \calN^\subu}{2N}\hat q^\ex(\bT,\calV^\subl,\calN^\subl).
\label{e:Fg-res-def}
\end{align}
Thus, \eqref{e:Fg-var-neq-re2} rewrites the variational free energy as the sum of the two terms defined in \eqref{e:Fg-eff-def} and \eqref{e:Fg-res-def}.
The first has the same configurational form as the equilibrium variational free energy in \eqref{e:Fg-var-eq}, with $g$ replaced by $g_\eff$.
The second term is the genuinely nonequilibrium contribution: it is proportional to the excess latent heat and vanishes at $\Xi=0$.

\subsection{Effective gravity as the macroscopic representative of the gravity-like component}

We now examine $\calF_{g,\eff}$, in which the temperature-difference contribution enters only through the combination $g_\eff$ defined in \eqref{e:g-eff}.

To make the meaning of $g_\eff$ explicit, we introduce the mechanical pressure drop and the saturation-pressure drop
\begin{align}
\Delta P_{\mech}\equiv \botP-\topP, \qquad
\Delta P_{\sat}\equiv \Ps(T(0)) - \Ps(T(L)).
\label{e:pressure-drops}
\end{align}
Using the hydrostatic force balance $mgL=\bar{v}(\botP-\topP)$ and the linear-response relation $\Ps(T(L)) - \Ps(T(0))=(d\Ps/d\bT)\Xi$, \eqref{e:g-eff} can be rewritten as
\begin{align}
mg_{\eff}L
=
\bar{v}(\Delta P_{\mech}-\Delta P_{\sat}).
\label{e:g-eff-Ps}
\end{align}
Thus, the sign of $g_\eff$ compares two macroscopic drives: the mechanical pressure drop generated by gravity and the saturation-pressure drop generated by the temperature bias.
Therefore, the condition $g_\eff=0$ indicates the temperature-difference strength at which the entropic saturation-pressure effect balances the mechanical pressure drop.
Here, the saturation-pressure drop $\Delta P_{\sat}$  is entropic in origin because
\begin{align}
\Delta P_{\sat}=-\frac{d\Ps}{d\bT}\Xi
=
-\frac{s(\bT,v_\subC^\subG(\bT)) - s(\bT,v_\subC^\subL(\bT))}{v_\subC^\subG(\bT)-v_\subC^\subL(\bT)}
\Xi
\end{align}
where the last equality follows from the Clausius--Clapeyron relation \eqref{e:CC-relation}.

By contrast, the residual contribution $\calF_{g,\res}$ is not reduced to $g_\eff$.
It is proportional to $\hat q^\ex$ and is responsible for nonequilibrium effects that survive after the gravity-like reduction, as will be discussed in the later sections.

\section{Steady state solutions governed by the effective-gravity component}
\label{s:sol-var-eq}

The nonequilibrium steady state corresponds to the state that minimizes the variational free energy $\calF_g$.
In this section, we identify the minimizing solution for given $(mgL,\Xi)$.
We first examine the solutions corresponding to liquid--gas coexistence and then compare them with the solutions corresponding to single-phase states.
These solutions are obtained from the variational equations \eqref{e:varV-neq} and \eqref{e:varN-neq}, which ultimately lead to the condition \eqref{e:mu-continuous-neq}.
For brevity, we call the former ``\textit{heterogeneous solutions}", where the phase assignment is $(\subl,\subu)=(\subL,\subG)$ or $(\subG,\subL)$, and the latter ``\textit{homogeneous solutions}", where the system remains locally in a single phase.
When necessary, we distinguish the liquid-like and gas-like homogeneous realizations by $(\subl,\subu)=(\subL,\subL)$ and $(\subG,\subG)$, respectively.

Below, $f(T,v)$ is the free energy per particle at $g=0$ and $\Xi=0$, which is obtained from the equation of state $p=p(T,v)$ and the specific heat $c_v=c_v(T,v)$. Note that $f(T,v)$  corresponds to the free energy for single-phase states, and, in principle, it can be determined regardless of whether the states are thermodynamically stable, metastable, or unstable.
At $g=0$ and $\Xi=0$, the global minimum state is the liquid-gas coexistence if $v$ is set as $v_\subC^\subL(T)<v<v_\subC^\subG(T)$.
This is the heterogeneous solution, and its free energy per particle is written as
\begin{align}
&f^\hetero(T,v) \equiv \frac{N_0^\subL}{N}f(T,v_\subC^\subL(T)) + \frac{N_0^\subG}{N}f(T,v_\subC^\subG(T)),
\label{e:F-hetero}
\end{align}
where $N_0^\subL$ and $V_0^\subL$ are determined by the lever rule in \eqref{e:NV0-L} and \eqref{e:NV0-G}, with $N_0^\subG=N-N_0^\subL$ and $V_0^\subG=V-V_0^\subL$.

\subsection{\texorpdfstring{Liquid-gas coexistence as heterogeneous solutions: $(\subl,\subu)=(\subL,\subG)$ or $(\subG,\subL)$}{Liquid-gas coexistence as heterogeneous solutions}}

We now evaluate the two liquid--gas coexistence configurations.
At the present order, they provide the clearest setting for identifying the global ordering governed by the effective-gravity component.
There are two configurations in the heterogeneous solution, one is $(\subl,\subu)=(\subL,\subG)$, and the other is $(\subl,\subu)=(\subG,\subL)$.
We distinguish these two heterogeneous solutions as $(N^\subl_{(\subL,\subG)}, V^\subl_{(\subL,\subG)})$ and $(N^\subl_{(\subG,\subL)}, V^\subl_{(\subG,\subL)})$,
which are expressed as
\begin{align}
\frac{V^\subl_{(\subL,\subG)}}{N^\subl_{(\subL,\subG)}} = v_\subC^\subL(T) + O(\ep), \qquad
\frac{V^\subl_{(\subG,\subL)}}{N^\subl_{(\subG,\subL)}} = v_\subC^\subG(T) + O(\ep).
\label{e:hetero}
\end{align}
Using $N_0^\subL$, $V_0^\subL$, $N_0^\subG$, $V_0^\subG$ in \eqref{e:NV0-L} and \eqref{e:NV0-G},
one of the heterogeneous solutions in $(\subl,\subu) = (\subL,\subG)$ is given as
\begin{align}
V^\subl_{(\subL,\subG)} = V_0^\subL + \Delta V, \quad
N^\subl_{(\subL,\subG)} = N_0^\subL + \Delta N,
\label{e:hetero-LG}
\end{align}
while the other in $(\subl,\subu) = (\subG,\subL)$ is
\begin{align}
V^\subl_{(\subG,\subL)} = V_0^\subG + \Delta V, \quad
N^\subl_{(\subG,\subL)} = N_0^\subG + \Delta N,
\label{e:hetero-GL}
\end{align}
where the shifts $\Delta V$ and $\Delta N$ are of $O(\ep)$ and account for the effects of gravity and the temperature difference.
$\Delta V$ and $\Delta N$ will be analyzed in Sec.~\ref{sec:steady_states}.

According to \eqref{e:qex-def}, the excess latent heat at $(V^\subl_{(\subL,\subG)},N^\subl_{(\subL,\subG)})$ is estimated as
\begin{align}
\hat{q}^\ex(\bT,V^\subl_{(\subL,\subG)},N^\subl_{(\subL,\subG)})
&=
\bT
\left(\frac{S_0^\subG}{N_0^\subG}-\frac{S_0^\subL}{N_0^\subL}\right)
-
\bT \der{\Ps}{T}
\left(\frac{V^\subG}{N^\subG}-\frac{V^\subL}{N^\subL}
\right)+O(\ep),\nm
&=O(\ep)
\label{e:qex-hetero-LG-order}
\end{align}
by applying \eqref{e:CC-relation}.
Similarly, for $(V^\subl_{(\subG,\subL)},N^\subl_{(\subG,\subL)})$, we have
\begin{align}
\hat{q}^\ex(\bT,V^\subl_{(\subG,\subL)},N^\subl_{(\subG,\subL)})
&=
\bT
\left(\frac{S_0^\subL}{N_0^\subL}-\frac{S_0^\subG}{N_0^\subG}\right)
-
\bT \der{\Ps}{T}
\left(\frac{V^\subL}{N^\subL}-\frac{V^\subG}{N^\subG}
\right)+O(\ep),\nm
&=O(\ep).
\label{e:qex-hetero-GL-order}
\end{align}
Thus, we conclude
\begin{align}
\calF_{g,\res}=O(\ep^2),
\end{align}
for both heterogeneous solutions, $(\subL,\subG)$ and $(\subG,\subL)$.

In other words, $\calF_{g,\res}$ does not affect their global ordering at the present order and
the stationary values of $\calF_g$ are determined by $\calF_{g,\eff}$.
Concretely, we write the free energies for the two heterogeneous solutions as
\begin{align}
&F_g^{(\subL,\subG)}\equiv \calF_{g,\eff}(N^\subl_{(\subL,\subG)}, V^\subl_{(\subL,\subG)};~\bT,V,N,mg_\eff L)+O(\ep^2),
\label{e:Fg-LG-def} \\
&F_g^{(\subG,\subL)}\equiv \calF_{g,\eff}(N^\subl_{(\subG,\subL)}, V^\subl_{(\subG,\subL)};~\bT,V,N,mg_\eff L)+O(\ep^2).
\label{e:Fg-GL-def}
\end{align}

Substituting \eqref{e:hetero-LG} and \eqref{e:hetero-GL} into the first two terms of \eqref{e:Fg-eff-def} and then expanding this, we have
\begin{align}
&F(\bT,V_0^\subL+\Delta V,N_0^\subL+\Delta N)+F(\bT,V_0^\subG-\Delta V,N_0^\subG-\Delta N)\nm
&=F(\bT,V_0^\subL,N_0^\subL)+F(\bT,V_0^\subG,N_0^\subG)-(\bP^\subL-\bP^\subG)\Delta V+(\mu^\subL-\mu^\subG)\Delta N+O(\ep^2)\nm
&=F(\bT,V_0^\subL,N_0^\subL)+F(\bT,V_0^\subG,N_0^\subG)+O(\ep^2)\nm
&=N f^\hetero(\bT,\bar{v})+O(\ep^2).
\label{e:Fg-hetero-common}
\end{align}
Here we applied the equilibrium balance conditions at the reference state ($g=0$ and $\Xi=0$), i.e., $p(\bT,v_\subC^\subL(\bT))=p(\bT,v_\subC^\subG(\bT))=\Ps(\bT)$ and $\mu(\bT,v_\subC^\subL(\bT))=\mu(\bT,v_\subC^\subG(\bT))$.

For the third term of \eqref{e:Fg-eff-def}, define
\begin{align}
&\psi_g^{(\subL, \subG)}\equiv \frac{1}{2}\left(\frac{{N}_0^\subL}{N}-\frac{{V}_0^\subL}{V}\right).
\label{e:Psig-LG0}
\end{align}
Here, we note that $\psi_g^{(\subL, \subG)}$ is positive definite for $v_\subC^\subL(\bT)<\bar{v}<v_\subC^\subG(\bT)$,
because substituting \eqref{e:NV0-L} into \eqref{e:Psig-LG0} leads to
\begin{align}
\psi_g^{(\subL,\subG)}=\frac{1}{2}\frac{(v^\subG_\subC(\bT)-\bar{v})(\bar{v}-v^\subL_\subC(\bT))}{\bar{v}(v^\subG_\subC(\bT)-v^\subL_\subC(\bT))}>0.
\label{e:Psig-LG}
\end{align}
The free energies for the two heterogeneous solutions are written as
\begin{align}
&F_g^{(\subL,\subG)}=N\left[f^\hetero(\bT,\bar{v})-mg_\eff L\psi_g^{(\subL, \subG)}\right]+O(\ep^2),
\label{e:Fg-LG} \\
&F_g^{(\subG,\subL)}=N\left[f^\hetero(\bT,\bar{v})+mg_\eff L\psi_g^{(\subL, \subG)}\right]+O(\ep^2).
\label{e:Fg-GL}
\end{align}
Their difference is
\begin{align}
F_g^{(\subG, \subL)}-F_g^{(\subL, \subG)}
=2Nmg_\eff L\psi_g^{(\subL, \subG)}+O(\ep^2),
\label{e:Fg-diff-twomin}
\end{align}
where $\psi_g^{(\subL,\subG)}>0$.
Thus, the sign of $g_\eff$ determines which of the two heterogeneous solutions has the lower value of $F_g$,
\begin{align}
&F_g^{(\subL, \subG)}<F_g^{(\subG, \subL)}, \qquad  (g_\eff >0), \\
&F_g^{(\subL, \subG)}=F_g^{(\subG, \subL)}, \qquad  (g_\eff=0), \\
&F_g^{(\subG, \subL)}<F_g^{(\subL, \subG)}, \qquad  (g_\eff <0).
\end{align}

\subsection{Heat conducting liquid-gas coexistence satisfying the variational principle}

Return to the variational equation \eqref{e:mu-continuous-neq}.
For the liquid-gas coexistence, i.e. the heterogeneous solutions $(\subl, \subu) = (\subL, \subG)$ or $(\subG, \subL)$,
\eqref{e:mu-continuous-neq} holds only when
\begin{align}
 T_{\mathrm{s}}=\Tc(p_\theta),
 \label{e:Ts-Tc}
\end{align}
where $\Tc(p_\theta)$ is the transition temperature at $p_\theta$ satisfying \eqref{e:mu-continuous-neq}.
Using \eqref{e:Ts}, we conclude that
\begin{align}
&\bT^\subl=\Tc(p_\theta)-\frac{\Xi}{2}\frac{N^\subl}{N},\quad
\bT^\subu=\Tc(p_\theta)+\frac{\Xi}{2}\frac{N^\subu}{N},\label{e:Tlu-st}\\
&\bT=\Tc(p_\theta)+\frac{\Xi}{2}\frac{N^\subu-N^\subl}{N}.\label{e:bT-st}
\end{align}
The last relation can be rewritten as an equation for $p_\theta$:
\begin{align}
p_\theta=\Ps\left(\bT-\frac{\Xi}{2}\frac{N^\subu-N^\subl}{N}\right).
\label{e:p_theta-neq}
\end{align}
These formulas are the same as those derived in \cite{NS17,NS19} for the heat conduction systems without gravity.
For a heterogeneous stationary state, the formal dividing position coincides with the actual interface position, namely $l=x_\theta$.
The actual interface temperature is formulated as
\begin{align}
\theta=\Tc(p_{\theta})+J\bar v\left(\frac{\rho_\subC^\subl}{\kappa^\subu}-\frac{\rho_\subC^\subu}{\kappa^\subl}\right)\frac{x_\theta(L-x_\theta)}{2L},
\label{e:theta-J}
\end{align}
which has been derived in \cite{NS17,NS19} by combining 
\eqref{e:Tlu-st} and a uniform heat-flux condition
\begin{align}
J=-\kappa^\subl\frac{\theta-T_1}{x_\theta}
=-\kappa^\subu\frac{T_2-\theta}{L-x_\theta}.
\end{align}
Here $\kappa^\subl$ and $\kappa^\subu$ are the heat conductivities in the lower and upper phases, and
$\rho_\subC^\subl$ and $\rho_\subC^\subu$ are the saturation number densities of the lower and upper phases at $\Tc(p_\theta)$, i.e., $\rho_\subC^\subl=1/v_\subC^\subl(\Tc(p_\theta))$ and $\rho_\subC^\subu=1/v_\subC^\subu(\Tc(p_\theta))$.
Because $\theta$ generally differs from $\Tc(p_\theta)$, the local chemical potential is discontinuous at the actual interface,
i.e., $\mu^\subl(x_\theta)\neq\mu^\subu(x_\theta)$. See also Fig.~\ref{fig:mu-neq}(b) in Sec.~\ref{s:chemical_potential} for a schematic reference.
As we will reveal in the subsequent stability analysis (Sec.~\ref{s:stability}) and the free-energy landscape (Sec.~\ref{sec:landscape}), this mismatch is not merely a local boundary effect.
Rather, it is the macroscopic signature of the residual nonequilibrium contribution $\calF_{g,\res}$ associated with heat flow.

\begin{table}[tb!]
\centering
\small
\begin{tabular}{|c|c|c|c|}
\hline
\textbf{Configuration} &
\textbf{Heat flow} &
\textbf{Lower value when} &
\textbf{Region near $x=x_\theta$} \\
\hline
$(\subL,\subG)$ & $T_1<T_2$ ($J<0$) & $g_\eff>0$ & supercooled gas \\
$(\subL,\subG)$ & $T_1>T_2$ ($J>0$) & $g_\eff>0$ & superheated liquid \\
$(\subG,\subL)$ & $T_1<T_2$ ($J<0$) & $g_\eff<0$ & superheated liquid \\
$(\subG,\subL)$ & $T_1>T_2$ ($J>0$) & $g_\eff<0$ & supercooled gas \\
\hline
\end{tabular}
\caption{Classification of the interfacial metastable pattern in heterogeneous stationary states for usual fluids satisfying $\rho_\subC^\subL \kappa^\subL > \rho_\subC^\subG \kappa^\subG$. The third column indicates which heterogeneous solution has the lower thermodynamic value away from the coexistence edges. At $g_\eff=0$, the two heterogeneous solutions are degenerate.}
\label{tab:interfacial-anomaly}
\end{table}

\begin{figure}[tb!]
\centering
\includegraphics[width=0.8\linewidth]{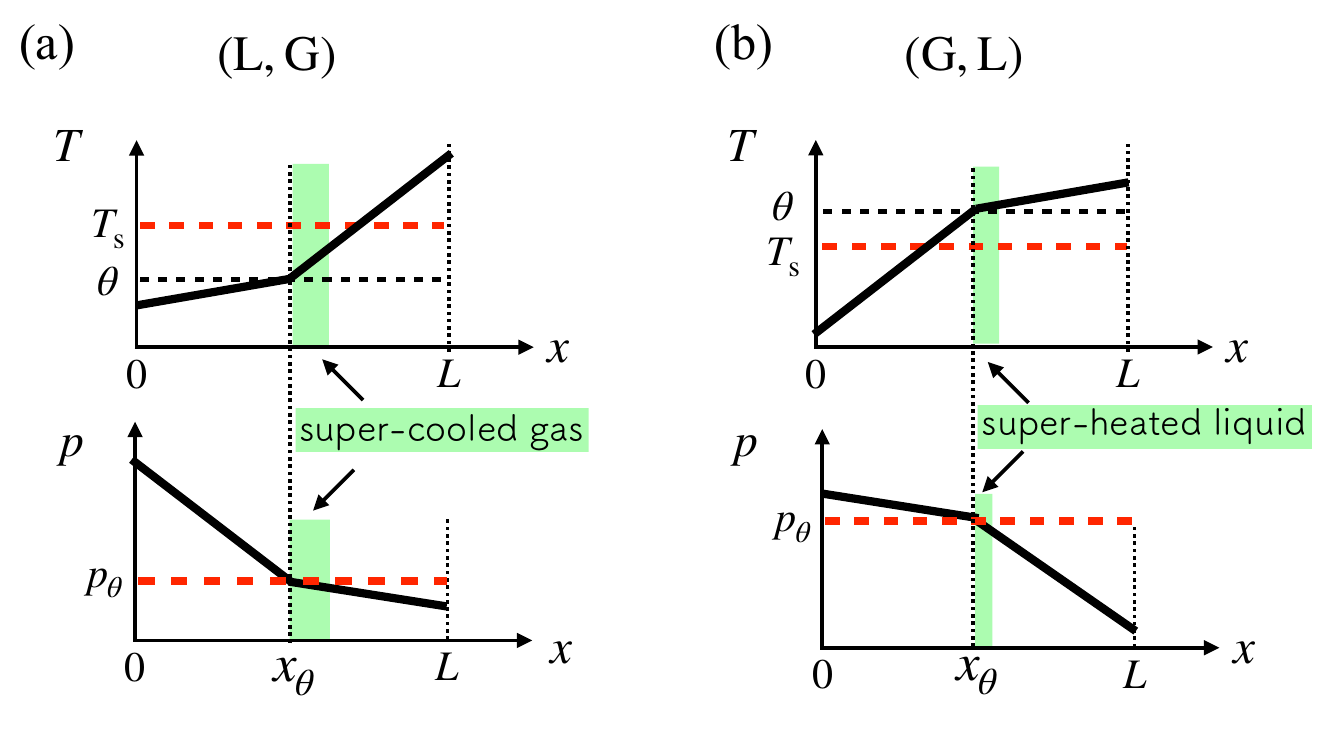}
\caption{Interfacial metastable patterns for $T_2>T_1$. In the lower-free-energy $(\subL,\subG)$ branch for $g_\eff>0$, $\theta<\Tc(p_\theta)$ and the gas adjacent to the interface is supercooled. In the lower-free-energy $(\subG,\subL)$ branch for $g_\eff<0$, $\theta>\Tc(p_\theta)$ and the liquid adjacent to the interface is superheated. The green shaded layers are schematic; reversing the heat-flow direction exchanges the two patterns.}
\label{fig:Ts-theta-anomaly}
\end{figure}

The sign of $\theta-\Tc(p_\theta)$ identifies the interfacial metastable pattern:
if $\theta<\Tc(p_\theta)$, the gas adjacent to the actual interface is supercooled, whereas
if $\theta>\Tc(p_\theta)$, the liquid adjacent to the actual interface is superheated.
For usual fluids satisfying
$\rho_\subC^\subL \kappa^\subL > \rho_\subC^\subG \kappa^\subG$,
Eq.~\eqref{e:theta-J} implies that $\theta-\Tc(p_\theta)$ has the same sign as $J$ for $(\subl,\subu)=(\subL,\subG)$ and the opposite sign for $(\subl,\subu)=(\subG,\subL)$.
With our sign convention, $J<0$ for $T_2>T_1$ and $J>0$ for $T_1>T_2$.
The resulting classification is shown in Table~\ref{tab:interfacial-anomaly}, where the third column also records which heterogeneous solution has the lower thermodynamic value away from the coexistence edges.
Equivalently, for the lower heterogeneous branch,
\begin{align}
\operatorname{sgn}\!\bigl(\theta-\Tc(p_\theta)\bigr)
=
-\operatorname{sgn}(\Xi g_\eff),
\end{align}
so that the interface-adjacent region is a superheated liquid for $\Xi g_\eff<0$ and a supercooled gas for $\Xi g_\eff>0$.

Figure~\ref{fig:Ts-theta-anomaly} shows the two lower-value heterogeneous configurations for the fixed heat-flow direction $T_2>T_1$.
For the $(\subL,\subG)$ branch, which has the lower value for $g_\eff>0$, the layer adjacent to the interface is a supercooled gas.
For the $(\subG,\subL)$ branch, which has the lower value for $g_\eff<0$, it is a superheated liquid.
The schematic chemical-potential profile in Fig.~\ref{fig:mu-neq}(b) in Sec.~\ref{s:chemical_potential} corresponds to the first case.

\subsection{Homogeneous solutions}

The homogeneous solutions are single-phase states without a liquid-gas interface.
For $v_\subC^\subL(\bT)<\bar{v}<v_\subC^\subG(\bT)$, the homogeneous states are either metastable or unstable.
For clarity, we conceptually divide the uniform state into $\subl$ and $\subu$,
and denote the number of particles and volumes in $\subl$ as $(N^\subl_\mathrm{homo}, V^\subl_\mathrm{homo})$.

When $g=0$ and $\Xi=0$, the homogeneous solution forms uniform profiles throughout the system.
We have
\begin{align}
\frac{V^\subl_\mathrm{homo}}{N^\subl_\mathrm{homo}} =
\frac{V^\subu_\mathrm{homo}}{N^\subu_\mathrm{homo}} =
\bar{v},
\label{e:homo0}
\end{align}
where $\bar{v} = V/N$ is the average specific volume.
$(N^\subl_\mathrm{homo}, V^\subl_\mathrm{homo})$ are unrelated to $(N_0^\subL, V_0^\subL)$ or $(N_0^\subG, V_0^\subG)$.

We extend the homogeneous solution to $\ep \neq 0$.
The profiles $p(x)$ and $\rho(x)$ in the homogeneous solution can have linear slopes, although the local phase remains the same throughout the system.
Thus, we have
\begin{align}
\frac{V^\subl_\mathrm{homo}}{N^\subl_\mathrm{homo}} = \bar{v} + O(\ep).
\label{e:homo}
\end{align}
Substituting this solution into $\calF_{g,\eff}$ in \eqref{e:Fg-eff-def}, the third term turns out to be $O(\ep^2)$ regardless of the value of $g_\eff$.
For the residual contribution $\calF_{g,\res}$, we use the definition of the excess latent heat $\hat{q}^\ex$ in \eqref{e:qex-def}.
The second term of \eqref{e:qex-def} is $O(\ep)$, and its first term is
\begin{align}
\frac{S(\bT,V^\subl_\mathrm{homo},N^\subl_\mathrm{homo})}{N^\subl_\mathrm{homo}}-\frac{S(\bT,V^\subu_\mathrm{homo},N^\subu_\mathrm{homo})}{N^\subu_\mathrm{homo}}
=O(\ep).
\end{align}
Thus, similarly to the heterogeneous solution, the homogeneous stationary value is controlled by $\calF_{g,\eff}$ to $O(\ep)$, whereas $\calF_{g,\res}$ contributes only beyond this order.
The free energy for the homogeneous solution is written as
\begin{align}
&F_g^\homo= F(\bT,V^\subl_\mathrm{homo}, N^\subl_\mathrm{homo})+F(\bT,V^\subu_\mathrm{homo}, N^\subu_\mathrm{homo}).
\label{e:Fg-homo-def}
\end{align}
Because any profile is linear in $x$ over the system, the trapezoidal rule results in
\begin{align}
F_g^\homo=N f(\bT,\bar{v})+O(\ep^2).
\end{align}

\subsection{Global minimum state}
\label{s:global-minimum-neq}

When $v_\subC^\subL(\bT)<\bar{v}<v_\subC^\subG(\bT)$, the second law of thermodynamics leads to
\begin{align}
f^\hetero(\bT,\bar{v})<f(\bT,\bar{v}).
\end{align}
Consequently, we find
\begin{align}
F_g^{(\subL, \subG)}-F_g^\homo
&=N\left[f^\hetero(\bT,\bar{v})-f(\bT,\bar{v})-mg_\eff L\psi_g^{(\subL, \subG)}\right]
<-Nmg_\eff L\psi_g^{(\subL, \subG)},
\label{e:Fg-diff-homo1} \\
F_g^{(\subG, \subL)}-F_g^\homo
&=N\left[f^\hetero(\bT,\bar{v})-f(\bT,\bar{v})+mg_\eff L\psi_g^{(\subL, \subG)}\right]
<Nmg_\eff L\psi_g^{(\subL, \subG)}.
\label{e:Fg-diff-homo2}
\end{align}
Because $\psi_g^{(\subL, \subG)}>0$, we have
\begin{align}
&F_g^{(\subL, \subG)}<F_g^{(\subG, \subL)}, \quad F_g^{(\subL, \subG)}<F_g^\homo \qquad (g_\eff >0), \\
&F_g^{(\subL, \subG)}=F_g^{(\subG, \subL)}<F_g^\homo \qquad \qquad \qquad ~(g_\eff =0), \\
&F_g^{(\subG, \subL)}<F_g^{(\subL, \subG)}, \quad F_g^{(\subG, \subL)}<F_g^\homo \qquad (g_\eff <0).
\end{align}
These inequalities give the primary ordering in liquid-gas coexistence.
The sign of $g_\eff$ determines which heterogeneous stationary state gives the lowest value among the stationary candidates.
According to \eqref{e:g-eff-Ps}, this means that the competition between the mechanical pressure drop $\Delta P_{\mech}$ and the thermally induced saturation-pressure drop $\Delta P_{\sat}$ defined in \eqref{e:pressure-drops} sets the ordering of the two-phase configurations.
The liquid floats above the gas when $\Delta P_{\mech}<\Delta P_{\sat}$, namely when the entropic effect inducing the saturation-pressure difference due to heat flow exceeds the mechanical pressure drop produced by gravity.

Because the thermodynamic free energy is
\begin{align}
F_g(\bT,V,N,mgL,\Xi)
=
\min_{\calN^\subl,\calV^\subl}
\calF_g(\calN^\subl,\calV^\subl;~\bT,V,N,mgL,\Xi),
\end{align}
the preceding inequalities show that, in the coexistence interval, it is obtained by taking the lower of the two heterogeneous stationary values \eqref{e:Fg-LG} and \eqref{e:Fg-GL}.
This gives
\begin{align}
F_g
=
Nf^\hetero(\bT,\bar v)
-N|mg_\eff L|\psi_g^{(\subL,\subG)}
+O(\ep^2).
\end{align}

\begin{figure}[bt]
\begin{center}
\includegraphics[width=0.85\linewidth]{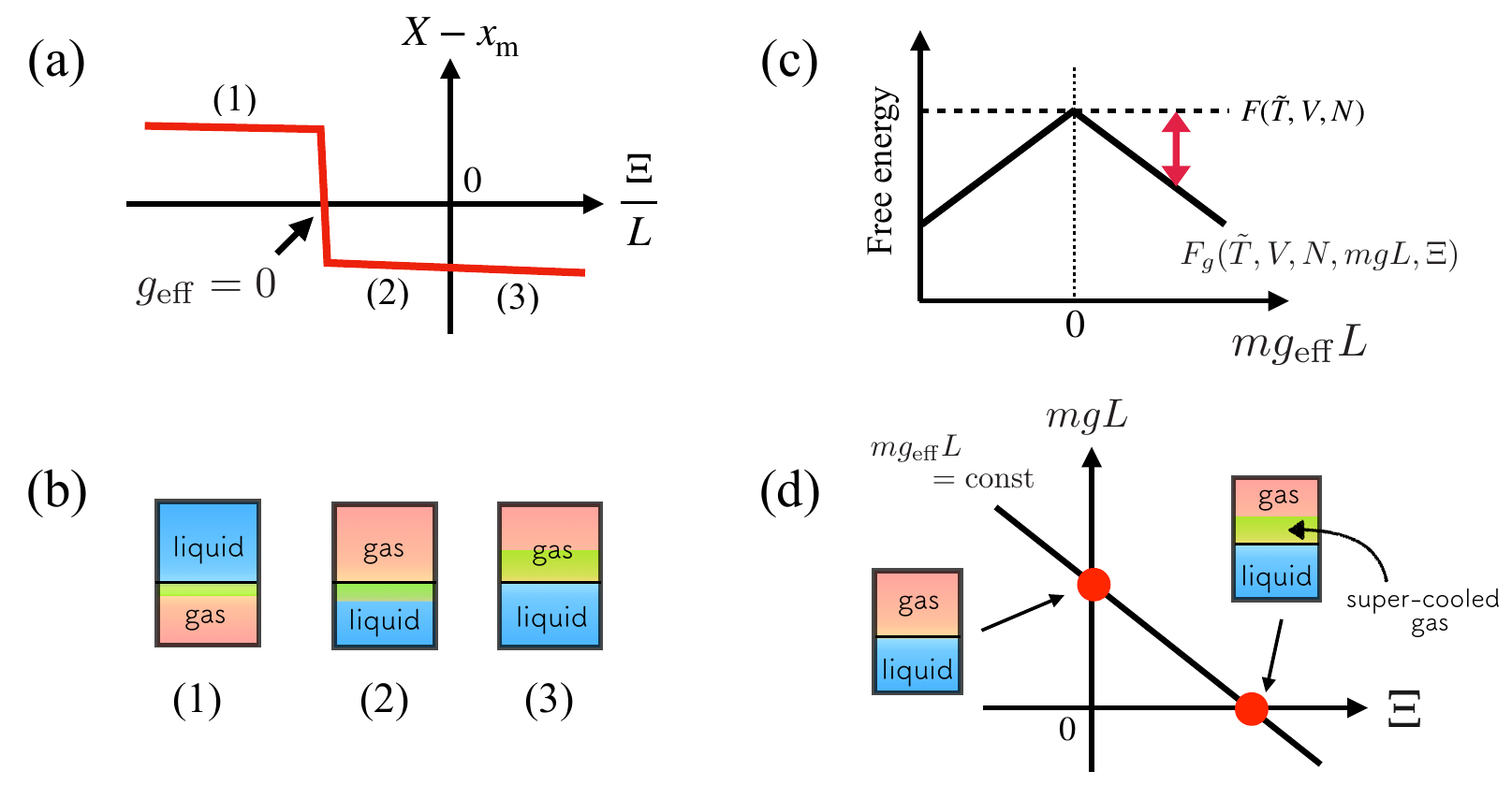}
\end{center}
\caption{Configurational transition reproduced and adapted from Figs.~2(a), 2(b), 3(a), and 3(b) of Ref.~\cite{NSgeff26} under the Creative Commons Attribution 4.0 International license. (a) Center-of-mass displacement under fixed gravity. (b) Representative configurations; light-green regions indicate interfacial metastable layers. (c) Cusp of $F_g$ at $g_\eff=0$. (d) Constant-$mg_\eff L$ lines in the $(mgL,\Xi)$ plane.}
\label{fig:Fg}
\end{figure}

Figure~\ref{fig:Fg} summarizes the configurational-transition scenario of Ref.~\cite{NSgeff26}, now expressed using the stationary values derived above and the equilibrium gravity formulation of Ref.~\cite{NSgravity25}.
Figure~\ref{fig:Fg}(a) plots the stationary center-of-mass displacement obtained by applying \eqref{e:xm-X-g} with $g$ replaced by $g_\eff$.
Figure~\ref{fig:Fg}(b) displays the corresponding phase configurations, using the ordering in \eqref{e:Fg-diff-twomin} and the interfacial metastable patterns classified by \eqref{e:theta-J} and Fig.~\ref{fig:Ts-theta-anomaly}.
The labels (1)--(3) mark the same representative states in panels (a) and (b): state (1) is the minimizing $(\subG,\subL)$ configuration, whereas states (2) and (3) are minimizing $(\subL,\subG)$ configurations.
The light-green layer represents a supercooled gas in states (1) and (3), and a superheated liquid in state (2).

Figure~\ref{fig:Fg}(c) plots this minimized thermodynamic free energy as a function of $mg_\eff L$, showing a cusp at $g_\eff=0$.
The dashed horizontal line, labeled $F(\bT,V,N)$, denotes the gravity-independent two-phase Helmholtz contribution $F^\subl+F^\subu$.
As shown in \eqref{e:Fg-hetero-common}, this contribution is $Nf^\hetero(\bT,\bar v)+O(\ep^2)$ for either heterogeneous configuration, independently of $mg_\eff L$.
The two inclined lines in Fig.~\ref{fig:Fg}(c) are obtained by adding the configurational terms $\mp Nmg_\eff L\psi_g^{(\subL,\subG)}$ to this common baseline, and their lower value gives $F_g$.
The one-sided slopes with respect to $mg_\eff L$ are $\mp N\psi_g^{(\subL,\subG)}$, and the cusp therefore corresponds to the jump of the center-of-mass displacement shown in Fig.~\ref{fig:Fg}(a).
Following the equilibrium gravity formulation in Ref.~\cite{NSgravity25} and its nonequilibrium extension in Ref.~\cite{NSgeff26}, we interpret this nonanalyticity as a first-order configurational transition in the sense of global thermodynamics.
The transition is not the ordinary liquid--gas phase transition between two bulk phases.
It means that, when $mg_\eff L$ changes sign, the lower value of $F_g$ switches between the two separated liquid--gas configurations, $(\subL,\subG)$ and $(\subG,\subL)$.
At the two heterogeneous stationary values, $\calF_{g,\res}=O(\ep^2)$, whereas the cusp term in \eqref{e:Fg-diff-twomin} is $O(\ep)$ when both phases occupy finite fractions. The residual contribution therefore does not remove this configurational transition in the present construction.

Figure~\ref{fig:Fg}(d) clarifies the scope of this correspondence.
Equations \eqref{e:Fg-LG} and \eqref{e:Fg-GL} have the same form as the equilibrium gravity result with the replacement $T\to\bT$ and $g\to g_\eff$.
Thus states on a line of constant $mg_\eff L$ have the same stationary value in this configurational comparison.
Along such a line, the representative states in Fig.~\ref{fig:Fg}(b) may be read as changing from state (2), through the equilibrium-gravity point with $\Xi=0$, to state (3).
They have the same macroscopic value of $mg_\eff L$, but they do not have the same local thermodynamic state.
Indeed, \eqref{e:theta-J} gives $\theta \neq \Tc(p_\theta)$ in heat conduction, so the local states near the interface differ from those in the corresponding equilibrium system under gravity.

\section{Phase redistribution driven by gravity and heat conduction}
\label{sec:steady_states}

In this section, we determine how gravity and heat conduction drive the redistribution of particle number and volume between the lower and upper regions.
Specifically, we calculate the shifts from the equilibrium two-phase reference state in the linear response regime.

Consider the shift $(\Delta N^\subl, \Delta V^\subl)$ in the lower region from the equilibrium values $(N_0^\subl, V_0^\subl)$ at $g=0$ and $\Xi=0$ given by either \eqref{e:NV0-L} or \eqref{e:NV0-G} according to $(\subl,\subu)=(\subL,\subG)$ or $(\subG,\subL)$.
Conservation of $N$ and $V$ yields $(\Delta N^\subu, \Delta V^\subu)=-(\Delta N^\subl, \Delta V^\subl)$.

\subsection{Response of distribution to gravity and heat flow}

We first formulate response relations.
We introduce the variational variable
\begin{align}
\vec{x} \equiv (c^\subl, \phi^\subl)^T,
\qquad c^\subl\equiv \frac{\calN^\subl}{N},
\qquad \phi^\subl\equiv\frac{\calV^\subl}{V}.
\end{align}
In the steady state at $(mg_\eff L,\Xi)$, this variable takes the value
\begin{align}
\vec{x}_*(mg_\eff L,\Xi) \equiv
\begin{pmatrix}
N^\subl/N \\
V^\subl/V
\end{pmatrix}.
\end{align}
The reference value at $g=0$ and $\Xi=0$ is
\begin{align}
\vec{x}_0 \equiv \vec{x}_*(0,0) =
\begin{pmatrix}
N_0^\subl/N \\
V_0^\subl/V
\end{pmatrix}.
\end{align}
We define the shift from the reference state by
\begin{align}
\Delta \vec{x} \equiv \vec{x}_* - \vec{x}_0 =
\begin{pmatrix}
\Delta N^\subl/N \\
\Delta V^\subl/V
\end{pmatrix}.
\end{align}
The steady-state shift is formulated as
\begin{align}
\Delta\vec{x}
=
\left.\pder{\vec{x}_*}{(mg_\eff L)}\right|_{mg_\eff L=0,\Xi=0} mg_\eff L
+
\left.\pder{\vec{x}_*}{\Xi}\right|_{mg_\eff L=0,\Xi=0}\Xi,
\end{align}
in the linear response regime.
We normalize the response coefficients by $v_\subC^\subG-v_\subC^\subL>0$.
The response coefficients are defined by
\begin{align}
\begin{pmatrix}
\eta_g \\
\zeta_g
\end{pmatrix}
&\equiv
-
\frac{v_\subC^\subG-v_\subC^\subL}{\bar v}
\left.\pder{\vec{x}_*}{(mg_\eff L)}\right|_{mg_\eff L=0,\Xi=0},
\nonumber\\
\begin{pmatrix}
\eta_\Xi \\
\zeta_\Xi
\end{pmatrix}
&\equiv
-
\frac{v_\subC^\subG-v_\subC^\subL}{\bar v}
\left.\pder{\vec{x}_*}{\Xi}\right|_{mg_\eff L=0,\Xi=0}.
\label{e:response-def}
\end{align}
With these coefficients, the shift from the reference state is obtained as
\begin{align}
\Delta\vec{x}
=
-\frac{\bar v}{v_\subC^\subG-v_\subC^\subL}
\left[
\begin{pmatrix}
\eta_g \\
\zeta_g
\end{pmatrix}
mg_\eff L
+
\begin{pmatrix}
\eta_\Xi \\
\zeta_\Xi
\end{pmatrix}
\Xi
\right].
\end{align}
The response coefficients separate according to the decomposition in \eqref{e:Fg-var-neq-re2}: $(\eta_g,\zeta_g)$ are generated by $\calF_{g,\eff}$, whereas $(\eta_\Xi,\zeta_\Xi)$ are generated by $\calF_{g,\res}$. Define
\begin{align}
H_0 \equiv \left.\nabla_{\vec{x}}\nabla_{\vec{x}} \mathcal{F}_g\right|_{\substack{\vec{x}=\vec{x}_0\\ mg_\eff L=0,\Xi=0}}.
\end{align}
Then the linear-response relations can be written as
\begin{align}
\begin{pmatrix} \eta_g \\ \zeta_g \end{pmatrix}
&=
\frac{v_\subC^\subG - v_\subC^\subL}{\bar{v}} H_0^{-1}
\left. \nabla_{\vec{x}}
\frac{\partial  \calF_{g,\eff} }{\partial (mg_\eff L)}
\right|_{\substack{\vec{x}=\vec{x}_0\\ mg_\eff L=0,\Xi=0}},
\\
\begin{pmatrix} \eta_\Xi \\ \zeta_\Xi \end{pmatrix}
&=
\frac{v_\subC^\subG - v_\subC^\subL}{\bar{v}} H_0^{-1}
\left. \nabla_{\vec{x}}
\frac{\partial \calF_{g,\res}}{\partial \Xi}
\right|_{\substack{\vec{x}=\vec{x}_0\\ mg_\eff L=0,\Xi=0}},
\end{align}
as will be derived in Sec.~\ref{s:response-Fg}.
Substituting the explicit forms of $\calF_{g,\eff}$ and $\calF_{g,\res}$, these are obtained as
\begin{align}
\begin{pmatrix} \eta_g \\ \zeta_g \end{pmatrix}
&=
\frac{N}{2}\frac{v_\subC^\subG - v_\subC^\subL}{\bar{v}}
H_0^{-1}
\begin{pmatrix}
-1\\
1
\end{pmatrix},
\label{e:response-shift-form1}\\
\begin{pmatrix} \eta_\Xi \\ \zeta_\Xi \end{pmatrix}
&=
-\frac{N}{2}\frac{v_\subC^\subG - v_\subC^\subL}{\bar{v}}
H_0^{-1}
\left.\nabla_{\vec{x}}
c^\subl(1-c^\subl)\frac{\hat{q}^\ex}{\bT}
\right|_{\substack{\vec{x}=\vec{x}_0\\ mg_\eff L=0,\Xi=0}}.
\label{e:response-shift-form2}
\end{align}

Carrying out the calculation described in Sec.~\ref{s:calc-DV-DN}, we obtain the explicit expressions for the response coefficients as
\begin{align}
\eta_g
&=
\frac{1}{2\bar{v}^2}
\left(
v_\subC^\subL(\bT)\alpha^\subL(c^\subL)^2
-
v_\subC^\subG(\bT)\alpha^\subG(c^\subG)^2
\right),
\label{e:response-g}\\
\zeta_g
&=
\frac{v_\subC^\subL(\bT)v_\subC^\subG(\bT)}{2\bar{v}^3}
\left(
\alpha^\subL(c^\subL)^2
-
\alpha^\subG(c^\subG)^2
\right),
\\
\eta_\Xi
&=
\frac{c^\subL c^\subG}{2\bar{v}}
\left(
\der{v_\subC^\subL}{T}
-
\der{v_\subC^\subG}{T}
\right),
\label{e:response-Xi}\\
\zeta_\Xi
&=
\frac{c^\subL c^\subG}{2\bar{v}^2}
\left(
v_\subC^\subG(\bT)\der{v_\subC^\subL}{T}
-
v_\subC^\subL(\bT)\der{v_\subC^\subG}{T}
\right),
\end{align}
where
\begin{align}
c^\subL \equiv \frac{N_0^\subL}{N},
\qquad
c^\subG \equiv \frac{N_0^\subG}{N}.
\end{align}
Here, $\alpha^\subL=\alpha(\bT,v_\subC^\subL(\bT))$ and $\alpha^\subG=\alpha(\bT,v_\subC^\subG(\bT))$
with the isothermal compressibility
\begin{align}
\alpha \equiv -\frac{1}{V} \pderf{V}{p}{\bT}.
\label{e:compressibility}
\end{align}

Note that the response coefficients in \eqref{e:response-g} and \eqref{e:response-Xi} are common to the two heterogeneous solutions $(\subl,\subu)=(\subL,\subG)$ and $(\subG,\subL)$.
This leads to the following symmetric response of the lower/upper-region redistribution.
We write the common lower-region shift as $(\Delta N,\Delta V)$, with
\begin{align}
\Delta N \equiv \Delta N^\subl,
\qquad
\Delta V \equiv \Delta V^\subl.
\label{e:Delta-common}
\end{align}
For the configuration $(\subl,\subu)=(\subL,\subG)$, we have
\begin{align}
&N^\subl_{(\subL,\subG)}=N_0^\subL + \Delta N, \qquad
V^\subl_{(\subL,\subG)}=V_0^\subL + \Delta V,
\nonumber\\
&N^\subu_{(\subL,\subG)}=N_0^\subG - \Delta N, \qquad
V^\subu_{(\subL,\subG)}=V_0^\subG - \Delta V,
\label{e:LG-NlVl}
\end{align}
whereas for $(\subl,\subu)=(\subG,\subL)$,
\begin{align}
&N^\subl_{(\subG,\subL)}=N_0^\subG + \Delta N, \qquad
V^\subl_{(\subG,\subL)}=V_0^\subG + \Delta V,
\nonumber\\
&N^\subu_{(\subG,\subL)}=N_0^\subL - \Delta N, \qquad
V^\subu_{(\subG,\subL)}=V_0^\subL - \Delta V.
\label{e:GL-NlVl}
\end{align}
Thus, the same pair $(\Delta N,\Delta V)$ characterizes the lower-region shift, with the upper region shifting complementarily, in both configurations.
What differs between the two solutions is only which phase occupies each region.

\subsection{Respective origin of the response coefficients}
\label{s:response-Fg}

Using the variational variable introduced above, we write the steady-state value as
\begin{align}
\vec{x}_*=\vec{x}_*(mg_\eff L,\Xi),
\label{e:x_*}
\end{align}
which is determined by the variational principle
\begin{align}
\left.\nabla_{\vec{x}} \mathcal{F}_g(\vec{x}; mg_\eff L, \Xi) \right|_{\vec{x}_*}= 0.
\label{e:Fg-var-nabla}
\end{align}

By differentiating \eqref{e:Fg-var-nabla} with respect to $mg_\eff L$ and $\Xi$, while noting the functional dependence of $\vec{x}_*$ in \eqref{e:x_*},
we obtain
\begin{align}
H_0 \left.\frac{\partial \vec{x}_*}{\partial (mg_\eff L)}\right|_{mg_\eff L=0,\Xi=0}
&= - \left.\frac{\partial \nabla_{\vec{x}} \calF_g}{\partial (mg_\eff L)}\right|_{\substack{\vec{x}=\vec{x}_0\\ mg_\eff L=0,\Xi=0}}, \\
H_0 \left.\frac{\partial \vec{x}_*}{\partial \Xi}\right|_{mg_\eff L=0,\Xi=0}
&= -\left. \frac{\partial  \nabla_{\vec{x}} \calF_g }{\partial \Xi}\right|_{\substack{\vec{x}=\vec{x}_0\\ mg_\eff L=0,\Xi=0}},
\end{align}
where $H_0$ is the Hessian matrix at $g=0$, $\Xi=0$, and $\vec{x}=\vec{x}_0$.
Recall $\calF_g=\calF_{g,\eff}+\calF_{g,\res}$
defined in \eqref{e:Fg-var-neq-re2}.
There, $\calF_{g,\res}$ is independent of $mg_\eff L$, whereas $\calF_{g,\eff}$ does not depend explicitly on $\Xi$ if we regard $g_\eff$ as an independent parameter of $\Xi$.
We then have
\begin{align}
H_0 \left.\frac{\partial \vec{x}_*}{\partial (mg_\eff L)}\right|_{mg_\eff L=0,\Xi=0}
&=- \left. \nabla_{\vec{x}}\frac{\partial  \calF_{g,\eff} }{\partial (mg_\eff L)}\right|_{\substack{\vec{x}=\vec{x}_0\\ mg_\eff L=0,\Xi=0}}, \\
H_0 \left.\frac{\partial \vec{x}_*}{\partial \Xi}\right|_{mg_\eff L=0,\Xi=0}
&=- \left. \nabla_{\vec{x}}\frac{\partial  \calF_{g,\res}}{\partial \Xi}\right|_{\substack{\vec{x}=\vec{x}_0\\ mg_\eff L=0,\Xi=0}}.
    \label{eq:matrix_response}
\end{align}
Solving \eqref{eq:matrix_response} and substituting into \eqref{e:response-def}, we have the formulas for the response coefficients,
\begin{align}
\begin{pmatrix} \eta_g \\ \zeta_g \end{pmatrix}
&=
\frac{v_\subC^\subG - v_\subC^\subL}{\bar{v}} H_0^{-1}
\left. \nabla_{\vec{x}}
\frac{\partial  \calF_{g,\eff} }{\partial (mg_\eff L)}
\right|_{\substack{\vec{x}=\vec{x}_0\\ mg_\eff L=0,\Xi=0}}, \nm
\begin{pmatrix} \eta_\Xi \\ \zeta_\Xi \end{pmatrix}
&=
\frac{v_\subC^\subG - v_\subC^\subL}{\bar{v}} H_0^{-1}
\left. \nabla_{\vec{x}}
\frac{\partial \calF_{g,\res}}{\partial \Xi}
\right|_{\substack{\vec{x}=\vec{x}_0\\ mg_\eff L=0,\Xi=0}}.
    \label{e:response-origin}
\end{align}
The formulas \eqref{e:response-origin} clarify that the response coefficients $(\eta_g,\zeta_g)$ quantify the response to the effective-gravity component, whereas $(\eta_\Xi,\zeta_\Xi)$ quantify the residual displacement that remains when $mg_\eff L$ is held fixed.
Substituting \eqref{e:Fg-eff-def} and \eqref{e:Fg-res-def} into \eqref{e:response-origin} gives the final expressions already displayed in \eqref{e:response-shift-form1} and \eqref{e:response-shift-form2}.

\subsection{\texorpdfstring{Explicit calculation of shifts $\Delta N$ and $\Delta V$}{Explicit calculation of shifts Delta N and Delta V}}
\label{s:calc-DV-DN}

Although \eqref{e:response-shift-form1} and \eqref{e:response-shift-form2} express the response coefficients through the inverse Hessian, it is more transparent to evaluate them by a direct perturbative expansion of the particle-number and volume constraints.
We now carry out this calculation.

Recalling the definitions of the specific volumes for the lower and upper regions, $v^\subl(\bT^\subl,\bP^\subl)$ and $v^\subu(\bT^\subu,\bP^\subu)$, we have $V^\subl=v^\subl(\bT^\subl,\bP^\subl)N^\subl$ and $V^\subu=v^\subu(\bT^\subu,\bP^\subu)N^\subu$.
From the conservation of particle number and volume, we have
\begin{align}
N^\subl+N^\subu=N,\quad v^\subl(\bT^\subl,\bP^\subl)N^\subl+v^\subu(\bT^\subu,\bP^\subu)N^\subu=\bar v N,
\end{align}
with $\bar v=V/N$.
Solving for $N^\subl$ and $N^\subu$, we obtain
\begin{align}
\frac{N^\subl}{N}=\frac{\bar v-v^\subu(\bT^\subu,\bP^\subu)}{v^\subl(\bT^\subl,\bP^\subl)-v^\subu(\bT^\subu,\bP^\subu)},\qquad
\frac{N^\subu}{N}=\frac{\bar v-v^\subl(\bT^\subl,\bP^\subl)}{v^\subu(\bT^\subu,\bP^\subu)-v^\subl(\bT^\subl,\bP^\subl)}.
\label{e:N-lu-neq}
\end{align}
We introduce the shift induced by $g$ and $\Xi$ as $\Delta N^\subl$ and $\Delta N^\subu$,
\begin{align}
\frac{\Delta N^\subl}{N}\equiv \frac{N^\subl}{N}-\frac{N_0^\subl}{N},\qquad
\frac{\Delta N^\subu}{N}\equiv \frac{N^\subu}{N}-\frac{N_0^\subu}{N},
\end{align}
with
\begin{align}
&\frac{N^\subl_0}{N}=\frac{\bar v-v^\subu_\subC(\bT)}{v^\subl_\subC(\bT)-v^\subu_\subC(\bT)},\qquad
\frac{N^\subu_0}{N}=\frac{\bar v-v^\subl_\subC(\bT)}{v^\subu_\subC(\bT)-v^\subl_\subC(\bT)}.
\label{e:N0-lu-neq}
\end{align}

We now focus on the linear response regime ($O(\ep)$).
We define the reference saturated state at $\ep=0$ using the global temperature $\bT$ as
\begin{align}
v^\subl_\subC(\bT)=v^\subl(\bT,\Ps(\bT)),\qquad
v^\subu_\subC(\bT)=v^\subu(\bT,\Ps(\bT)).
\end{align}
All derivatives in this paragraph are evaluated at the corresponding saturated reference state.
Expanding $v^\subl(\bT^\subl, \bP^\subl)$ around the reference state $(\bT, \Ps(\bT))$, we have
\begin{align}
v^\subl(\bT^\subl,\bP^\subl)&=v_\subC^\subl(\bT)+\pderf{v^\subl}{T}{p}(\bT^\subl-\bT)+\pderf{v^\subl}{p}{T}(\bP^\subl-\Ps(\bT))\nm
&=v_\subC^\subl(\bT)+\pderf{v^\subl}{p}{T}\left[(\bP^\subl-\Ps(\bT))-\pderf{p}{T}{v^\subl}(\bT^\subl-\bT)\right].
\label{e:vc-0}
\end{align}
Using the thermodynamic relation
\begin{align}
 \pderf{v^\subl}{T}{p} = -\pderf{v^\subl}{p}{T} \pderf{p}{T}{v^\subl},
\end{align}
and the chain rule with the definition of saturation pressure,
\begin{align}
\der{\Ps}{T}=\pderf{p}{T}{v^\subl}+\pderf{p}{v}{T}\der{v_\subC^\subl}{T},
\end{align}
\eqref{e:vc-0} is transformed as
\begin{align}
v^\subl(\bT^\subl,\bP^\subl)=v_\subC^\subl(\bT)+\der{v_\subC^\subl}{T}(\bT^\subl-\bT)+\pderf{v^\subl}{p}{T}\left[\bP^\subl-\Ps(\bT)-\der{\Ps}{T}(\bT^\subl-\bT)\right].
\label{e:vl-neq-pre}
\end{align}
Since $p_\theta-\Ps(\bT)=(\Tc(p_\theta)-\bT)d\Ps/dT+O(\ep^2)$, we have
\begin{align}
&\bP^\subl-\Ps(\bT)-\der{\Ps}{T}(\bT^\subl-\bT)=\bP^\subl-p_\theta-\der{\Ps}{T}(\bT^\subl-\Tc(p_\theta))\nm
&\qquad\qquad
=\frac{N^\subl mgL}{2V}+\frac{\Xi}{2}\frac{N^\subl}{N}\der{\Ps}{T}
=\frac{mg_\eff L}{2}\frac{N^\subl}{N}\frac{1}{\bar v},
\label{e:bP-l-geff}
\end{align}
where we applied the force balance relation and \eqref{e:Tlu-st} to obtain the second line, and then used the definition of $g_\eff$ in \eqref{e:g-eff} to obtain the last form.
Substituting \eqref{e:bP-l-geff} into \eqref{e:vl-neq-pre}, we have
\begin{align}
&v^\subl(\bT^\subl,\bP^\subl)=v_\subC^\subl(\bT)
-
\frac{mg_\eff L}{2}\frac{v_\subC^\subl(\bT)\alpha^\subl}{\bar v}\frac{N^\subl}{N}
-\frac{\Xi}{2}\frac{N^\subu}{N}\der{v_\subC^\subl}{T}.
\label{e:vl-neq}
\end{align}
Repeating the same procedure for $v^\subu(\bT^\subu,\bP^\subu)$,
we obtain
\begin{align}
&v^\subu(\bT^\subu,\bP^\subu)=v_\subC^\subu(\bT)
+
\frac{mg_\eff L}{2}
\frac{v_\subC^\subu(\bT)\alpha^\subu }{\bar v}
\frac{N^\subu}{N}
+\frac{\Xi}{2}\frac{N^\subl}{N}\der{v_\subC^\subu}{T}.
\label{e:vu-neq}
\end{align}

Substituting \eqref{e:vl-neq} and \eqref{e:vu-neq} into \eqref{e:N-lu-neq}, replacing
$N^\subl/N$ and $N^\subu/N$ on the right-hand sides of \eqref{e:vl-neq} and \eqref{e:vu-neq}
with $N_0^\subl/N$ and $N_0^\subu/N$, and expanding in $\ep$, we conclude

\begin{align}
\frac{\Delta N^\subl}{N}
&=-\frac{\bar{v}}{v_\subC^\subu(\bT)-v_\subC^\subl(\bT)}
\left[\frac{mg_\eff L}{2\bar{v}^2}\left(
v_\subC^\subl(\bT)\alpha^\subl\left(\frac{N_0^\subl}{N}\right)^2
-
v_\subC^\subu(\bT)\alpha^\subu\left(\frac{N_0^\subu}{N}\right)^2\right)
\right. \nm
&\quad\left.
+\frac{\Xi}{2\bar{v}}
\frac{N_0^\subl N_0^\subu}{N^2}
\left(\der{v_\subC^\subl}{T}-\der{v_\subC^\subu}{T}\right)
\right]
+O(\ep^2).
\label{e:Delta-Nl}
\end{align}

Next, we consider the volume fraction of the lower region
\begin{align}
\frac{V^\subl}{V}=\frac{v^\subl(\bT^\subl,\bP^\subl)}{\bar{v}}\frac{N^\subl}{N}.
\end{align}
Applying \eqref{e:vl-neq} and \eqref{e:Delta-Nl}, and expanding in $\ep$, we have
\begin{align}
\frac{V^\subl}{V}
&=
\frac{1}{\bar v}
\left[
v_\subC^\subl(\bT)
-
\frac{mg_\eff L}{2}\frac{v_\subC^\subl(\bT)\alpha^\subl}{\bar v}\frac{N_0^\subl}{N}
-
\frac{\Xi}{2}\frac{N_0^\subu}{N}\der{v_\subC^\subl}{T}
\right]
\left[
\frac{N_0^\subl}{N}
+
\frac{\Delta N^\subl}{N}
\right]
+O(\ep^2)
\nonumber\\
&=
\frac{V_0^\subl}{V}
-\frac{\bar v}{v_\subC^\subu(\bT)-v_\subC^\subl(\bT)}
\Biggl[
\frac{mg_\eff L}{2\bar v^3}
v_\subC^\subl(\bT)v_\subC^\subu(\bT)
\left(
\alpha^\subl\left(\frac{N_0^\subl}{N}\right)^2
-
\alpha^\subu\left(\frac{N_0^\subu}{N}\right)^2
\right)
\nonumber\\
&\qquad\qquad\qquad\qquad
+\frac{\Xi}{2\bar v^2}
\frac{N_0^\subl N_0^\subu}{N^2}
\left(
v_\subC^\subu(\bT)\der{v_\subC^\subl}{T}
-
v_\subC^\subl(\bT)\der{v_\subC^\subu}{T}
\right)
\Biggr]
+O(\ep^2),
\end{align}
where
\begin{align}
\frac{V_0^\subl}{V}=\frac{v_\subC^\subl(\bT)}{\bar v}\frac{N_0^\subl}{N}.
\end{align}
Therefore, we conclude
\begin{align}
\frac{\Delta V^\subl}{V}
&=
-\frac{\bar v}{v_\subC^\subu(\bT)-v_\subC^\subl(\bT)}
\Biggl[
\frac{mg_\eff L}{2\bar v^3}
v_\subC^\subl(\bT)v_\subC^\subu(\bT)
\left(
\alpha^\subl\left(\frac{N_0^\subl}{N}\right)^2
-
\alpha^\subu\left(\frac{N_0^\subu}{N}\right)^2
\right)
\nonumber\\
&\qquad\qquad\qquad\qquad
+\frac{\Xi}{2\bar v^2}
\frac{N_0^\subl N_0^\subu}{N^2}
\left(
v_\subC^\subu(\bT)\der{v_\subC^\subl}{T}
-
v_\subC^\subl(\bT)\der{v_\subC^\subu}{T}
\right)
\Biggr]
+O(\ep^2).
\label{e:Delta-Vl}
\end{align}

Note that \eqref{e:Delta-Nl} and \eqref{e:Delta-Vl} are symmetric under the interchange of $(\subl,\subu)$.
This fact indicates that the lower-region shifts have the same form in the two configurations $(\subL,\subG)$ and $(\subG,\subL)$, with the upper-region shifts fixed by conservation.
We write the common lower-region shifts as $(\Delta N,\Delta V)$, so that
\begin{align}
&N^\subl=N_0^\subl+\Delta N,\qquad V^\subl=V_0^\subl+\Delta V,\\
&N^\subu=N_0^\subu-\Delta N,\qquad V^\subu=V_0^\subu-\Delta V.
\end{align}
Accordingly,
\begin{align}
\frac{\Delta N}{N}
&=
-\frac{\bar{v}}{v_\subC^\subG(\bT)-v_\subC^\subL(\bT)}
\left[
(mg_\eff L)\eta_g+\Xi\eta_\Xi
\right]
+O(\ep^2),\\
\frac{\Delta V}{V}
&=
-\frac{\bar{v}}{v_\subC^\subG(\bT)-v_\subC^\subL(\bT)}
\left[
(mg_\eff L)\zeta_g+\Xi\zeta_\Xi
\right]
+O(\ep^2),
\end{align}
where
\begin{align}
\eta_g
&=
\frac{1}{2\bar{v}^2}
\left(
v_\subC^\subL(\bT)\alpha^\subL(c^\subL)^2
-
v_\subC^\subG(\bT)\alpha^\subG(c^\subG)^2
\right),\\
\eta_\Xi
&=
\frac{c^\subL c^\subG}{2\bar{v}}
\left(
\der{v_\subC^\subL}{T}
-
\der{v_\subC^\subG}{T}
\right),\\
\zeta_g
&=
\frac{v_\subC^\subL(\bT)v_\subC^\subG(\bT)}{2\bar{v}^3}
\left(
\alpha^\subL(c^\subL)^2
-
\alpha^\subG(c^\subG)^2
\right),\\
\zeta_\Xi
&=
\frac{c^\subL c^\subG}{2\bar{v}^2}
\left(
v_\subC^\subG(\bT)\der{v_\subC^\subL}{T}
-
v_\subC^\subL(\bT)\der{v_\subC^\subG}{T}
\right).
\end{align}

\section{Fundamental relations of global thermodynamics in heat conduction under gravity}
\label{s:fundamental-relations}

We now derive the fundamental relation for the thermodynamic free energy
obtained from the variational function.  By definition,
\begin{align}
F_g(\bT,V,N,mgL,\Xi)
\equiv
\min_{\calN^\subl,\calV^\subl}
\calF_g(\calN^\subl,\calV^\subl;\bT,V,N,mgL,\Xi).
\label{e:Fg_min}
\end{align}

The preceding sections suggest a simpler description.  In
Sec.~\ref{s:sol-var-eq}, the variational function was decomposed as
\eqref{e:Fg-var-neq-re2}, with the effective and residual parts defined
in \eqref{e:Fg-eff-def} and \eqref{e:Fg-res-def}.  For the liquid--gas
coexistence states, $\calF_{g,\res}=O(\ep^2)$ at the stationary values,
and the lower free-energy arrangement is therefore determined by
\eqref{e:Fg-LG-def} and \eqref{e:Fg-GL-def}.  Equations \eqref{e:Fg-LG}
and \eqref{e:Fg-GL} have the same form as the equilibrium gravity free
energies under the replacement $mgL\to mg_\eff L$.  Thus, for separated
liquid--gas states,
\begin{align}
F_g(\bT,V,N,mgL,\Xi)
=
F_g^\eq(\bT,V,N,mg_\eff L)+O(\ep^2),
\label{e:Fg-geff-memory}
\end{align}
where $F_g^\eq(\bT,V,N,mg_\eff L)$ denotes the gravity-only state
evaluated at the field value $mgL=mg_\eff L$.
The differential relation below is understood on each nondegenerate
stationary branch.  At $g_\eff=0$, the cusp shown in
Fig.~\ref{fig:Fg}(c) and derived from \eqref{e:Fg-diff-twomin} implies
that the conjugates are interpreted as one-sided values.

This observation naturally leads one to expect a gravity-only
thermodynamic relation.  If the heat-conducting free energy is, at the
stationary value, the gravity-only free energy with $mgL$ replaced by
$mg_\eff L$, one may also expect its differential to be obtained by the
same replacement.  In analogy with the equilibrium relation
\eqref{e:Fg-relation-main}, the fundamental relation for
$F_g^\eq(\bT,V,N,mg_\eff L)$ has the form
\begin{align}
dF_g^\eq
=
-S^\eq d\bT-\bP^\eq dV+\bmu_g^\eq dN
-\Psi_g^\eq d(mg_\eff L).
\end{align}
Here $S^\eq$, $\bP^\eq$, $\bmu_g^\eq$, and $\Psi_g^\eq$ are the
conjugate variables appearing in the gravity-only relation.  When these
coefficients are compared with the laboratory ones below, the
corresponding effective-part derivatives are evaluated at the same
stationary point $x_*$ with $mg_\eff L$ held fixed.  The
surprising point is that this expectation is not correct.  The effective
gravity controls which liquid--gas arrangement has the lower stationary
free-energy value, but it is not the natural work variable of the
thermodynamic relation in the laboratory variables.  The actual result, derived below
from the variational function, is
\begin{align}
dF_g
=
-(S^\subl+S^\subu)d\bT
-\bP dV
+\bmu_g dN
-\Psi_g d(mgL)
-\Psi_\Xi d\Xi .
\label{e:Fg-relation-main-neq}
\end{align}
The conjugate variables $\Psi_g$ and $\Psi_\Xi$ are identified below from
the $mgL$ and $\Xi$ derivatives.
The effective part gives the gravity-only structure of the liquid--gas
arrangements, while the residual part changes the derivatives with
respect to the laboratory variables $(\bT,V,N,mgL,\Xi)$.

\subsection{Differentiating the minimized value}

We write
\begin{align}
x\equiv(\calN^\subl,\calV^\subl).
\end{align}
Using the definition \eqref{e:g-eff}, the split form
\eqref{e:Fg-var-neq-re2} is written as
\begin{align}
\calF_g(x;\bT,V,N,mgL,\Xi)
=
\calF_{g,\eff}(x;\bT,V,N,mg_\eff L)
+
\calF_{g,\res}(x;\bT,V,N,\Xi).
\label{e:split-calF-repeat}
\end{align}
The minimizing point $x_*$ satisfies
\begin{align}
\left.\frac{\partial \calF_g}{\partial \calN^\subl}\right|_{x_*}=0,
\qquad
\left.\frac{\partial \calF_g}{\partial \calV^\subl}\right|_{x_*}=0.
\end{align}
Therefore, when $F_g$ is differentiated with respect to an external
control variable $\lambda\in\{\bT,V,N,mgL,\Xi\}$, the terms proportional
to $dx_*/d\lambda$ vanish:
\begin{align}
\left(\frac{\partial F_g}{\partial \lambda}\right)
=
\left.
\left(\frac{\partial \calF_g}{\partial \lambda}\right)_{x}
\right|_{x=x_*},
\label{e:split-envelope}
\end{align}
where the derivative on the right-hand side is taken at fixed variational coordinates $x$.
When this formula is used for a derivative of $F_g$, the fixed-$x$ derivative is evaluated at
$x=x_*$ after the differentiation.

\subsection{\texorpdfstring{$V$ derivative}{V derivative}}

We first consider the $V$ derivative, because it gives the simplest
test of the distinction stated above.  The laboratory pressure
coefficient is obtained by differentiating the full variational function
at fixed $(x,\bT,N,mgL,\Xi)$.  Using the unseparated form
\eqref{e:Fg-var-neq}, \eqref{e:split-envelope} gives
\begin{align}
\left(\frac{\partial F_g}{\partial V}\right)_{\bT,N,mgL,\Xi}
&=
\left[
\left(\frac{\partial \calF_g}{\partial V}\right)_{x,\bT,N,mgL,\Xi}
\right]_{x_*}
\nonumber\\
&=
-\bP^\subu
-(\bP^\subl-\bP^\subu)\frac{V^\subl}{V}
\nonumber\\
&=
-\left(
\bP^\subl\frac{V^\subl}{V}
+\bP^\subu\frac{V^\subu}{V}
\right)
=-\bP .
\label{e:dFg_dV}
\end{align}
Here we used the mechanical balance \eqref{e:varV-neq-1} in the second
line and the definition of the spatially averaged pressure in the last
line.

This direct calculation gives the laboratory coefficient.  It should be
distinguished from the gravity-only pressure coefficient obtained from
the effective part.  Define
\begin{align}
\bP^\eq
\equiv
\left[
\left(-\frac{\partial \calF_{g,\eff}}{\partial V}\right)_{x,\bT,N,mg_\eff L}
\right]_{x_*}
\label{e:Peq-def}
\end{align}
where $mg_\eff L$ is held fixed as the field variable of
$\calF_{g,\eff}$.  Evaluating \eqref{e:Peq-def} at the same stationary
point gives
\begin{align}
\bP^\eq
&=
\bP
-\frac{\Xi}{2}
\left[
\frac{N^\subl}{N}
\left(
\frac{\partial p^\subu}{\partial T}
\right)_{T=\bT,V^\subu,N^\subu}
-
\frac{V^\subl}{V}\frac{d\Ps}{d\bT}
\right].
\label{e:Peq-split}
\end{align}
Thus,
\begin{align}
\bP\neq \bP^\eq.
\end{align}
The equality of the stationary free-energy values with the gravity-only
reference therefore does not imply equality of their $V$ derivatives.

The split form explains how the laboratory coefficient is recovered.
When the variables $(mgL,\Xi)$ are held fixed, the combination
$mg_\eff L=mgL+(V/N)(d\Ps/d\bT)\Xi$ changes with $V$.  This gives the
additional contribution
\begin{align}
\frac{\Xi}{N}\frac{d\Ps}{d\bT}
\left[
\left(\frac{\partial \calF_{g,\eff}}{\partial(mg_\eff L)}\right)_{x,\bT,V,N}
\right]_{x_*}
=
\frac{\Xi}{2}
\left(
\frac{V^\subl}{V}-\frac{N^\subl}{N}
\right)
\frac{d\Ps}{d\bT}.
\label{e:split-eff-V-chain}
\end{align}
The residual contribution supplies the remaining term.  Indeed, using
\eqref{e:Fg-res-def} with \eqref{e:qex-def} and the Maxwell relation
$(\partial S^\subu/\partial V^\subu)_{\bT,N^\subu}
=(\partial p^\subu/\partial T)_{V^\subu,N^\subu}$,
\begin{align}
\left[
\left(\frac{\partial \calF_{g,\res}}{\partial V}\right)_{x,\bT,N,mgL,\Xi}
\right]_{x_*}
&=
-\frac{\Xi}{2}
\frac{N^\subl}{N}
\left[
\left(
\frac{\partial p^\subu}{\partial T}
\right)_{T=\bT,V^\subu,N^\subu}
\!-\!
\frac{d\Ps}{d\bT}
\right].
\label{e:split-res-V-at-star}
\end{align}
Adding \eqref{e:split-eff-V-chain} and \eqref{e:split-res-V-at-star}
to the fixed-$mg_\eff L$ derivative in \eqref{e:Peq-def}, we recover
\begin{align}
\left(\frac{\partial F_g}{\partial V}\right)_{\bT,N,mgL,\Xi}
=
-\bP^\eq
+\frac{\Xi}{2}
\left(
\frac{V^\subl}{V}-\frac{N^\subl}{N}
\right)
\frac{d\Ps}{d\bT}
-\frac{\Xi}{2}
\frac{N^\subl}{N}
\left[
\left(
\frac{\partial p^\subu}{\partial T}
\right)_{T=\bT,V^\subu,N^\subu}
\!-\!
\frac{d\Ps}{d\bT}
\right]
=-\bP .
\end{align}
Thus the effective-gravity part by itself gives the gravity-only
coefficient $\bP^\eq$, whereas the laboratory pressure coefficient
$\bP$ is obtained only after the $V$ dependence of $mg_\eff L$ and the
residual derivative are both included.

\subsection{\texorpdfstring{$\bT$ derivative}{T derivative}}

We next consider the $\bT$ derivative.  As in the $V$ derivative, we
first identify the laboratory coefficient directly from the full
variational function.  At fixed $(x,V,N,mgL,\Xi)$, \eqref{e:bT-Tlu}
gives
\begin{align}
\left(\frac{\partial \bT^\subl}{\partial \bT}\right)_{x,V,N,mgL,\Xi}
=
\left(\frac{\partial \bT^\subu}{\partial \bT}\right)_{x,V,N,mgL,\Xi}
=1 .
\end{align}
Using \eqref{e:split-envelope} and the unseparated form
\eqref{e:Fg-var-neq}, we obtain
\begin{align}
\left(\frac{\partial F_g}{\partial \bT}\right)_{V,N,mgL,\Xi}
&=
\left[
\left(\frac{\partial \calF_g}{\partial \bT}\right)_{x,V,N,mgL,\Xi}
\right]_{x_*}
\nonumber\\
&=
-S^\subl-S^\subu .
\label{e:dFg_dT}
\end{align}
Here $S^\subl=S(\bT^\subl,V^\subl,N^\subl)$ and
$S^\subu=S(\bT^\subu,V^\subu,N^\subu)$.

This entropy coefficient is not the gravity-only coefficient.  Define
\begin{align}
S^\eq
&\equiv
\left[
\left(-\frac{\partial \calF_{g,\eff}}{\partial \bT}\right)_{x,V,N,mg_\eff L}
\right]_{x_*}
\nonumber\\
&=
S(\bT,V^\subl,N^\subl)+S(\bT,V^\subu,N^\subu),
\label{e:Seq-split}
\end{align}
where $mg_\eff L$ is held fixed as the field variable of
$\calF_{g,\eff}$.  Thus $S^\eq$ is the entropy coefficient of the
gravity-only reference state.  Using \eqref{e:bT-Tlu}, the actual
entropy coefficient is written as
\begin{align}
S^\subl+S^\subu
&=
S^\eq
+
\frac{\Xi}{2}
\left[
\frac{N^\subl}{N}
\left(
\frac{\partial S^\subu}{\partial T}
\right)_{T=\bT,V^\subu,N^\subu}
-
\frac{N^\subu}{N}
\left(
\frac{\partial S^\subl}{\partial T}
\right)_{T=\bT,V^\subl,N^\subl}
\right].
\label{e:Seq-compare}
\end{align}
Thus, in general,
\begin{align}
S^\eq\neq S^\subl+S^\subu .
\end{align}
The equality of the stationary free-energy values with the gravity-only
reference therefore does not imply equality of their $\bT$ derivatives.

The split form explains how the laboratory coefficient is recovered.
When the variables $(mgL,\Xi)$ are held fixed, the combination
$mg_\eff L=mgL+(V/N)(d\Ps/d\bT)\Xi$ changes with $\bT$.  This gives the
additional contribution
\begin{align}
\frac{V\Xi}{N}\frac{d^2\Ps}{d\bT^2}
\left[
\left(\frac{\partial \calF_{g,\eff}}{\partial(mg_\eff L)}\right)_{x,\bT,V,N}
\right]_{x_*}
=
\frac{\Xi}{2}V
\left(
\frac{V^\subl}{V}-\frac{N^\subl}{N}
\right)
\frac{d^2\Ps}{d\bT^2}.
\label{e:split-eff-T-chain}
\end{align}
Combining this with the fixed-$mg_\eff L$ derivative in
\eqref{e:Seq-split}, the effective contribution to the laboratory
$\bT$ derivative is
\begin{align}
-S^\eq
+\frac{\Xi}{2}V
\left(
\frac{V^\subl}{V}-\frac{N^\subl}{N}
\right)
\frac{d^2\Ps}{d\bT^2}.
\label{e:split-eff-T-at-star}
\end{align}
The residual contribution supplies the complementary term.  Differentiating
\eqref{e:Fg-res-def} with respect to $\bT$ at fixed
$(x,V,N,mgL,\Xi)$, and using the definition of $\hat q^\ex$ in
\eqref{e:qex-def}, gives the difference between the common-temperature
entropies and the actual entropies, together with the derivative of the
Clausius--Clapeyron contribution:
\begin{align}
\left[
\left(\frac{\partial \calF_{g,\res}}{\partial \bT}\right)_{x,V,N,mgL,\Xi}
\right]_{x_*}
&=
S(\bT,V^\subl,N^\subl)+S(\bT,V^\subu,N^\subu)
-(S^\subl+S^\subu)
-\frac{\Xi}{2}V
\left(
\frac{V^\subl}{V}-\frac{N^\subl}{N}
\right)
\frac{d^2\Ps}{d\bT^2}.
\label{e:split-res-T-at-star}
\end{align}
Adding \eqref{e:split-eff-T-at-star} and \eqref{e:split-res-T-at-star},
we recover
\begin{align}
\left(\frac{\partial F_g}{\partial \bT}\right)_{V,N,mgL,\Xi}
=
-(S^\subl+S^\subu).
\end{align}
Thus the effective-gravity part by itself gives the gravity-only entropy
coefficient $S^\eq$, whereas the laboratory entropy coefficient
$S^\subl+S^\subu$ is obtained only after the $\bT$ dependence of
$mg_\eff L$ and the residual derivative are both included.

\subsection{Derivatives with respect to \texorpdfstring{$mgL$}{mgL} and \texorpdfstring{$\Xi$}{Xi}}

At fixed $(\bT,V,N,\Xi)$, one has $\partial(mg_\eff L)/\partial(mgL)=1$ and
$\calF_{g,\res}$ has no explicit $mgL$ dependence.  Therefore, from
\eqref{e:Fg-eff-def},
\begin{align}
\left(\frac{\partial F_g}{\partial(mgL)}\right)_{\bT,V,N,\Xi}
&=
\left.
\left(\frac{\partial \calF_g}{\partial(mgL)}\right)_{x,\bT,V,N,\Xi}
\right|_{x_*}
\nonumber\\
&=
\left.
\left(\frac{\partial \calF_{g,\eff}}{\partial(mg_\eff L)}\right)_{x,\bT,V,N}
\right|_{x_*}
\nonumber\\
&=
\frac{N}{2}
\left(
\frac{\calV^\subl}{V}-\frac{\calN^\subl}{N}
\right)_{x=x_*}
=
-\Psi_g,
\label{e:split-eff-M-derivative}
\end{align}
where
\begin{align}
\Psi_g
\equiv
N\frac{\xm-X}{L}
=
\frac{N}{2}\left(
\frac{N^\subl}{N}
-
\frac{V^\subl}{V}
\right).
\label{e:Psi-g}
\end{align}

For the $\Xi$ derivative, it is useful to first compute the full
fixed-$x$ derivative rather than separating it into the effective and
residual pieces.  From the unseparated form \eqref{e:Fg-var-neq-re1},
\begin{align}
\left(\frac{\partial \calF_g}{\partial \Xi}\right)_{x,\bT,V,N,mgL}
&=
\frac{\calN^\subl\calN^\subu}{2N}
\left(
\frac{\mathcal{S}^\subl}{\calN^\subl}
-
\frac{\mathcal{S}^\subu}{\calN^\subu}
\right)
\nonumber\\
&=
-\frac{V}{N}\frac{d\Ps}{d\bT}
\frac{N}{2}
\left(
\frac{\calN^\subl}{N}
-
\frac{\calV^\subl}{V}
\right)
-
\frac{\calN^\subl \calN^\subu}{2N\bT}\hat q^\ex ,
\label{e:split-Xi-full}
\end{align}
where \eqref{e:qex-def} was used in the second equality.
Equation~\eqref{e:split-Xi-full} shows that the fixed-$x$ partial
derivative contains the excess-latent-heat term.
The first term in \eqref{e:split-Xi-full} is the chain contribution of
the effective part through $mg_\eff L$, since
$\partial(mg_\eff L)/\partial\Xi=(V/N)d\Ps/d\bT$ at fixed
$(\bT,V,N,mgL)$.  Evaluating this contribution at the minimizing point
and using \eqref{e:split-eff-M-derivative}, we define
\begin{align}
\Psi_\Xi
\equiv
\frac{V}{N}\frac{d\Ps}{d\bT}\Psi_g,
\label{e:Psi-Xi}
\end{align}
so that the first term in \eqref{e:split-Xi-full} gives $-\Psi_\Xi$.
The second term in \eqref{e:split-Xi-full} is estimated using
\eqref{e:qex-hetero-LG-order} and \eqref{e:qex-hetero-GL-order},
which show that $\hat q^\ex=O(\ep)$ for the heterogeneous stationary
states.  Hence this term is $O(\ep)$ at $x=x_*$.
Thus, as a fixed-$x$ partial derivative evaluated at the stationary
point,
\begin{align}
\left[
\left(\frac{\partial \calF_g}{\partial \Xi}\right)_{x,\bT,V,N,mgL}
\right]_{x_*}
=
-\Psi_\Xi
+O(\ep).
\label{e:split-Xi-stationary}
\end{align}
For weak-driving variations with $d\Xi=O(\ep)$, this gives
\begin{align}
(dF_g)_\Xi
=
-\Psi_\Xi d\Xi +O(\ep^2).
\label{e:dFg_dXi}
\end{align}
Together with \eqref{e:split-eff-M-derivative}, this gives, at fixed
$(\bT,V,N)$,
\begin{align}
dF_g\big|_{\bT,V,N}
=
-\Psi_g d(mgL)-\Psi_\Xi d\Xi
=
-\Psi_g d(mg_\eff L)
\label{e:dFg-work-geff}
\end{align}
with an error of $O(\ep^2)$.

\subsection{\texorpdfstring{$N$ derivative}{N derivative}}

The $N$ derivative gives the global chemical potential.  The gravity-only
chemical-potential coefficient is obtained from the effective part at
fixed $mg_\eff L$:
\begin{align}
\bmu_g^\eq
&=
\left[
\left(\frac{\partial \calF_{g,\eff}}{\partial N}\right)_{x,\bT,V,mg_\eff L}
\right]_{x_*}
\nonumber\\
&=
\left(\frac{\partial F^\subu}{\partial N^\subu}\right)_{\bT,V^\subu}
+
\frac{mg_\eff L}{2}\frac{V^\subl}{V}.
\label{e:mueq-split}
\end{align}
At fixed laboratory variables $(x,\bT,V,mgL,\Xi)$, the effective part gives
\begin{align}
\left[
\left(\frac{\partial \calF_{g,\eff}}{\partial N}\right)_{x,\bT,V,mgL,\Xi}
\right]_{x_*}
&=
\left(\frac{\partial F^\subu}{\partial N^\subu}\right)_{\bT,V^\subu}
+
\frac{mgL}{2}\frac{V^\subl}{V}
+
\frac{\Xi}{2}\frac{V N^\subl}{N^2}\frac{d\Ps}{d\bT}.
\label{e:split-eff-N-at-star}
\end{align}
Here and below in this paragraph, $\mu^\alpha$ denotes
$\mu^\alpha(\bT^\alpha,\bP^\alpha)$, which represents
$\bmu_g^\alpha$ to the present order by \eqref{e:mu-lu-def}.
The residual part supplies the complementary term,
\begin{align}
\left[
\left(\frac{\partial \calF_{g,\res}}{\partial N}\right)_{x,\bT,V,mgL,\Xi}
\right]_{x_*}
&=
\mu^\subu-\left(\frac{\partial F^\subu}{\partial N^\subu}\right)_{\bT,V^\subu}
+
\frac{\Xi}{2}\frac{N^\subl}{N^2}
\left[
S^\subl+S^\subu
-V\frac{d\Ps}{d\bT}
\right].
\label{e:split-res-N-at-star}
\end{align}
Adding \eqref{e:split-eff-N-at-star} and \eqref{e:split-res-N-at-star},
we obtain
\begin{align}
\left(\frac{\partial F_g}{\partial N}\right)_{\bT,V,mgL,\Xi}
&=
\left[
\left(\frac{\partial \calF_g}{\partial N}\right)_{x,\bT,V,mgL,\Xi}
\right]_{x_*}
\nonumber\\
&=
\mu^\subu+\frac{mgL}{2}\frac{V^\subl}{V}
+\frac{\Xi}{2}\frac{N^\subl}{N^2}(S^\subl+S^\subu),
\label{e:dFg_dN-intermediate-split}
\end{align}
or, equivalently,
\begin{align}
\left(\frac{\partial F_g}{\partial N}\right)_{\bT,V,mgL,\Xi}
&=
\bmu_g^\eq
+
\left[
\mu^\subu
-
\left(\frac{\partial F^\subu}{\partial N^\subu}\right)_{\bT,V^\subu}
\right]
+
\frac{\Xi}{2}
\left[
\frac{N^\subl}{N^2}(S^\subl+S^\subu)
-
\frac{V^\subl}{N}\frac{d\Ps}{d\bT}
\right].
\label{e:mueq-compare}
\end{align}
With this notation, \eqref{e:bT-Tlu} gives
$(\partial\bT^\subl/\partial N)_{\calN^\subl}
=(\partial\bT^\subu/\partial N)_{\calN^\subl}
=-\Xi N^\subl/(2N^2)$ at fixed $x$, $\bT$, $V$, $mgL$, and $\Xi$.
Using the variational relation \eqref{e:varN-neq-1}, \eqref{e:dFg_dN-intermediate-split} is rewritten as
\begin{align}
\left(\frac{\partial F_g}{\partial N}\right)_{\bT,V,mgL,\Xi}
&=
\frac{N^\subl}{N}\mu^\subl
+\frac{N^\subu}{N}\mu^\subu
+\frac{mgL}{2}
\left(
\frac{V^\subl}{V}-\frac{N^\subl}{N}
\right)
\nonumber\\
&=
\frac{A}{N}
\int_0^L[\mu(x)+mg(x-\xm)]\rho(x)\,dx
=\bmu_g .
\label{e:mug-neq}
\end{align}
Thus the residual part is essential here, not because it adds
a new state variable, but because it cancels the $\bT,V,N$
derivatives generated when the Clausius--Clapeyron piece is absorbed
into $mg_\eff L$.

\subsection{Resulting fundamental relation}

Combining \eqref{e:Psi-Xi}, \eqref{e:dFg_dT}, \eqref{e:dFg_dV}, \eqref{e:split-eff-M-derivative}, \eqref{e:dFg_dXi}, and \eqref{e:mug-neq}, the thermodynamic
state-function differential in the laboratory variables is
\begin{align}
dF_g
=
-(S^\subl+S^\subu)d\bT
-\bP dV
+\bmu_g dN
-\Psi_g d(mgL)
-\Psi_\Xi d\Xi .
\label{e:Fg-relation-neq-Xi-result}
\end{align}
The split derivation clarifies its meaning.
When $\bT$, $V$, and $N$ are held fixed, \eqref{e:dFg-work-geff}
gives the gravity-type form with $g$ replaced by $g_\eff$.
This projection explains the macroscopic configurational ordering.
It is not, however, the laboratory fundamental relation.
Once $\bT$, $V$, or $N$ is varied,
$mg_\eff L=mgL+(V/N)(d\Ps/d\bT)\Xi$ also varies through
$\bT$, $V$, and $N$.
The residual derivative in \eqref{e:split-Xi-full} belongs to the
$O(\ep^2)$ stationary contribution to the fixed-$\bT$ variational free-energy structure.
The natural nonequilibrium control variables are therefore $(mgL,\Xi)$,
while $mg_\eff L$ is the effective combination for the
fixed-$(\bT,V,N)$ variation.

Using $\Xi/\bT$ instead of $\Xi$, define
\begin{align}
S\equiv S^\subl+S^\subu+\frac{\Xi}{\bT}\Psi_\Xi .
\label{e:split-S-def}
\end{align}
Then \eqref{e:Fg-relation-neq-Xi-result} is equivalently written as
\begin{align}
dF_g
=
-S d\bT
-\bP dV
+\bmu_g dN
-\Psi_g d(mgL)
-\bT\Psi_\Xi d\left(\frac{\Xi}{\bT}\right),
\label{e:Fg-relation-neq-phi-result}
\end{align}
which is the form connected to the entropy representation in Refs.~\cite{NS19,NS22}.

The lesson of this derivation is that the stationary value alone is not
enough to reconstruct the thermodynamic relation.  At heterogeneous
stationary points, $\calF_{g,\res}$ does not determine the macroscopic
ordering of separated states; its derivatives are needed to recover
the physical laboratory coefficients, such as the pressure coefficient
$-\bP$, from the minimized state function.

\section{Local--global consistency of the chemical potential}
\label{s:chemical_potential}

Equation \eqref{e:mug-neq} identifies $\bmu_g$ as the quantity conjugate to $N$ in the global free-energy relation.
For a liquid--gas heterogeneous stationary state, we recover the same quantity directly from the local fields $T(x)$, $p(x)$, $\rho(x)$, and $\mu(x)$.
The key point is that local and global notions of chemical potential are not identical under heat conduction.
In the sharp-interface description, $T(x)$ and $p(x)$ remain continuous across the interface, whereas the liquid- and gas-side limits of
\begin{align}
\mu(x)\equiv \mu(T(x),p(x))
\end{align}
generally do not coincide at the liquid-gas interface.
The mismatch follows from the interface-temperature shift $\theta\neq\Tc(p_\theta)$ in \eqref{e:theta-J}, which was obtained in \cite{NS17,NS19}.
The calculation below shows how this local discontinuity is reconciled with the density-weighted global conjugate variable \eqref{e:bmug-def}.

\begin{figure}[bt]
\begin{center}
\includegraphics[width=0.7\linewidth]{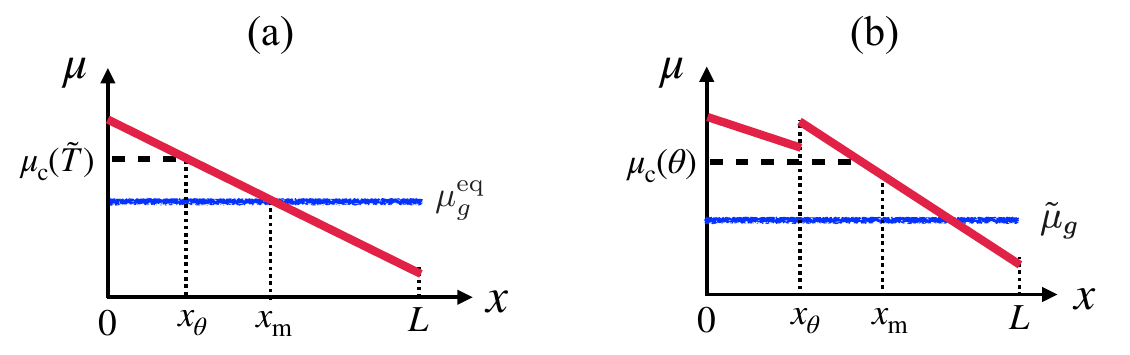}
\end{center}
\caption{
Schematic comparison of the chemical-potential profile in (a) equilibrium under gravity and (b) a heat-conducting steady state under gravity.
(a) In equilibrium, $\mu(x)$ is continuous and linear in $x$ with slope $-mg$.
(b) In heat conduction, shown for $(\subl,\subu)=(\subL,\subG)$ and $\Xi>0$, $\mu(x)$ is generally discontinuous at $x_\theta$ because $\theta \neq \Tc(p_\theta)$.
The dashed line indicates $\mu_\subC(\theta)$, and the blue line schematically represents the global chemical potential $\bmu_g$.
}
\label{fig:mu-neq}
\end{figure}

\subsection{\texorpdfstring{Equilibrium reference ($g \neq 0$, $\Xi = 0$)}{Equilibrium reference}}
\label{subsec:equilibrium_comparison}

In equilibrium under gravity, the generalized chemical potential
\begin{align}
\mu_g(x)\equiv \mu(x)+mg(x-\xm)
\end{align}
is spatially uniform \cite{NSgravity25}.
Denote this uniform local value by $\mu_g^\eq$; the non-bold notation
distinguishes it from the global coefficient $\bmu_g^\eq$ used in
Sec.~\ref{s:fundamental-relations}.
The chemical potential at phase coexistence is
\begin{align}
\mu_\subC(\bT)
=
\mu^\subL(\bT,\Ps(\bT))
=
\mu^\subG(\bT,\Ps(\bT)).
\end{align}
At the liquid--gas interface, $\mu(\xint)=\mu_\subC(\bT)$.
Therefore, the constant value of $\mu_g(x)$ is
\begin{align}
\mu_g^\eq
=
\mu_\subC(\bT)+mg(\xint-\xm).
\label{e:mueq_equilibrium}
\end{align}
Since $\mu_g(x)$ is spatially uniform, its density-weighted global average coincides with the same constant value, $\bmu_g=\mu_g^\eq$.
The local profile is therefore $\mu(x)=\mu_g^\eq-mg(x-\xm)$ to the present order, and hence
\begin{align}
\mu(\xm)=\mu_g^\eq.
\end{align}
Figure~\ref{fig:mu-neq}(a) illustrates this equilibrium correspondence.

\subsection{\texorpdfstring{Non-equilibrium steady state ($g\neq0$, $\Xi\neq0$)}{Non-equilibrium steady state}}
\label{subsec:nonequilibrium_comparison}

We now derive the local-field expression for $\bmu_g$ in the heat-conducting steady state.
Figure~\ref{fig:mu-neq}(b) shows the corresponding nonequilibrium profile.
The local chemical potential is discontinuous at $x_\theta$, whereas $\bmu_g$ remains a global conjugate variable and is not obtained by reading off $\mu(x)$ at a reference position such as $\xm$.
The assumed local thermodynamic relations allow us to evaluate $\mu(x)$
from the appropriate single-phase equation of state,
\begin{align}
\mu(x)=\mu(T(x),p(x)).
\end{align}
Expanding this local chemical potential around the saturation chemical potential
\begin{align}
\mu_\subC(T(x))\equiv \mu(T(x),\Ps(T(x))),
\end{align}
we have
\begin{align}
\mu(x)
=
\mu_\subC(T(x))
+
\frac{p(x)-\Ps(T(x))}{\rho(x)}
+O(\ep^2).
\label{e:mu_expand_again}
\end{align}

Introduce the density-weighted saturation chemical potential
\begin{align}
\bmu_\subC
\equiv
\frac{A}{N}\int_0^L
\mu_\subC(T(x))\rho(x)\,dx.
\label{e:bmuC-def}
\end{align}
By the definition of $\bT$ in \eqref{e:def-bT},
\begin{align}
\bmu_\subC
=
\mu_\subC(\bT)+O(\ep^2).
\label{e:bmuC-bT}
\end{align}

Substituting \eqref{e:mu_expand_again} into \eqref{e:bmug-def}, we decompose $\bmu_g$ as
\begin{align}
\bmu_g
&=
\bmu_\subC+\mathcal I_2+O(\ep^2),
\label{e:bmug-decomp}\\
\mathcal I_2
&\equiv
\frac{A}{N}\int_0^L
\left[
p(x)-\Ps(T(x))+mg\rho(x)(x-\xm)
\right]dx.
\label{e:I2_def}
\end{align}

Expanding $\Ps(T(x))$ around $\bT$ gives
\begin{align}
\Ps(T(x))
=
\Ps(\bT)+\der{\Ps}{\bT}\left(T(x)-\bT\right)
+O(\ep^2).
\end{align}
To the present order, the spatial average of $T(x)$ in each subregion can be replaced by $\bT^\subl$ or $\bT^\subu$.
Thus,
\begin{align}
\mathcal I_2
&=
\frac{V^\subl}{N}
\left[
\bP^\subl-\Ps(\bT)-\der{\Ps}{\bT}(\bT^\subl-\bT)
\right]
\nonumber\\
&\qquad
+
\frac{V^\subu}{N}
\left[
\bP^\subu-\Ps(\bT)-\der{\Ps}{\bT}(\bT^\subu-\bT)
\right]
+
mg(X-\xm)
+O(\ep^2).
\label{e:mug-second-split}
\end{align}

For the lower region, \eqref{e:bP-l-geff} gives
\begin{align}
\bP^\subl-\Ps(\bT)-\der{\Ps}{\bT}(\bT^\subl-\bT)
=
\frac{mg_\eff L}{2}\frac{N^\subl}{N}\frac{1}{\bar v}.
\label{e:mug-lower-geff}
\end{align}
The corresponding relation for the upper region is
\begin{align}
\bP^\subu-\Ps(\bT)-\der{\Ps}{\bT}(\bT^\subu-\bT)
=
-\frac{mg_\eff L}{2}\frac{N^\subu}{N}\frac{1}{\bar v}.
\label{e:mug-upper-geff}
\end{align}
Substituting \eqref{e:mug-lower-geff} and \eqref{e:mug-upper-geff} into \eqref{e:mug-second-split}, we obtain
\begin{align}
\mathcal I_2
&=
\frac{mg_\eff L}{2}
\frac{N^\subl V^\subl-N^\subu V^\subu}{NV}
+
mg(X-\xm)
+O(\ep^2).
\label{e:mug-second-1-2}
\end{align}
Using $x_\theta=l$ for a heterogeneous stationary state and the geometric relation \eqref{e:X-xm}, the first term is rewritten as
\begin{align}
\frac{L}{2}
\frac{N^\subl V^\subl-N^\subu V^\subu}{NV}
=
x_\theta-X.
\end{align}
Therefore,
\begin{align}
\mathcal I_2
=
mg_\eff(x_\theta-X)+mg(X-\xm)+O(\ep^2).
\label{e:mug-second-2}
\end{align}

Combining \eqref{e:bmug-decomp} and \eqref{e:mug-second-2}, we finally obtain
\begin{align}
\bmu_g
&=
\bmu_\subC
+
mg_\eff(x_\theta-X)
+
mg(X-\xm)
+O(\ep^2)
\nonumber\\
&=
\bmu_\subC
+
mg_\eff(x_\theta-\xm)
+
(mg-mg_\eff)(X-\xm)
+O(\ep^2).
\label{e:bmug-noneq_final}
\end{align}
Equivalently, using \eqref{e:bmuC-bT}, one may replace $\bmu_\subC$ by $\mu_\subC(\bT)$ within the present accuracy.

\subsection{Relation to the equilibrium reference}

Equation \eqref{e:bmug-noneq_final} is the local-field evaluation of the same global chemical potential that appears as the conjugate variable to $N$ in \eqref{e:mug-neq}.
The first form separates the effective-gravity contribution, measured from the center of mass to the interface, from the physical gravitational contribution, measured from the reference point $\xm$ to the center of mass.
The second form shows that the result does not reduce, in general, to the equilibrium expression \eqref{e:mueq_equilibrium} by the replacement
\begin{align}
g\longrightarrow g_\eff.
\end{align}
Such a reduction occurs only in special cases, for example when $\Xi=0$ so that $g_\eff=g$, or when $X=\xm$.

Thus, $\bmu_g$ is the global chemical potential conjugate to $N$ at fixed $(mgL,\Xi)$, as stated in \eqref{e:mug-neq}, and it retains the mixing between the physical gravity $g$ and the imposed-temperature-difference contribution contained in $g_\eff$.
At the same time, the local chemical potential remains discontinuous at the interface because $\theta\neq\Tc(p_\theta)$.
The global averaging in \eqref{e:bmug-def} therefore does not remove the local nonequilibrium anomaly; it gives the corresponding global conjugate variable in the laboratory basis.

\section{Hierarchy of the stationary solutions}
\label{s:hierarchy}

The fundamental relation and the global chemical potential derived in
Secs.~\ref{s:fundamental-relations} and \ref{s:chemical_potential}
clarify the thermodynamic meaning of the stationary value.
We now return to the set of stationary solutions and compare their
values throughout the coexistence interval.
We focus on the case $g_\eff >0$ in this section.
The heterogeneous solution $(\subl,\subu)=(\subL,\subG)$ has the lower thermodynamic value relative to the inverted heterogeneous solution $(\subG,\subL)$ when both phases occupy macroscopic fractions.
The purpose of this section is to compare this heterogeneous ordering with the homogeneous stationary state and to identify the part of the coexistence interval where the ordering is separated at the level of the stationary free-energy values.

To compare the inverted heterogeneous state and the homogeneous state, we consider
\begin{align}
\frac{F_g^{(\subG, \subL)}-F_g^\homo}{N}
=
f^\hetero(\bT,\bar{v})-f(\bT,\bar{v})
+mg_\eff L\psi_g^{(\subL, \subG)}.
\label{e:diff-second}
\end{align}
Here,
\begin{align}
f^\hetero(\bT,\bar{v})-f(\bT,\bar{v})
&=
\frac{N_0^\subL}{N}
\left[
f(\bT,v_\subC^\subL(\bT))-f(\bT,\bar{v})
\right]
+
\frac{N_0^\subG}{N}
\left[
f(\bT,v_\subC^\subG(\bT))-f(\bT,\bar{v})
\right].
\label{e:hetero-homo}
\end{align}

First, we consider the gas-side edge of coexistence, where $\bar v$ is close to $v_\subC^\subG(\bT)$, and define
\begin{align}
z\equiv
\frac{v_\subC^\subG(\bT)-\bar{v}}
{v_\subC^\subG(\bT)-v_\subC^\subL(\bT)},
\label{e:z-def}
\end{align}
as the normalized distance from the gas coexistence volume.
Thus $z=0$ corresponds to $\bar v=v_\subC^\subG(\bT)$, and small positive $z$ measures how far $\bar v$ is shifted from the gas coexistence edge into the coexistence interval.
Recall the equilibrium relation at $g=0$,
\begin{align}
f(\bT,v_\subC^\subL(\bT))
=
f(\bT,v_\subC^\subG(\bT))
+
\Ps(\bT)
\left[
v_\subC^\subG(\bT)-v_\subC^\subL(\bT)
\right].
\end{align}
Substituting this relation into \eqref{e:hetero-homo} and using the lever rule, the right-hand side of \eqref{e:hetero-homo} becomes
\begin{align}
f^\hetero(\bT,\bar{v})-f(\bT,\bar{v})
=
\left(
v_\subC^\subG(\bT)-\bar{v}
\right)
\left(
\Ps(\bT)
+
\frac{
f(\bT,v_\subC^\subG(\bT))-f(\bT,\bar{v})
}
{
v_\subC^\subG(\bT)-\bar{v}
}
\right).
\end{align}
Expanding this in $z$, we have
\begin{align}
f^\hetero(\bT,\bar{v})-f(\bT,\bar{v})
=
-\frac{z^2}{2}
\frac{
\left[
v_\subC^\subG(\bT)-v_\subC^\subL(\bT)
\right]^2
}
{
v_\subC^\subG(\bT)\alpha^\subG
}
+O(z^3).
\label{e:expansion-z}
\end{align}
Thus the free-energy difference between the heterogeneous and homogeneous branches starts at order $z^2$ near the gas-side coexistence edge.
We next rewrite $\psi_g^{(\subL,\subG)}$ in terms of $z$.
Since \eqref{e:z-def} yields
\begin{align}
\bar v=v_\subC^\subG(\bT)-z\left[v_\subC^\subG(\bT)-v_\subC^\subL(\bT)\right],
\end{align}
\eqref{e:Psig-LG} is expanded as
\begin{align}
\psi_g^{(\subL, \subG)}
=
\frac{1}{2}
\frac{
v_\subC^\subG(\bT)-v_\subC^\subL(\bT)
}
{
v_\subC^\subG(\bT)
}
z
\left(
1-\frac{v_\subC^\subL(\bT)}{v_\subC^\subG(\bT)}z
\right)
+O(z^3).
\label{e:expansion2-z}
\end{align}
Substituting \eqref{e:expansion-z} and \eqref{e:expansion2-z} into \eqref{e:diff-second}, we obtain
\begin{align}
\frac{F_g^{(\subG, \subL)}-F_g^\homo}{N}
&=
z\frac{mg_\eff L}{2}
\frac{
v_\subC^\subG(\bT)-v_\subC^\subL(\bT)
}
{
v_\subC^\subG(\bT)
}
-\frac{z^2}{2}
\frac{
(v_\subC^\subG(\bT)-v_\subC^\subL(\bT))^2
}
{
v_\subC^\subG(\bT)\alpha^\subG
}
+O(z^3,\ep z^2).
\label{e:edge-gas-diff}
\end{align}
The difference between the two heterogeneous values is, from \eqref{e:Fg-diff-twomin} and \eqref{e:expansion2-z},
\begin{align}
\frac{F_g^{(\subG,\subL)}-F_g^{(\subL,\subG)}}{N}
=
z mg_\eff L
\frac{
v_\subC^\subG(\bT)-v_\subC^\subL(\bT)
}
{
v_\subC^\subG(\bT)
}
+O(\ep z^2,\ep^2).
\label{e:edge-gas-hetero-diff}
\end{align}
Equations \eqref{e:edge-gas-diff} and \eqref{e:edge-gas-hetero-diff} show that, in the gas-side edge layer $z=O(\ep)$, all three stationary values are separated only by $O(\ep^2)$.
Thus the present $O(\ep)$ construction does not resolve their hierarchy in this edge layer.
Balancing the $O(\ep z)$ and $O(z^2)$ terms in \eqref{e:edge-gas-diff} gives the gas-side edge-layer scale
\begin{align}
z_{\mathrm{edge}}^{\subG}
\sim
\frac{\alpha^\subG mg_\eff L}
{v_\subC^\subG(\bT)-v_\subC^\subL(\bT)}.
\label{e:vrev-def}
\end{align}
Next, we set $\bar v$ close to $v_\subC^\subL(\bT)$ and introduce
\begin{align}
z'\equiv
\frac{\bar{v}-v_\subC^\subL(\bT)}
{v_\subC^\subG(\bT)-v_\subC^\subL(\bT)}.
\end{align}
Then,
\begin{align}
f^\hetero(\bT,\bar{v})-f(\bT,\bar{v})
=
-\frac{z'^2}{2}
\frac{
\left[
v_\subC^\subG(\bT)-v_\subC^\subL(\bT)
\right]^2
}
{
v_\subC^\subL(\bT)\alpha^\subL
}
+O(z'^3),
\end{align}
and
\begin{align}
\psi_g^{(\subL, \subG)}
=
\frac{1}{2}
\frac{
v_\subC^\subG(\bT)-v_\subC^\subL(\bT)
}
{
v_\subC^\subL(\bT)
}
z'
\left(
1-\frac{v_\subC^\subG(\bT)}{v_\subC^\subL(\bT)}z'
\right)
+O(z'^3).
\end{align}
The difference in the free energies is estimated as
\begin{align}
\frac{F_g^{(\subG, \subL)}-F_g^\homo}{N}
&=
z'\frac{mg_\eff L}{2}
\frac{
v_\subC^\subG(\bT)-v_\subC^\subL(\bT)
}
{
v_\subC^\subL(\bT)
}
-\frac{z'^2}{2}
\frac{
(v_\subC^\subG(\bT)-v_\subC^\subL(\bT))^2
}
{
v_\subC^\subL(\bT)\alpha^\subL
}
+O(z'^3,\ep z'^2),
\label{e:edge-liquid-diff}
\end{align}
and
\begin{align}
\frac{F_g^{(\subG,\subL)}-F_g^{(\subL,\subG)}}{N}
=
mg_\eff L
\frac{
v_\subC^\subG(\bT)-v_\subC^\subL(\bT)
}
{
v_\subC^\subL(\bT)
}
z'
+O(\ep z'^2,\ep^2).
\label{e:edge-liquid-hetero-diff}
\end{align}
Similarly, in the liquid-side edge layer $z'=O(\ep)$, the three stationary values are again separated only by $O(\ep^2)$.
The corresponding edge-layer scale is
\begin{align}
z_{\mathrm{edge}}^{\subL}
\sim
\frac{\alpha^\subL mg_\eff L}
{v_\subC^\subG(\bT)-v_\subC^\subL(\bT)}.
\label{e:rank-rev-L}
\end{align}
We now summarize the stationary ordering resolved by the present theory.
The notation $\sim$ in the edge relations below means that the differences between the indicated stationary values are $O(\ep^2)$.
For $g_\eff>0$, we have
\begin{align}
&F_g^{(\subL,\subG)}\sim F_g^{(\subG,\subL)}\sim F_g^\homo,
\qquad
(z'=O(\ep)),
\label{e:rank-left}\\
&F_g^{(\subL,\subG)}<F_g^{(\subG,\subL)}<F_g^\homo,
\qquad
(z\gg\ep,\quad z'\gg\ep),
\label{e:rank-center}\\
&F_g^{(\subL,\subG)}\sim F_g^{(\subG,\subL)}\sim F_g^\homo,
\qquad
(z=O(\ep)).
\label{e:rank-right}
\end{align}
Away from the coexistence edges, the resolved ordering is therefore \eqref{e:rank-center}.
The edge layers express the ordinary coexistence-edge degeneracy: at $g=0$ and $\Xi=0$ the three candidates collapse to the same limiting state, and weak gravity and heat conduction split them only at $O(\ep^2)$ inside an $O(\ep)$ volume-fraction layer.
For $g_\eff<0$, the roles of $(\subL,\subG)$ and $(\subG,\subL)$ are exchanged.

\begin{figure}[tb]
\centering
\includegraphics[width=0.9\linewidth]{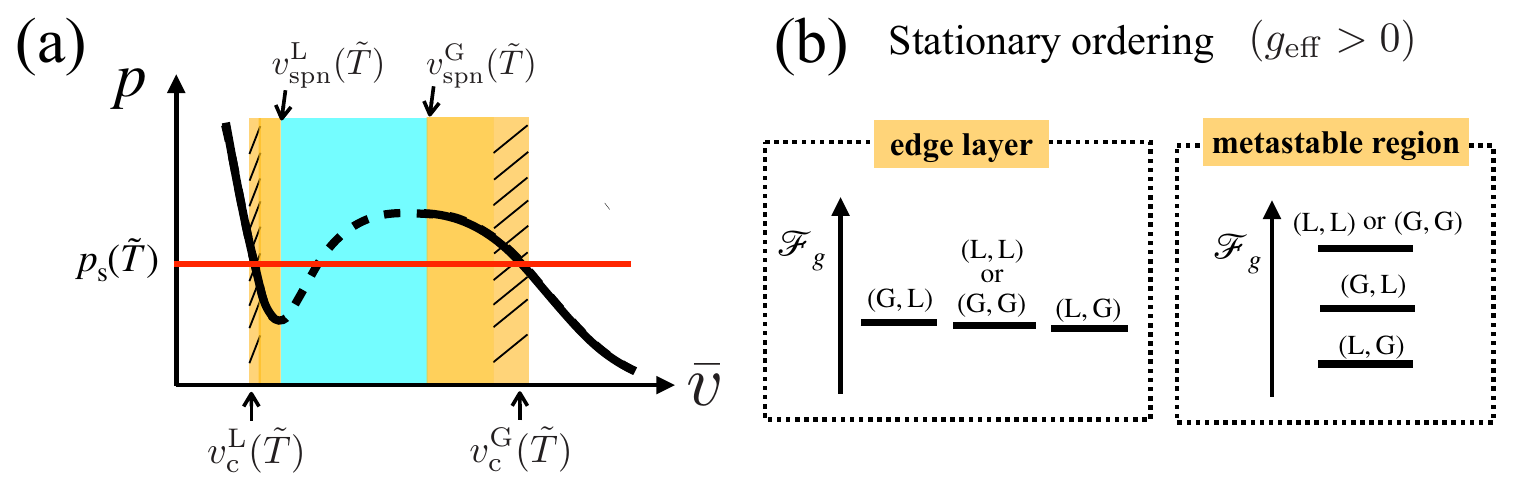}
\caption{Specific-volume subdivision and stationary ordering in the coexistence interval for $g_\eff>0$. (a) Isothermal pressure curve at $\bT$. The orange regions are metastable homogeneous continuations outside the spinodal interval, and the cyan region is the spinodal interval. The hatched strips inside the orange regions indicate the $O(\ep)$ edge layers adjacent to $v_\subC^\subL(\bT)$ and $v_\subC^\subG(\bT)$. (b) Schematic ordering in an edge layer and in a metastable region. In the edge layer, the three stationary values are not resolved at first order. In the metastable region, $(\subL,\subG)$ has the lowest value, followed by $(\subG,\subL)$, while the homogeneous continuation remains locally metastable. The spinodal interval in (a), where the homogeneous continuation is locally unstable, is not shown in (b). For $g_\eff<0$, the roles of $(\subL,\subG)$ and $(\subG,\subL)$ are exchanged.}
\label{fig:global-hierarchy}
\end{figure}

Figure~\ref{fig:global-hierarchy} summarizes the stationary ordering in the edge layer and in the metastable region, and also indicates the spinodal subdivision of the homogeneous branch used in the stability analysis below.
Figure~\ref{fig:global-hierarchy}(a) shows an equation of state far below the critical temperature.
The equation of state is plotted with a solid line when the corresponding homogeneous state is locally stable, whereas the dotted line indicates the spinodal-unstable continuation.
The spinodal boundaries shown in the figure are defined by
\begin{align}
\left.
\pderf{p}{v}{\bT}
\right|_{v=v_{\mathrm{spn}}^{\subL}(\bT)}
=0,
\qquad
\left.
\pderf{p}{v}{\bT}
\right|_{v=v_{\mathrm{spn}}^{\subG}(\bT)}
=0.
\label{e:vspn-def}
\end{align}
Away from the edge layers, the coexistence interval is divided by these spinodal boundaries into locally stable and locally unstable homogeneous continuations.
The metastable regions are the parts of the coexistence interval that are away from the edge layers and outside the spinodal interval,
\begin{align}
z\gg\ep,
\quad z'\gg\ep,
\quad
\bar v<v_{\mathrm{spn}}^{\subL}(\bT)
\quad\text{or}\quad
\bar v>v_{\mathrm{spn}}^{\subG}(\bT),
\end{align}
where the homogeneous state remains locally stable.
The spinodal interval is
\begin{align}
v_{\mathrm{spn}}^{\subL}(\bT)<\bar v<v_{\mathrm{spn}}^{\subG}(\bT),
\end{align}
where the homogeneous state is locally unstable.
Thus the metastable/spinodal distinction is not a difference in the ordering of the resolved stationary values in \eqref{e:rank-center}; it is a difference in the local stability of the homogeneous branch.
In the liquid-side superheated-liquid region, the locally stable homogeneous state is liquid-like.
In the gas-side supercooled-gas region, it is gas-like.

\section{Local stability of the stationary states}
\label{s:stability}

In this section, we investigate the stability of the homogeneous and heterogeneous solutions obtained in the previous sections.
To do this, we examine the second-order variation of the free energy, $\delta^2 \calF_g$, around each solution.
A solution is considered stable if $\delta^2 \calF_g > 0$ for all infinitesimal deviations. This is equivalent to requiring that the Hessian matrix of $\calF_g$ with respect to the variational parameters be positive definite.
Throughout this section, $\delta^2 \calF_g$ is used as a variational measure of local stability on the fixed-$(\bT,V,N,mgL,\Xi)$ manifold.
Its role here is to characterize the local geometry of the variational landscape in the fixed-$\bT$ description.

The variational function $\calF_g$, introduced in \eqref{e:Fg-var-neq} and rewritten as \eqref{e:Fg-var-neq-re2}, is
\begin{align}
{\calF_g}(c^\subl, \phi^\subl; \bT, V, N, mgL, \Xi)
&= F^\subl(\bT^\subl, V\phi^\subl, Nc^\subl)
+ F^\subu(\bT^\subu, V(1-\phi^\subl), N(1-c^\subl))
- \frac{mgL}{2}N\left(c^\subl - \phi^\subl\right),
\label{e:Fg-var-neq-stab}
\end{align}
where \(c^\subl = \calN^\subl / N\) and \(\phi^\subl = \calV^\subl / V\) are the variational parameters, and the temperatures $\bT^\subl$ and $\bT^\subu$, as defined in \eqref{e:bT-Tlu}, are given by:
\begin{align}
\bT^\subl = \bT - \frac{\Xi}{2}(1 - c^\subl),\qquad \bT^\subu = \bT + \frac{\Xi}{2}c^\subl.
\label{e:bT-lu-stab}
\end{align}

Consider a small deviation $(\delta c^\subl, \delta \phi^\subl)$ around $(c^\subl, \phi^\subl)$. The corresponding change in the variational free energy is
\begin{align}
\delta \calF_g(\delta c^\subl, \delta \phi^\subl)
= \calF_g(c^\subl+\delta c^\subl, \phi^\subl+\delta \phi^\subl; \bT, V, N, mgL,\Xi)
- \calF_g(c^\subl, \phi^\subl; \bT, V, N, mgL,\Xi).
\label{e:delta-Fg}
\end{align}
At a stationary point, the linear terms of $O(\delta)$ vanish because the variational equations are satisfied. Therefore,
\begin{align}
\delta\calF_g(\delta c^\subl, \delta \phi^\subl) = \delta^2 \calF_g,
\label{e:delta2-F-eq}
\end{align}
where $\delta^2 \calF_g$ is the second-order variation at that stationary point, given by
\begin{align}
\delta^2\calF_g=
\frac{1}{2}
\begin{pmatrix}
\delta c^\subl& \delta\phi^\subl
\end{pmatrix}
H
\begin{pmatrix}
\delta c^\subl \\
\delta\phi^\subl
\end{pmatrix}.
\label{e:matrix-stability}
\end{align}
$H$ is the Hessian matrix:
\begin{align}
H\equiv \nabla_{\vec{x}}\nabla_{\vec{x}} \mathcal{F}_g =
\begin{pmatrix}
A & C \\
C & B
\end{pmatrix}
=
\begin{pmatrix}
\pdert{\calF_g}{c^\subl} & \pderc{\calF_g}{c^\subl}{\phi^\subl} \\
\pderc{\calF_g}{c^\subl}{\phi^\subl} & \pdert{\calF_g}{\phi^\subl}
\end{pmatrix}.
\label{e:Hessian_def}
\end{align}
Here the derivatives are taken with respect to $\vec{x}=(c^\subl,\phi^\subl)^T$ at fixed $(\bT,V,N,mgL,\Xi)$.
The Hessian matrix $H(c^\subl,\phi^\subl)$ is defined at any point on the variational landscape. The local stability criterion below is applied by evaluating it at a stationary point.
A stationary solution is locally stable if
\begin{align}
\delta^2 \calF_g > 0,
\label{e:stability-matrix}
\end{align}
which is equivalent to the condition that the Hessian matrix $H$ is positive definite. This, in turn, is equivalent to the conditions that the trace and determinant of $H$ are both positive:
\begin{align}
\mathrm{Tr}(H) = A + B > 0, \quad
\mathrm{Det}(H) = AB - C^2 > 0.
\label{e:stability_conditions}
\end{align}

\subsection{Hessian matrix for subregions}

We calculate the elements of the Hessian matrix.
In \eqref{e:Fg-var-neq-stab}, the gravity term is linear in the combination $(c^\subl-\phi^\subl)$ and therefore does not contribute to the Hessian.

Therefore, the Hessian matrix $H$ is composed solely of the free-energy contributions from the two subregions, $F^\subl$ and $F^\subu$:
\begin{align}
H = H^\subl + H^\subu =
\begin{pmatrix}
A^\subl & C^\subl \\
C^\subl & B^\subl
\end{pmatrix}
+
\begin{pmatrix}
A^\subu & C^\subu \\
C^\subu & B^\subu
\end{pmatrix}.
\end{align}
This additive structure allows us to evaluate the contribution of each subsystem explicitly.
Here
\begin{align}
\begin{pmatrix}
A^\subl & C^\subl \\
C^\subl & B^\subl
\end{pmatrix}
=
\begin{pmatrix}
\pdert{F^\subl}{c^\subl} & \pderc{F^\subl}{c^\subl}{\phi^\subl} \\
\pderc{F^\subl}{c^\subl}{\phi^\subl} & \pdert{F^\subl}{\phi^\subl}
\end{pmatrix},
\quad
\begin{pmatrix}
A^\subu & C^\subu \\
C^\subu & B^\subu
\end{pmatrix}
=
\begin{pmatrix}
\pdert{F^\subu}{c^\subu} & \pderc{F^\subu}{c^\subu}{\phi^\subu} \\
\pderc{F^\subu}{c^\subu}{\phi^\subu} & \pdert{F^\subu}{\phi^\subu}
\end{pmatrix}.
\end{align}

To calculate each component, we should note that $F^\subl=F^\subl(\bT^\subl(c^\subl),V\phi^\subl, N c^\subl)$
and $F^\subu=F^\subu(\bT^\subu(c^\subu),V\phi^\subu, N c^\subu)$ with $c^\subu=1-c^\subl$ and $\phi^\subu=1-\phi^\subl$.
We have
\begin{align}
&\pderf{F^\subl}{\phi^\subl}{c^\subl}=-p^\subl V,\qquad
\pderf{F^\subu}{\phi^\subu}{c^\subu}=-p^\subu V,\\
&\pderf{F^\subl}{c^\subl}{\phi^\subl}=\pderf{F^\subl}{\bT^\subl}{N^\subl,V^\subl}\der{\bT^\subl}{c^\subl}+N\pderf{F^\subl}{N^\subl}{\bT^\subl,V^\subl}=-\frac{\Xi}{2} S^\subl+\mu^\subl N ,\\
&\pderf{F^\subu}{c^\subu}{\phi^\subu}=\pderf{F^\subu}{\bT^\subu}{N^\subu,V^\subu}\der{\bT^\subu}{c^\subu}+N\pderf{F^\subu}{N^\subu}{\bT^\subu,V^\subu}=\frac{\Xi}{2} S^\subu+\mu^\subu N,
\end{align}
where $\bP^\subl=\bP^\subl(\bT^\subl(c^\subl),V\phi^\subl,N c^\subl)$,
$\bP^\subu=\bP^\subu(\bT^\subu(c^\subu),V\phi^\subu,N c^\subu)$,
$S^\subl=S^\subl(\bT^\subl(c^\subl),V\phi^\subl,N c^\subl)$,
and
$S^\subu=S^\subu(\bT^\subu(c^\subu),V\phi^\subu,N c^\subu)$.
Using $\phi^\subl =v^\subl c^\subl N/V$, the second derivatives are
\begin{align}
&\pdertf{F^\subl}{\phi^\subl}{c^\subl}
=-V\pderf{\bP^\subl}{\phi^\subl}{c^\subl}
=-\frac{V^2}{Nc^\subl}\pderf{\bP^\subl}{v^\subl}{c^\subl}
=\frac{V^2}{N c^\subl v^\subl \alpha^\subl}
=\frac{V}{\alpha^\subl \phi^\subl},\\
&\pdertf{F^\subu}{\phi^\subu}{c^\subu}=\frac{V}{\alpha^\subu \phi^\subu},\\
&\pderc{F^\subl}{c^\subl}{\phi^\subl}
=-V\pderf{\bP^\subl}{c^\subl}{\phi^\subl}
= -V\! \left[
\pderf{\bP^\subl}{c^\subl}{T,\phi^\subl}
+ \pderf{\bP^\subl}{T}{v^\subl,\phi^\subl}
\der{\bT^\subl}{c^\subl}
\right]\nm
&\qquad \qquad
= -V \left[ \frac{1}{\alpha^\subl c^\subl} +  \frac{\Xi}{2} \pderf{p^\subl}{T}{v^\subl}\right],\\
&\pderc{F^\subu}{c^\subu}{\phi^\subu}=-V\pderf{\bP^\subu}{c^\subu}{\phi^\subu}
= -V \left[
\pderf{\bP^\subu}{c^\subu}{T,\phi^\subu}
+ \pderf{\bP^\subu}{T}{v^\subu,\phi^\subu}
\der{\bT^\subu}{c^\subu}
\right]\nm
&\qquad\qquad
= -V \left[ \frac{1}{\alpha^\subu c^\subu} -  \frac{\Xi}{2} \pderf{p^\subu}{T}{v^\subu}\right],\\
&\pdertf{F^\subl}{c^\subl}{\phi^\subl}
=N\pderf{\mu^\subl}{c^\subl}{\phi^\subl}
-\frac{\Xi}{2}\pderf{S^\subl}{c^\subl}{\phi^\subl} \nm
&\qquad\qquad
=N\left(-s^\subl \pderf{\bT^\subl}{c^\subl}{\phi^\subl}
+v^\subl\pderf{\bP^\subl}{c^\subl}{\phi^\subl}\right)
-\frac{\Xi}{2}N\pderf{S^\subl}{N^\subl}{\bT^\subl,V^\subl} \nm
&\qquad\qquad
=-\frac{N^2{v^\subl}^3 }{\phi^\subl V}
\pderf{\bP^\subl}{v^\subl}{\phi^\subl}
-\frac{\Xi}{2}N
\left(s^\subl-\pderf{\mu^\subl}{\bT^\subl}{v^\subl}\right) \nm
&\qquad\qquad
=-\frac{N^2{v^\subl}^3 }{\phi^\subl V}
\left[\!
\pderf{\bP^\subl}{v^\subl}{\bT^\subl,\phi^\subl}
\!+ \!\pderf{\bP^\subl}{\bT^\subl}{v^\subl,\phi^\subl}
\!\pderf{\bT^\subl}{v^\subl}{\phi^\subl}
\!\right]
-\frac{\Xi}{2}N
\!\left(s^\subl-\pderf{\mu^\subl}{\bT^\subl}{v^\subl}\!\right) \nm
&\qquad\qquad
=\frac{V\phi^\subl}{\alpha^\subl{c^\subl}^2}
+ \frac{\Xi}{2}N v^\subl
\pderf{\bP^\subl}{\bT^\subl}{v^\subl,\phi^\subl}
-\frac{\Xi}{2}N
\left(s^\subl-\pderf{\mu^\subl}{\bT^\subl}{v^\subl}\right) \nm
&\qquad\qquad
=\frac{V\phi^\subl}{\alpha^\subl{c^\subl}^2}
+\Xi N\pderf{\mu^\subl}{\bT^\subl}{v^\subl},\\
&\pdertf{F^\subu}{c^\subu}{\phi^\subu}
=\frac{V\phi^\subu}{\alpha^\subu{c^\subu}^2}
-\Xi N\pderf{\mu^\subu}{\bT^\subu}{v^\subu}.
\end{align}

We thus obtain the Hessian matrices for the lower and upper regions as
\begin{align}
\frac{H^\subl}{V}&=\frac{1}{\alpha^\subl c^\subl}
\begin{pmatrix}
\frac{\phi^\subl}{c^\subl} & -1\\
-1 & \frac{c^\subl}{\phi^\subl}
\end{pmatrix}
+\frac{\Xi}{2}
\begin{pmatrix}
2\frac{N}{V}\pderf{\mu^\subl}{\bT^\subl}{v^\subl} & -\pderf{\bP^\subl}{\bT^\subl}{v^\subl}\\
-\pderf{\bP^\subl}{\bT^\subl}{v^\subl} & 0
\end{pmatrix}
,\label{e:Hessian-l}\\
\frac{H^\subu}{V}&=\frac{1}{\alpha^\subu c^\subu}
\begin{pmatrix}
\frac{\phi^\subu}{c^\subu} & -1\\
-1 & \frac{c^\subu}{\phi^\subu}
\end{pmatrix}
-\frac{\Xi}{2}
\begin{pmatrix}
2\frac{N}{V}\pderf{\mu^\subu}{\bT^\subu}{v^\subu} & -\pderf{\bP^\subu}{\bT^\subu}{v^\subu}\\
-\pderf{\bP^\subu}{\bT^\subu}{v^\subu} & 0
\end{pmatrix}.
\label{e:Hessian-u}
\end{align}

To examine the stability condition \eqref{e:stability_conditions}, we formulate the traces and determinants.
For the subregions $\subl$ and $\subu$, \eqref{e:Hessian-l} and \eqref{e:Hessian-u} lead to
\begin{align}
&\mathrm{Tr}(H^\subl/V) = \frac{1}{\alpha^\subl c^\subl} \left( \frac{\phi^\subl}{c^\subl} + \frac{c^\subl}{\phi^\subl} \right) + \frac{\Xi}{\bar{v}} \pderf{\mu^\subl}{\bT^\subl}{v^\subl} ,\label{e:tr-Hl}\\
&\mathrm{Tr}(H^\subu/V) = \frac{1}{\alpha^\subu c^\subu} \left( \frac{\phi^\subu}{c^\subu} + \frac{c^\subu}{\phi^\subu} \right) - \frac{\Xi}{\bar{v}} \pderf{\mu^\subu}{\bT^\subu}{v^\subu} ,\label{e:tr-Hu}\\
&\mathrm{Det}(H^\subl/V) = \frac{\Xi}{\alpha^\subl c^\subl} \left[ \frac{1}{v^\subl} \pderf{\mu^\subl}{\bT^\subl}{v^\subl} - \pderf{\bP^\subl}{\bT^\subl}{v^\subl} \right] -\frac{\Xi^2}{4}\left(\pderf{\bP^\subl}{\bT^\subl}{v^\subl}\right)^2 \nm
&\qquad\qquad\quad
= -\frac{\Xi  s^\subl}{\alpha^\subl \phi^\subl \bar{v}}
 -\frac{\Xi^2}{4}\left(\pderf{\bP^\subl}{\bT^\subl}{v^\subl}\right)^2,\label{e:det-Hl}\\
&\mathrm{Det}(H^\subu/V) =  \frac{\Xi s^\subu}{\alpha^\subu \phi^\subu \bar{v}}-\frac{\Xi^2}{4}\left(\pderf{\bP^\subu}{\bT^\subu}{v^\subu}\right)^2,\label{e:det-Hu}
\end{align}
where we have utilized the thermodynamic identity $\pderf{\mu}{T}{v} = -s + v\pderf{p}{T}{v}$ and $c^\subl v^\subl=\phi^\subl\bar{v}$.

\subsection{Traces and determinants for Hessian matrices}

We next combine the two subregion contributions, $H=H^\subl+H^\subu$.
For compact notation, we introduce
\begin{align}
p_T^\subl\equiv \pderf{\bP^\subl}{\bT^\subl}{v^\subl},
\qquad
p_T^\subu\equiv \pderf{\bP^\subu}{\bT^\subu}{v^\subu},
\end{align}
and use the identity
\begin{align}
\pderf{\mu^\alpha}{\bT^\alpha}{v^\alpha}
=
-s^\alpha+v^\alpha p_T^\alpha
\qquad
(\alpha=\subl,\subu).
\end{align}
Then the total Hessian can be written as
\begin{align}
\frac{H}{V}
=
\begin{pmatrix}
A_0+\dfrac{\Xi}{\bar v}D_T
&
-B_0-\dfrac{\Xi}{2}\Delta p_T
\\[4pt]
-B_0-\dfrac{\Xi}{2}\Delta p_T
&
C_0
\end{pmatrix},
\label{e:H-total-matrix}
\end{align}
where
\begin{align}
A_0
&\equiv
\frac{1}{\alpha^\subl c^\subl}\frac{\phi^\subl}{c^\subl}
+
\frac{1}{\alpha^\subu c^\subu}\frac{\phi^\subu}{c^\subu},
\qquad
C_0
\equiv
\frac{1}{\alpha^\subl c^\subl}\frac{c^\subl}{\phi^\subl}
+
\frac{1}{\alpha^\subu c^\subu}\frac{c^\subu}{\phi^\subu},
\nonumber\\
B_0
&\equiv
\frac{1}{\alpha^\subl c^\subl}
+
\frac{1}{\alpha^\subu c^\subu},
\qquad
\Delta p_T
\equiv
p_T^\subl-p_T^\subu,
\nonumber\\
D_T
&\equiv
s^\subu-s^\subl
+
v^\subl p_T^\subl
-
v^\subu p_T^\subu .
\end{align}
Therefore,
\begin{align}
\mathrm{Tr}(H/V)
&=
A_0+C_0+\frac{\Xi}{\bar v}D_T,
\label{e:trace-middle}
\\
\mathrm{Det}(H/V)
&=
\left(A_0+\frac{\Xi}{\bar v}D_T\right)C_0
-
\left(B_0+\frac{\Xi}{2}\Delta p_T\right)^2
\nonumber\\
&=
A_0C_0-B_0^2
+
\Xi\left(
\frac{C_0}{\bar v}D_T
-
B_0\Delta p_T
\right)
-
\frac{\Xi^2}{4}(\Delta p_T)^2.
\label{e:det-middle}
\end{align}
Using $c^\subl v^\subl=\phi^\subl\bar v$ and $c^\subu v^\subu=\phi^\subu\bar v$,
the zeroth-order part of $\mathrm{Det}(H/V)$ is simplified as
\begin{align}
A_0C_0-B_0^2
=
\frac{1}{\alpha^\subl\alpha^\subu c^\subl c^\subu}
\left(
\sqrt{\frac{v^\subl}{v^\subu}}
-
\sqrt{\frac{v^\subu}{v^\subl}}
\right)^2.
\end{align}
The linear term in \(\Xi\) can be written as
\begin{align}
\frac{C_0}{\bar v}D_T
-
B_0\Delta p_T
=
-
\frac{1}{\alpha^\subl\alpha^\subu c^\subl c^\subu}\mathcal T,
\end{align}
where
\begin{align}
\mathcal T
&=
(v^\subu-v^\subl)
\left[
\frac{\alpha^\subl c^\subl}{v^\subu}
\left(
p_T^\subl
-
\frac{s^\subl-s^\subu}{v^\subl-v^\subu}
\right)
+
\frac{\alpha^\subu c^\subu}{v^\subl}
\left(
p_T^\subu
-
\frac{s^\subl-s^\subu}{v^\subl-v^\subu}
\right)
\right].
\label{e:thermal_coupling_T}
\end{align}
Combining these results, we obtain
\begin{align}
&\mathrm{Tr}(H/V)
=
\frac{1}{\alpha^\subl c^\subl}
\left(
\frac{\phi^\subl}{c^\subl}
+
\frac{c^\subl}{\phi^\subl}
\right)
+
\frac{1}{\alpha^\subu c^\subu}
\left(
\frac{\phi^\subu}{c^\subu}
+
\frac{c^\subu}{\phi^\subu}
\right)
+
\frac{\Xi}{\bar v}
\left(
s^\subu-s^\subl
+
v^\subl p_T^\subl
-
v^\subu p_T^\subu
\right),
\label{e:trace}
\\
&\mathrm{Det}(H/V)
=
\frac{1}{\alpha^\subl\alpha^\subu c^\subl c^\subu}
\left[
\left(
\sqrt{\frac{v^\subl}{v^\subu}}
-
\sqrt{\frac{v^\subu}{v^\subl}}
\right)^2
-
\Xi\mathcal T
\right]
-
\frac{\Xi^2}{4}
\left(p_T^\subl-p_T^\subu\right)^2.
\label{e:det}
\end{align}

\subsection{Stability of the heterogeneous solutions}

For the heterogeneous solutions corresponding to the phase coexistence, the subsystems represent the liquid and gas phases.
The compressibilities $\alpha^\subl$ and $\alpha^\subu$ correspond to $\alpha^\subL$ and $\alpha^\subG$, which are positive and finite due to the intrinsic thermodynamic stability of each phase.
Thus, $\mathrm{Tr}(H)$ is a positive finite value at $\Xi=0$, and this positivity is not disturbed by sufficiently small $\Xi$.

The stability is thus determined by the determinant in \eqref{e:det}.
When the temperature $\bT$ is set far below the critical temperature, the specific volumes of the two phases, $v_\subC^\subL(\bT)$ and $v_\subC^\subG(\bT)$, are significantly different.
Consequently, we have
\begin{align}
\mathrm{Det}(H/V) \approx
\frac{1}{\alpha^\subL\alpha^\subG c^\subL c^\subG}
\left(
\sqrt{\frac{v_\subC^\subL}{v_\subC^\subG}}
-
\sqrt{\frac{v_\subC^\subG}{v_\subC^\subL}}
\right)^2.
\end{align}
The significant density difference between liquid and gas provides a large positive value of $\mathrm{Det}(H)$.
Thus, within the present linear-response regime, the heterogeneous solutions satisfy the stability conditions $\mathrm{Tr}(H)>0$ and $\mathrm{Det}(H)>0$ for sufficiently small heat flow.

\subsection{Stability of the homogeneous solution}
\label{s:stability-homo}

In this subsection, we analyze the Hessian stability of the homogeneous solution.
For a homogeneous state, the two subregions are not distinct phases; they are two formal parts of the same single-phase fluid.
The Hessian test below therefore characterizes the local geometry of $\calF_g$ with respect to this formal split, not an instability associated with a liquid-gas interface.
For the homogeneous state, we therefore use
\begin{align}
c^\subl=\phi^\subl+O(\ep),\quad
c^\subu=\phi^\subu+O(\ep),\quad
\alpha^\subl=\alpha^\homo+O(\ep),\quad
\alpha^\subu=\alpha^\homo+O(\ep).
\end{align}
Using these relations in \eqref{e:trace}, we obtain
\begin{equation}
\mathrm{Tr}(H/V)
=
\frac{2}{\alpha^\homo\Lambda}\left(1+O(\ep)\right),
\end{equation}
where
\begin{align}
\Lambda\equiv \phi^\subl\phi^\subu=\phi^\subl(1-\phi^\subl).
\label{e:Lambda-def}
\end{align}
For any formal split with $\Lambda>0$, the trace condition is therefore reduced to
\begin{align}
\alpha^\homo>0.
\end{align}

We next estimate the determinant \eqref{e:det}.
For the homogeneous state, the specific volumes in the two formal subregions differ only by $O(\ep)$.
We define the dimensionless specific-volume contrast by
\begin{align}
\mathcal A\equiv \frac{v^\subu-v^\subl}{\bar v}.
\label{e:A-homo-def}
\end{align}
Then $\mathcal A=O(\ep)$, and
\begin{align}
\left(
\sqrt{\frac{v^\subl}{v^\subu}}
-
\sqrt{\frac{v^\subu}{v^\subl}}
\right)^2
=
\mathcal A^2+O(\ep^3).
\label{e:homo-sqrt-term}
\end{align}
The temperature difference between the two formal subregions is
\begin{align}
\bT^\subu-\bT^\subl=\frac{\Xi}{2}.
\label{e:homo-T-diff}
\end{align}

We now evaluate the leading contribution of the thermal coupling term $\mathcal T$ in \eqref{e:thermal_coupling_T}.
For the homogeneous state,
\begin{align}
p_T^\subl=p_T+O(\ep),\qquad
p_T^\subu=p_T+O(\ep),\qquad
v^\subl=\bar v+O(\ep),\qquad
v^\subu=\bar v+O(\ep),
\end{align}
where $p_T=(\partial p/\partial T)_v$ is evaluated at $(\bT,\bar v)$.
Thus, \eqref{e:thermal_coupling_T} gives, to $O(\ep)$,
\begin{align}
\mathcal T
&=
(v^\subu-v^\subl)\frac{\alpha^\homo}{\bar v}
\left[
p_T
-
\frac{s^\subl-s^\subu}{v^\subl-v^\subu}
\right]
+O(\ep^2).
\label{e:T-homo-1}
\end{align}
Using
\begin{align}
s^\subl-s^\subu
=
p_T(v^\subl-v^\subu)
+
\frac{c_v}{\bT}(\bT^\subl-\bT^\subu)
+O(\ep^2),
\end{align}
where
\begin{align}
c_v\equiv \bT\left(\frac{\partial s}{\partial T}\right)_v
\end{align}
is evaluated at $(\bT,\bar v)$, and using \eqref{e:homo-T-diff}, we find
\begin{align}
\mathcal T
=
-\frac{\alpha^\homo c_v}{2\bar v\bT}\Xi
+O(\ep^2).
\label{e:T-homo}
\end{align}
The last term in \eqref{e:det}, proportional to $\Xi^2(p_T^\subl-p_T^\subu)^2$, is of higher order.
Substituting \eqref{e:homo-sqrt-term} and \eqref{e:T-homo} into \eqref{e:det}, we obtain
\begin{equation}
\mathrm{Det}(H/V)
=
\frac{1}{(\alpha^\homo)^2\Lambda}
\left[
\mathcal A^2
+
\frac{\alpha^\homo c_v}{2\bar v\bT}\Xi^2
\right]
+O(\ep^3).
\label{e:det-homo-structural-final}
\end{equation}
For any fixed formal split with $\Lambda>0$, this expression is nonnegative when
$\alpha^\homo>0$ and $c_v>0$.
Within the present order, it vanishes only in the degenerate equilibrium limit
$\mathcal A=0$ and $\Xi=0$.
Thus, for a fixed formal split, the homogeneous state satisfies the Hessian stability conditions in the present expansion.

Because this conclusion holds for any $O(\ep)$ value of $\mathcal A$, it is unchanged when the formal split is specialized to the homogeneous stationary solution.
Equations \eqref{e:det-Hl} and \eqref{e:det-Hu} show that the individual subregion determinants vanish at $\Xi=0$ and acquire opposite linear contributions in $\Xi$.
These layer-wise determinants are not, by themselves, the stability criterion for the constrained two-variable problem.
For the total Hessian, the opposite linear contributions do not produce a term proportional to $\mathcal A\Xi$ in \eqref{e:det-homo-structural-final}.
Consequently, within the homogeneous-state Hessian calculation, no linear $O(\Xi)$ bulk instability appears.
The leading heat-conduction contribution to $\mathrm{Det}(H/V)$ is the positive quadratic term proportional to $c_v\Xi^2$.
Thus, the local stability of the homogeneous branch is controlled, to the present order, by the ordinary isothermal compressibility condition $\alpha^\homo>0$.
Equivalently, the loss of Hessian stability occurs at the spinodal boundaries defined in \eqref{e:vspn-def}, whereas heat conduction does not generate an additional linear-$\Xi$ bulk instability.

\section{Free-energy landscape}
\label{sec:landscape}

We now examine the variational free energy $\calF_g$ as a two-variable function on the fixed-$(\bT,V,N,mgL,\Xi)$ manifold.
The preceding sections determined the stationary values and the Hessian stability of the stationary solutions; here we use this two-dimensional landscape to show how the heterogeneous minima and the homogeneous continuation are embedded in the surrounding barrier structure.
This representation makes visible the part of the fixed-$\bT$ structure that is not captured by the stationary ordering alone.
The landscape is drawn at fixed $(\bT,V,N,mgL,\Xi)$, and should not be identified with a fluctuation landscape for a boundary-temperature experiment at fixed $(T_1,T_2)$.
Its role here is to display the thermodynamic geometry of the fixed-$\bT$ variational problem.

We use the variational function $\calF_g$ in \eqref{e:Fg-var-neq-stab}.
It is a function of the variational parameters $(c^\subl,\phi^\subl)$.
The ordering of the stationary values has already been derived in \eqref{e:rank-left}--\eqref{e:rank-right}, and the local stability of the homogeneous state is controlled by the spinodal boundaries $v_{\mathrm{spn}}^{\subL}$ and $v_{\mathrm{spn}}^{\subG}$ introduced in \eqref{e:vspn-def}.
We use the landscapes below to show how the spinodal interval, gas-side supercooled-gas region, and gas-side edge layer appear in $\calF_g$.

For plotting and for the later local expansion, we use the following linear coordinates on the same variational plane:
\begin{align}
\xi_1\equiv \phi^\subl-\frac{1}{2},
\qquad
\xi_2\equiv \frac{1}{2}\left(\phi^\subl-c^\subl\right),
\label{e:xi-landscape-def}
\end{align}
With the argument order of \eqref{e:Fg-var-neq-stab}, we write
\begin{align}
\calF_g(\xi_1,\xi_2;\bT,V,N,mgL,\Xi)
&\equiv
\calF_g\left(
\frac{1}{2}+\xi_1-2\xi_2,\frac{1}{2}+\xi_1;
\bT,V,N,mgL,\Xi
\right).
\label{e:Fg-xi-landscape}
\end{align}
The physical domain is determined by
\begin{align}
0<c^\subl<1,\qquad
0<c^\subu<1,\qquad
0<\phi^\subl<1,\qquad
0<\phi^\subu<1,
\end{align}
which corresponds to
\begin{align}
-\frac{1}{2}<\xi_1<\frac{1}{2},
\qquad
\frac{\xi_1}{2}-\frac{1}{4}<\xi_2<\frac{\xi_1}{2}+\frac{1}{4}.
\end{align}
In these variables, the local specific volumes are
\begin{align}
v^\subl=\bar v\frac{\phi^\subl}{c^\subl},
\qquad
v^\subu=\bar v\frac{\phi^\subu}{c^\subu}.
\label{e:vlu-xi-landscape}
\end{align}

For the numerical illustration, we use a van der Waals model with standard CO$_2$ parameters,
\begin{align}
&f_\vdW(v,T)=-RT\ln(v-b)-\frac{a}{v}-c_vT\ln T,\\
&s_\vdW(v,T)=R\ln(v-b)+c_v\ln T,
\end{align}
with $R=8.31\,\mathrm{J/(K\,mol)}$, $c_v=5R$, and
\begin{align}
a=0.365\,\mathrm{Pa\cdot m^6/mol^2},\quad
b=4.28\times10^{-5}\,\mathrm{m^3/mol}.
\end{align}
In the numerical examples below, the van der Waals thermodynamic functions are evaluated in molar units. Thus $v$ is the molar volume, $\kB$ is replaced by $R$, and the quantity denoted by $mgL$ in the numerical plots represents the molar gravitational work $MgL$.
Equivalently, this numerical convention uses the replacements $N\to n$, $m\to M$, and $\kB\to R$.
Here $n$ is the number of moles and $M$ is the molar mass.
We keep the symbol $N$ in the normalized free energies to match the theoretical notation; accordingly, $F_g/N$ in the figures denotes the corresponding molar specific free energy in units of J/mol.
With these parameters, the liquid--gas coexistence at $\bT=260\,\mathrm{K}$ occurs at a pressure of the order of MPa, as expected for CO$_2$.
These numerical landscapes provide representative illustrations of the variational structure.

\begin{figure*}[t]
\centering
\includegraphics[width=0.92\linewidth]{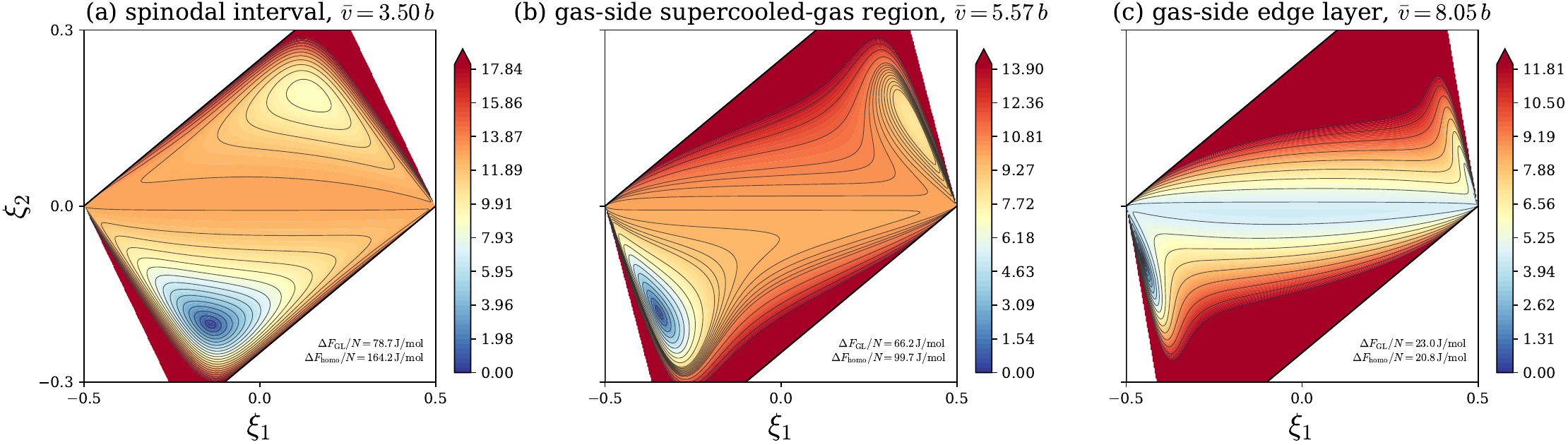}
\caption{
Representative variational landscapes on the gas-rich side of the coexistence region for the van der Waals model at $\bT=260.0\,\mathrm{K}$, $\Xi=5.0\,\mathrm{K}$, and fixed effective gravity $mg_\eff L=200\,\mathrm{J/mol}>0$.
Panels (a), (b), and (c) correspond to $\bar v=3.50b$, $5.57b$, and $8.05b$.
The color bars denote $\sqrt{\Delta\calF_g/N}$ with panelwise normalization.
Panels (a), (b), and (c) illustrate the spinodal interval, the gas-side supercooled-gas region, and a finite-parameter edge-scale case, respectively; the stationary-value gaps and local stability are stated in the text.
}
\label{fig:landscape-gas-side-three}
\end{figure*}

We focus on the case $g_\eff>0$, for which the heterogeneous configuration $(\subL,\subG)$ has the lower value away from the coexistence edges.
When the physical control parameter $mgL$ is needed in the numerical evaluation of \eqref{e:Fg-var-neq-stab}, it is chosen as
\begin{align}
mgL=mg_\eff L-\bar v\frac{d\Ps}{d\bT}\Xi.
\label{e:mgL-landscape-fixed-geff}
\end{align}
This ensures that all panels in Fig.~\ref{fig:landscape-gas-side-three} are compared at the same value of $g_\eff$.

Figure~\ref{fig:landscape-gas-side-three} shows representative landscapes on the gas-rich side.
We use
\begin{align}
\bT=260.0\,\mathrm{K},\qquad
\Xi=5.0\,\mathrm{K},\qquad
mg_\eff L=200\,\mathrm{J/mol}.
\end{align}
The three values of $\bar v$ are chosen to represent the spinodal interval, the gas-side supercooled-gas region, and the gas-side edge layer:
\begin{align}
\bar v=3.50b
\quad&\text{(spinodal interval)},\\
\bar v=5.57b
\quad&\text{(gas-side supercooled-gas region)},\\
\bar v=8.05b
\quad&\text{(gas-side edge layer)}.
\end{align}
For these parameters, $v_\subC^\subL=1.673b$ and $v_\subC^\subG=9.116b$.
The spinodal volumes are $v_{\mathrm{spn}}^{\subL}=2.03b$ and $v_{\mathrm{spn}}^{\subG}=5.10b$.
The gas-side distance
\begin{align}
z=\frac{v_\subC^\subG-\bar v}{v_\subC^\subG-v_\subC^\subL}
\end{align}
takes the values $0.755$, $0.476$, and $0.143$ in panels (a), (b), and (c), respectively.
The imposed temperature difference gives $|\Xi|/\bT=0.0192$.
Although $mg_\eff L/(R\bT)$ is fixed at $0.0926$, the small parameter is evaluated with the physical value of $mgL$ determined by \eqref{e:mgL-landscape-fixed-geff}.
This gives $\ep=0.070$, $0.057$, and $0.041$ for panels (a), (b), and (c), respectively.
Panel (a) shows the landscape between the two spinodal boundaries in \eqref{e:vspn-def}.
Panel (b) lies outside the spinodal interval and away from the coexistence edge on the scale set by $\ep$.
Panel (c) is an edge-scale case, with $z$ comparable to $mg_\eff L/(R\bT)$, although it is not the limiting case $z\to0$.

The plotted quantity is
\begin{align}
\frac{\Delta\calF_g(\xi_1,\xi_2)}{N}
\equiv
\frac{\calF_g(\xi_1,\xi_2)-F_g^{(\subL,\subG)}}{N},
\label{e:DeltaFg-landscape}
\end{align}
where $F_g^{(\subL,\subG)}$ is the value at the $(\subL,\subG)$ stationary point.
Each panel shows $\sqrt{\Delta\calF_g/N}$ in molar units.
The color normalization is chosen independently in each panel to resolve the local minima and barrier geometry.
Therefore, the color intensity should not be compared quantitatively between different panels.
The stationary-value ordering should be read from the explicit gaps stated below, while its asymptotic interpretation is given by
\eqref{e:rank-left}--\eqref{e:rank-right}.

For the spinodal-interval case shown in Fig.~\ref{fig:landscape-gas-side-three}(a), the ordering is
\begin{align}
F_g^{(\subL,\subG)}
<
F_g^{(\subG,\subL)}
<
F_g^\homo,
\end{align}
as in \eqref{e:rank-center}, and the homogeneous state is locally unstable, as discussed in Sec.~\ref{s:stability-homo}.
For the gas-side supercooled-gas region in Fig.~\ref{fig:landscape-gas-side-three}(b), the same ordering
\begin{align}
F_g^{(\subL,\subG)}
<
F_g^{(\subG,\subL)}
<
F_g^\homo
\end{align}
holds, again corresponding to \eqref{e:rank-center}; the homogeneous state is locally stable because $\bar v$ lies outside the spinodal interval.
For the gas-side edge-scale case in Fig.~\ref{fig:landscape-gas-side-three}(c), the numerical stationary values still show
\begin{align}
F_g^{(\subL,\subG)}
<
F_g^\homo
<
F_g^{(\subG,\subL)},
\end{align}
with $F_g^\homo-F_g^{(\subL,\subG)}=20.8\,\mathrm{J/mol}$ and $F_g^{(\subG,\subL)}-F_g^{(\subL,\subG)}=23.0\,\mathrm{J/mol}$.
A homogeneous state is not represented as a single point in Fig.~\ref{fig:landscape-gas-side-three}.
In the present variables, it is associated with the formal division of a single-phase state into lower and upper subregions.
Marking it as an isolated stationary point would therefore be misleading.
The figure instead shows the landscape structure together with the relative stationary values
$F_g^{(\subG,\subL)}-F_g^{(\subL,\subG)}$ and
$F_g^\homo-F_g^{(\subL,\subG)}$.

\subsection{Role of heat flow at fixed effective gravity}

At fixed $mg_\eff L$, the effective-gravity contribution that orders the separated heterogeneous states is held fixed.
Changing $\Xi$ then probes the part of the variational landscape that is not absorbed into the effective-gravity combination.
In the full two-dimensional landscapes of Fig.~\ref{fig:landscape-gas-side-three}, this heat-flow dependence is less visible than the change caused by moving $\bar v$ across the spinodal interval, the gas-side supercooled-gas region, and the gas-side edge layer.

\begin{figure*}[t]
\centering
\includegraphics[width=0.9\linewidth]{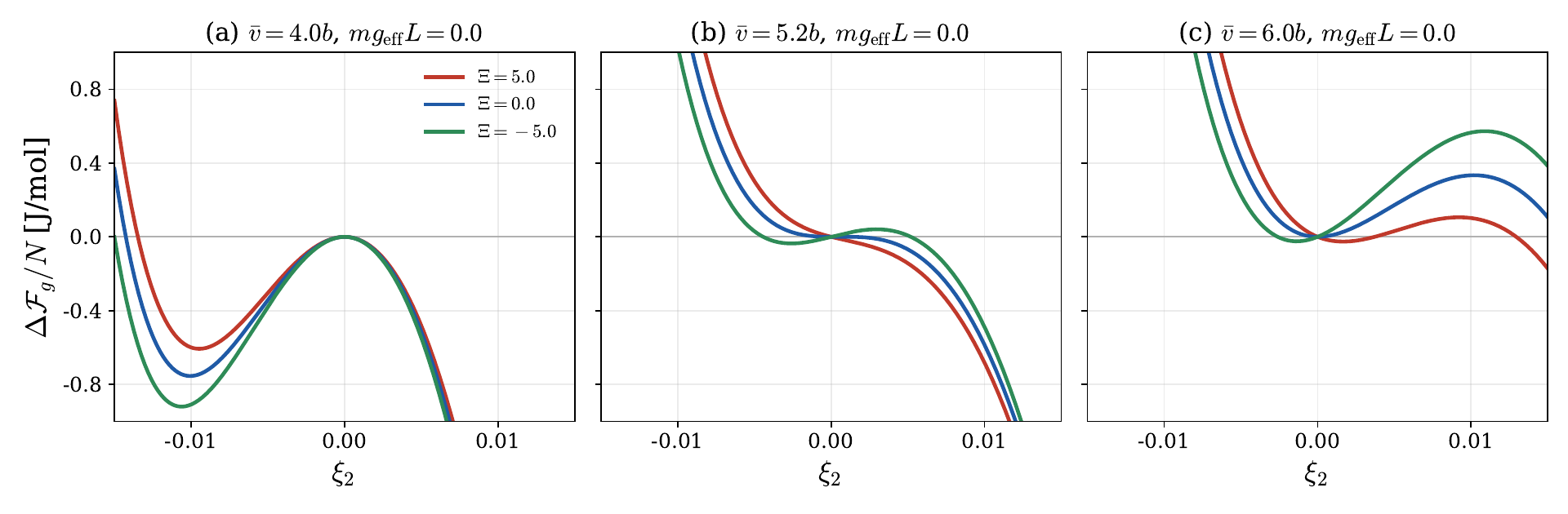}
\caption{
One-dimensional sections of the variational landscape at fixed $\xi_1=0.45$ for the van der Waals model with $mg_\eff L=0$.
The plotted quantity is
$\Delta\calF_g/N=[\mathcal{F}_g(\xi_1,\xi_2)-\mathcal{F}_g(\xi_1,0)]/N$
as a function of $\xi_2$.
Panels (a), (b), and (c) use $\bar v=4.0b$, $5.2b$, and $6.0b$.
The curves compare $\Xi=5.0\,\mathrm{K}$, $0.0$, and $-5.0\,\mathrm{K}$.
The figure magnifies local barrier deformation at fixed effective gravity; panel (b), near the gas-side spinodal boundary, is especially sensitive to the sign of $\Xi$.
}
\label{fig:vdw_1d_sections_pm5}
\end{figure*}

The effect becomes clearer in one-dimensional sections.
Figure~\ref{fig:vdw_1d_sections_pm5} shows cuts at fixed $\xi_1=0.45$ for the van der Waals model with $mg_\eff L=0$.
With $mg_\eff L=0$, the effective-gravity term does not favor either separated heterogeneous configuration.
The cuts therefore emphasize the deformation caused by the imposed temperature difference.
The three panels compare the spinodal interval, a point close to the gas-side spinodal boundary, and the gas-side supercooled-gas region.
In panel (a), changing $\Xi$ shifts the nearby extrema, whereas the cut remains shaped by the spinodal instability.
In panel (c), the local minimum near $\xi_2=0$ remains visible for all three values of $\Xi$, while the neighboring maximum moves and changes height.
Panel (b) is the most sensitive case: the near-center minimum--maximum structure is strongly displaced and can disappear when the sign of $\Xi$ is changed.
This last observation is not a Hessian instability of the homogeneous state.
The Hessian criterion in Sec.~\ref{s:stability-homo} tests the local quadratic form at the homogeneous stationary point; for the parameters of Fig.~\ref{fig:vdw_1d_sections_pm5}(b), this local curvature remains positive.
The change seen in the cut is instead a finite-shape change of the coarse-grained landscape.
The near-homogeneous expansion in Sec.~\ref{sec:saddle_approx} describes it in terms of the local valley and its companion curve.

\section{Saddle ridge and valley in the free energy landscape}
\label{sec:saddle_approx}

The representative landscapes in Fig.~\ref{fig:landscape-gas-side-three} show the overall structure of the fixed-$\bT$ variational free energy in the spinodal interval, the gas-side supercooled-gas region, and the gas-side edge layer.
The one-dimensional sections in Fig.~\ref{fig:vdw_1d_sections_pm5} show that, at fixed $mg_\eff L$, the imposed temperature difference $\Xi$ deforms the local barrier shape.
We analyze this deformation by expanding $\calF_g$ near a homogeneous state.
We use the variables $(\xi_1,\xi_2)$ introduced in \eqref{e:xi-landscape-def}.
The factor $\Lambda$ has been defined in \eqref{e:Lambda-def}; in the present variables,
\begin{align}
\Lambda=\phi^\subl\phi^\subu=\frac{1}{4}-\xi_1^2.
\end{align}
The specific volumes and the global temperatures of the two formal subregions are
\begin{align}
v^\subl
&=
\bar v+\bar v\frac{2\xi_2}{\phi^\subl-2\xi_2},
\qquad
v^\subu
=
\bar v-\bar v\frac{2\xi_2}{\phi^\subu+2\xi_2},
\label{e:delta-v}\\
\bT^\subl
&=
\bT-\frac{\Xi}{2}(\phi^\subu+2\xi_2),
\qquad
\bT^\subu
=
\bT+\frac{\Xi}{2}(\phi^\subl-2\xi_2),
\end{align}
where $\phi^\subl=1/2+\xi_1$ and $\phi^\subu=1/2-\xi_1$.
Therefore,
\begin{align}
\frac{v^\subl-v^\subu}{\bar v}
&=
\frac{2\xi_2}{(\phi^\subl-2\xi_2)(\phi^\subu+2\xi_2)}
=
\frac{2\xi_2}{\Lambda}
\left(1-\frac{2\xi_2}{\phi^\subl}\right)^{-1}
\left(1+\frac{2\xi_2}{\phi^\subu}\right)^{-1},
\label{e:vl-vu}\\
\bT^\subu-\bT^\subl
&=
\frac{\Xi}{2}.
\end{align}

The specific-volume contrast $\mathcal A$ has been defined in \eqref{e:A-homo-def}.
Using \eqref{e:vl-vu}, we have
\begin{align}
\mathcal A
=
-2\frac{\xi_2}{\Lambda}
+O\!\left[
(\xi_2/\Lambda)^2
\right].
\end{align}
Because configurations close to a homogeneous state are characterized by $\mathcal A=O(\ep)$, we use
\begin{align}
\frac{\xi_2}{\Lambda}=O(\ep)
\label{e:near-homo-scaling}
\end{align}
as the near-homogeneous scaling.
It is therefore not uniform in the limit $\Lambda\to0$: when one of the two formal subregions becomes very small, the condition $\xi_2/\Lambda=O(\ep)$ becomes more restrictive than $\xi_2=O(\ep)$ itself.
In the following calculation we keep $\xi_2/\Lambda$ as the expansion variable, because this is the combination that appears directly in \eqref{e:vl-vu} and in the resulting quadratic equation.

Using the Jacobian matrix for the transformation from
$\vec{x}=(c^\subl,\phi^\subl)^T$ to $\boldsymbol{\xi}=(\xi_1,\xi_2)^T$,
\begin{align}
J
=
\frac{\partial \vec{x}}{\partial \boldsymbol{\xi}}
=
\begin{pmatrix}
1 & -2\\
1 & 0
\end{pmatrix},
\end{align}
the gradient vector and the Hessian matrix are transformed as
\begin{align}
\tilde{\nabla}\calF_g
&=
J^T\nabla_{\vec{x}}\calF_g,
\label{e:nablaFg-tr}\\
\tilde H
&=
J^T HJ,
\label{e:Hessian-tr}
\end{align}
where $H$ is the Hessian matrix in the variables $(c^\subl,\phi^\subl)$.

To locate the near-homogeneous valley and its companion curve, we impose stationarity in the direction transverse to the homogeneous line.
This condition is expressed as
\begin{align}
\tilde H\tilde{\mathbf e}_\perp
=
\lambda_\perp\tilde{\mathbf e}_\perp,
\qquad
\tilde{\nabla}\calF_g\cdot \tilde{\mathbf e}_\perp=0,
\label{e:ridge-def}
\end{align}
where $\tilde{\mathbf e}_\perp$ is the Hessian eigenvector that approaches the $\xi_2$ direction near the homogeneous line, and $\lambda_\perp$ is the corresponding eigenvalue.
The subscript $\perp$ denotes this transverse direction; the curvature is positive on a local valley and negative on the companion saddle curve.

\subsection{Quadratic expansion and the master equation}

We treat gravity $g$ and the imposed temperature difference $\Xi$ as first-order perturbations of $O(\ep)$.
Using $\calF_g$ in \eqref{e:Fg-var-neq-stab}, the transformed gradient is
\begin{align}
\tilde{\nabla}\calF_g
=
N
\begin{pmatrix}
\Delta\mu-\bar v\Delta p-\bar s\,\Xi/2\\
-2\Delta\mu+\bar s\,\Xi+mgL
\end{pmatrix},
\label{e:tilde-grad-Fg}
\end{align}
where
\begin{align}
\bar s\equiv\frac{S^\subl+S^\subu}{N},
\qquad
s_0\equiv s(\bT,\bar v),
\qquad
\Delta p\equiv \bP^\subl-\bP^\subu,
\qquad
\Delta\mu\equiv \mu^\subl-\mu^\subu.
\end{align}
Here \(\bar s\) is the averaged entropy per particle of the formal split, whereas \(s_0\) is the local entropy per particle at the homogeneous reference state.
The transformed Hessian is then written as
\begin{align}
\tilde H
=
N
\begin{pmatrix}
\pder{}{\xi_1}
\left(
\Delta\mu-\bar v\Delta p-\bar s\,\Xi/2
\right)
&
\pder{}{\xi_2}
\left(
\Delta\mu-\bar v\Delta p-\bar s\,\Xi/2
\right)
\\
-2\pder{}{\xi_1}
\left(
\Delta\mu-\bar s\,\Xi/2
\right)
&
-2\pder{}{\xi_2}
\left(
\Delta\mu-\bar s\,\Xi/2
\right)
\end{pmatrix}.
\label{e:Hessian-tr0}
\end{align}

To evaluate \eqref{e:ridge-def} consistently, we expand the local thermodynamic variables up to quadratic order in $\xi_2/\Lambda$.
The pressure in each subregion is expanded around the homogeneous reference state $(\bar v,\bT)$ as
\begin{align}
p(v^\alpha,\bT^\alpha)
&=
p(\bar v,\bT)
+p_v(v^\alpha-\bar v)
+p_T(\bT^\alpha-\bT)
+\frac{p_{vv}}{2}(v^\alpha-\bar v)^2
\nonumber\\
&\qquad
+p_{vT}(v^\alpha-\bar v)(\bT^\alpha-\bT)
+O\!\left[(\xi_2/\Lambda)^3,\ep(\xi_2/\Lambda)^2,\ep^2\right],
\qquad
(\alpha=\subl,\subu).
\label{e:p-expansion}
\end{align}
Here all derivatives such as $p_v$, $p_T$, $p_{vv}$, and $p_{vT}$ are evaluated at $(\bar v,\bT)$.
We substitute \eqref{e:delta-v} into the pressure expansion \eqref{e:p-expansion} to obtain $\Delta p$.
In parallel, we expand $\mu(v^\alpha,\bT^\alpha)$ around $(\bar v,\bT)$, using the thermodynamic identities $\mu_v=vp_v$ and $\mu_T=-s_0+vp_T$, to obtain $\Delta\mu$.
The identity \eqref{e:vl-vu} is then used to collect the volume-contrast terms in powers of $\xi_2/\Lambda$.
We obtain
\begin{align}
\Delta p
&=
\frac{2\xi_2}{\Lambda}\bar v p_v
-\frac{4\xi_2^2}{\Lambda^2}\xi_1
\left.
\frac{\partial(v^2p_v)}{\partial v}
\right|_{\bar v}
-\frac{\Xi}{2}p_T
+2\bar v\xi_1\Xi p_{vT}\frac{\xi_2}{\Lambda}
+O\!\left[(\xi_2/\Lambda)^3,\ep(\xi_2/\Lambda)^2,\ep^2\right],
\label{e:Delta_P_general_expansion}\\
\Delta\mu
&=
\frac{2\xi_2}{\Lambda}\bar v^2p_v
-\frac{4\xi_2^2}{\Lambda^2}\xi_1
\left.
\frac{\partial(v^3p_v)}{\partial v}
\right|_{\bar v}
+\left(s_0-\bar v p_T\right)\frac{\Xi}{2}
+2\bar v^2\xi_1\Xi p_{vT}\frac{\xi_2}{\Lambda}
+O\!\left[(\xi_2/\Lambda)^3,\ep(\xi_2/\Lambda)^2,\ep^2\right].
\label{e:Delta_mu_general_expansion}
\end{align}
Their combination appearing in the first component of \eqref{e:tilde-grad-Fg} is
\begin{align}
\Delta\mu-\bar s\frac{\Xi}{2}-\bar v\Delta p
=
-\frac{4\xi_2^2}{\Lambda^2}\xi_1\bar v^2p_v
-\bar v^2\Xi p_{vT}
\frac{\xi_2^2}{\Lambda^2}
\left(\frac{1}{2}+2\xi_1^2\right)
+O\!\left[(\xi_2/\Lambda)^3,\ep(\xi_2/\Lambda)^3,\ep^2\right].
\label{e:delta_mu_v_delta_p_difference}
\end{align}
The \(O[\Xi(\xi_2/\Lambda)^2]\) term in this combination is evaluated by retaining the corresponding mixed terms in the individual Taylor expansions of \(p\) and \(\mu\). The higher-derivative contributions cancel in \(\Delta\mu-\bar v\Delta p\), and the remaining part, together with the expansion of \(\bar s=(S^\subl+S^\subu)/N\), gives the coefficient shown in \eqref{e:delta_mu_v_delta_p_difference}.

By differentiating the expanded forms with respect to $\xi_1$ and $\xi_2$, the Hessian matrix \eqref{e:Hessian-tr0} is evaluated as
\begin{align}
\tilde H
&=
V
\begin{pmatrix}
O(\xi_2^2,\ep) &
-\dfrac{8\xi_2}{\Lambda^2}\xi_1\bar v p_v\\[4pt]
-\dfrac{8\xi_2}{\Lambda^2}\xi_1\bar v p_v &
-\dfrac{4}{\Lambda}\bar v p_v
+
\dfrac{16\xi_2}{\Lambda^2}
\dfrac{\xi_1}{\bar v}
\left.
\frac{\partial(v^3p_v)}{\partial v}
\right|_{\bar v}
\end{pmatrix}
\nm
&\quad
-2V\bar v\Xi p_{vT}
\begin{pmatrix}
0 &
\dfrac{\xi_2}{\Lambda^2}\left(\frac{1}{2}+2\xi_1^2\right)
\\[4pt]
\dfrac{\xi_2}{\Lambda^2}\left(\frac{1}{2}+2\xi_1^2\right)
&
\dfrac{2\xi_1}{\Lambda}
\end{pmatrix}
+O(\xi_2^2,\ep^2).
\label{e:approximated_tilde_H}
\end{align}
The transverse eigenvector $\tilde{\mathbf e}_\perp$ is
\begin{align}
\tilde{\mathbf e}_\perp
=
\begin{pmatrix}
\left(
\dfrac{2\xi_1}{\Lambda}
+\Xi\dfrac{p_{vT}}{p_v}
\right)\xi_2\\[4pt]
1
\end{pmatrix}
+O(\xi_2^2,\ep^2).
\label{e:tilde-e-}
\end{align}
In the substitution below, $\xi_1$ is kept finite and $\xi_2/\Lambda$ is the small parameter.
Thus \eqref{e:approximated_tilde_H} and \eqref{e:tilde-e-} retain the terms needed to obtain the transverse condition through quadratic order in $\xi_2/\Lambda$.
Substituting \eqref{e:tilde-grad-Fg} and \eqref{e:tilde-e-} into the transverse-stationarity condition \eqref{e:ridge-def}, we find
\begin{align}
&-2
\left(
\Delta\mu-\bar s\frac{\Xi}{2}-\frac{mgL}{2}
\right)
+\left(
\frac{2\xi_1}{\Lambda}
+\Xi\frac{p_{vT}}{p_v}
\right)\xi_2
\left(
\Delta\mu-\bar v\Delta p-\bar s\frac{\Xi}{2}
\right)
=
O\!\left[(\xi_2/\Lambda)^3,\ep(\xi_2/\Lambda)^2,\ep^2\right].
\label{e:saddle-line-eval}
\end{align}
Using \eqref{e:delta_mu_v_delta_p_difference}, the second term on the left-hand side is of $O[(\xi_2/\Lambda)^3,\ep(\xi_2/\Lambda)^3]$.
Dropping the \(O[\ep(\xi_2/\Lambda)^2]\) correction to the quadratic coefficient and retaining the leading quadratic terms, \eqref{e:saddle-line-eval} reduces to
\begin{align}
\Delta\mu-\bar s\frac{\Xi}{2}-\frac{mgL}{2}=0.
\label{e:saddle-line-condition}
\end{align}

Substituting \eqref{e:Delta_mu_general_expansion} into
\eqref{e:saddle-line-condition}, retaining the leading quadratic terms, and dividing the resulting equation by
$2$, we obtain the quadratic master equation
\begin{align}
2\xi_1
\left.
\pderf{v^2K^\homo}{v}{T}
\right|_{\bar v,\bT}
\left(
\frac{\xi_2}{\Lambda}
\right)^2
-
\bar v K_\Xi^\homo
\left(
\frac{\xi_2}{\Lambda}
\right)
-
\frac{m g_{\bias}L}{4}
=
0.
\label{e:quadratic_master}
\end{align}
Here, the effective stiffness $K_{\Xi}^\homo$ is
\begin{align}
K_{\Xi}^\homo
\equiv
K^\homo
+\xi_1\Xi
\left.
\pderf{K^\homo}{T}{v}
\right|_{\bar v,\bT},
\label{e:effective_stiffness}
\end{align}
with
\begin{align}
K^\homo
\equiv
-\left.
\bar v\pderf{p}{v}{T}
\right|_{\bar v,\bT}
=
\frac{1}{\alpha^\homo}.
\end{align}
$g_{\bias}$ is the bias field
\begin{align}
m g_{\bias}
\equiv
mg+\bar v p_T\frac{\Xi}{L}.
\label{e:g-bias}
\end{align}
The subscript ``bias'' indicates that this combination controls the
linear bias of the near-homogeneous landscape.
It should not be confused with the effective gravity $g_\eff$.

Equation \eqref{e:quadratic_master} shows that the local bias and the local stiffness of the near-homogeneous landscape are controlled by different combinations.
The constant term is proportional to $g_{\bias}$, whereas the coefficient of $\xi_2/\Lambda$ is $K_\Xi^\homo$.
Thus, $g_{\bias}$ gives the constant bias term in the near-homogeneous equation.
It controls the regular displacement of the homogeneous root when the stiffness term is not small, whereas the existence and merging of the two local roots are determined by the full quadratic equation.

The two roots of \eqref{e:quadratic_master} define the asymptotic near-homogeneous curves.
We denote them by $\xi_2^\homo(\xi_1)$ and $\xi_2^\saddle(\xi_1)$:
\begin{align}
\frac{\xi_2^\homo}{\Lambda}
&=
\frac{\bar v K_\Xi^\homo-\sqrt{\Delta(\xi_1)}}
{4\xi_1
\left.
\pderf{v^2K^\homo}{v}{T}
\right|_{\bar v,\bT}},
\quad
\frac{\xi_2^\saddle}{\Lambda}
=
\frac{\bar v K_\Xi^\homo+\sqrt{\Delta(\xi_1)}}
{4\xi_1
\left.
\pderf{v^2K^\homo}{v}{T}
\right|_{\bar v,\bT}},
\label{e:xi2_solutions}
\end{align}
where
\begin{align}
&\Delta(\xi_1)
\equiv
(\bar vK_\Xi^\homo)^2
+
2\xi_1
\left.
\pderf{v^2K^\homo}{v}{T}
\right|_{\bar v,\bT}
m g_{\bias}L,
\label{e:discriminant_main}
\\
&=
\bar v^2(K^\homo)^2
+
2\xi_1 mgL
\left.
\pderf{v^2K^\homo}{v}{T}
\right|_{\bar v,\bT}
+
2\xi_1\Xi K^\homo
\left.
\pderf{v^2K^\homo}{T}{p}
\right|_{\bar v,\bT}.
\label{e:discriminant-Khomo}
\end{align}
The second equality follows from \eqref{e:effective_stiffness} and
\eqref{e:g-bias}, with terms of $O(\ep^2)$ omitted.
The derivative at fixed $p$ is taken along the local equation of state $p=p(T,v)$.

When the $O(\ep)$ terms in \eqref{e:discriminant-Khomo} are small compared with
$\bar v^2(K^\homo)^2$, the two roots can be expanded explicitly.
Choosing the square-root branch so that
$\sqrt{\Delta(\xi_1)}\to \bar v K^\homo$ as $mgL,\Xi\to0$, we have
\begin{align}
\sqrt{\Delta(\xi_1)}
=
\bar vK^\homo
+
\frac{\xi_1mgL}{\bar vK^\homo}
\left.
\pderf{v^2K^\homo}{v}{T}
\right|_{\bar v,\bT}
+
\frac{\xi_1\Xi}{\bar v}
\left.
\pderf{v^2K^\homo}{T}{p}
\right|_{\bar v,\bT}
+O(\ep^2).
\end{align}
Substituting this expansion and \eqref{e:effective_stiffness} into
\eqref{e:xi2_solutions}, we obtain
\begin{align}
\frac{\xi_2^\homo}{\Lambda}
&=
-\frac{mgL}{4\bar vK^\homo}
-
\frac{\Xi}{4\bar v}
\left.
\pderf{v}{T}{p}
\right|_{\bar v,\bT}
+O(\ep^2),
\label{e:xi2-homo-expanded}\\
\frac{\xi_2^\saddle}{\Lambda}
&=
\frac{\bar vK^\homo}
{2\xi_1
\left.
\pderf{v^2K^\homo}{v}{T}
\right|_{\bar v,\bT}}
+
\frac{mgL}{4\bar vK^\homo}
+
\frac{\Xi}{4\bar v}
\left.
\pderf{v}{T}{p}
\right|_{\bar v,\bT}
+
\frac{\Xi}{2\bar v}
\frac{
\left.
\pderf{v^2K^\homo}{T}{v}
\right|_{\bar v,\bT}}
{\left.
\pderf{v^2K^\homo}{v}{T}
\right|_{\bar v,\bT}}
+O(\ep^2).
\label{e:xi2-saddle-expanded}
\end{align}
The root $\xi_2^\homo$ is chosen so that
$\xi_2^\homo\to0$ as $mgL,\Xi\to0$ for fixed $\xi_1$.
It gives the perturbative displacement of the homogeneous stationary point.
The second root is not necessarily small; the notation $\xi_2^\saddle$
is used only in the parameter range where $|\xi_2^\saddle|/\Lambda\ll1$.

When $\bar v^2(K^\homo)^2$ is comparable to the $O(\ep)$ terms in
\eqref{e:discriminant-Khomo}, the perturbative expansions
\eqref{e:xi2-homo-expanded} and \eqref{e:xi2-saddle-expanded} are no longer the useful form.
The discriminant itself then controls the local structure: for $\Delta(\xi_1)>0$ the two real roots $\xi_2^\homo$ and $\xi_2^\saddle$ exist, at $\Delta(\xi_1)=0$ they merge, and for $\Delta(\xi_1)<0$ they do not appear as real stationary roots in this quadratic approximation.
The calculation remains a local expansion in $\xi_2/\Lambda\ll1$; therefore this disappearance of the two roots should not be identified with a sharp spinodal or barrier-loss boundary of the full two-dimensional landscape.

Thus the driving fields have two different roles in the near-homogeneous landscape.
In the first regime, they only perturb the two roots: they generate the regular displacement $\xi_2^\homo=O(\ep)$ and give $O(\ep)$ corrections to the companion curve.
In the second regime, the same $O(\ep)$ terms compete with the small stiffness-square term in $\Delta(\xi_1)$ and can change the number of local roots.
This is the sense in which heat conduction and gravity can produce a qualitative change near the spinodal neighborhood.

This is the near-homogeneous interpretation of the sensitivity seen in Fig.~\ref{fig:vdw_1d_sections_pm5}(b): the minimum--maximum pair in that cut corresponds, in the present expansion, to the local valley $\xi_2^\homo$ and its companion curve $\xi_2^\saddle$.
For the parameters of Fig.~\ref{fig:vdw_1d_sections_pm5}(b), namely
$\bar v=5.2b$, $\xi_1=0.45$, and $mg_\eff L=0$, the discriminant changes sign at
$\Xi\simeq0.166\,\mathrm{K}$.
Thus the plotted values $\Xi=5.0\,\mathrm{K}$ and $-5.0\,\mathrm{K}$ lie on opposite sides of the near-merging condition.
The observed approach and disappearance of the minimum--maximum pair are therefore described by the vanishing of $\Delta(\xi_1)$ in the quadratic approximation, not by the Hessian spinodal criterion of Sec.~\ref{s:stability-homo}.
\subsection{Residual modulation at fixed effective gravity}

For a heat-conducting landscape specified by the laboratory controls $(mgL,\Xi)$, we compare it with the pure-gravity landscape at $\Xi=0$ having the same value
\begin{align}
mg_\eff L
=
mgL+\bar v\frac{d\Ps}{d\bT}\Xi .
\label{e:mgL-fixed-geff-general}
\end{align}
By the decomposition \eqref{e:Fg-var-neq-re2}, their difference is given by the residual contribution:
\begin{align}
\calF_g(\calN^\subl,\calV^\subl;\bT,V,N,mgL,\Xi)
-
\calF_g(\calN^\subl,\calV^\subl;\bT,V,N,mg_\eff L,0)
&=
\calF_{g,\res}(\calN^\subl,\calV^\subl;\bT,V,N,\Xi)
\nm&=
-\frac{\Xi}{\bT}
\frac{\calN^\subl\calN^\subu}{2N}
\hat q^\ex .
\label{e:landscape-difference-res}
\end{align}

We now derive the local form of \eqref{e:landscape-difference-res} from the
residual term \eqref{e:Fg-res-def}.  We expand the function
$\hat q^\ex(\bT,\calV^\subl,\calN^\subl)$ defined in \eqref{e:qex-def}
near a homogeneous state.  At fixed $\bT$, set
$v^\subl=\calV^\subl/\calN^\subl$ and
$v^\subu=\calV^\subu/\calN^\subu$.  Then
$\mathcal{S}^\alpha/\calN^\alpha=s(\bT,v^\alpha)$
$(\alpha=\subl,\subu)$, and Taylor expansion around $v=\bar v$ gives,
using $(\partial s/\partial v)_T=(\partial p/\partial T)_v$,
\begin{align}
\hat q^\ex
=
\bT
\left[
\left.\pderf{p}{T}{v}\right|_{(\bT,\bar v)}
-
\frac{d\Ps}{d\bT}
\right]
(v^\subu-v^\subl)
+O\!\left((v^\subl-\bar v)^2+(v^\subu-\bar v)^2\right).
\label{e:qex-homo-landscape}
\end{align}
The variables $(\xi_1,\xi_2)$ give the identity
\begin{align}
\frac{\calN^\subl\calN^\subu}{2N}
(v^\subu-v^\subl)
=
-V\xi_2 .
\label{e:qex-homo-geom}
\end{align}
Substituting \eqref{e:qex-homo-landscape} and \eqref{e:qex-homo-geom}
into \eqref{e:Fg-res-def}, we obtain
\begin{align}
\calF_{g,\res}
&=
N\bar v\Xi
\left[
\left.\pderf{p}{T}{v}\right|_{(\bT,\bar v)}
-
\frac{d\Ps}{d\bT}
\right]\xi_2
+O\!\left(\xi_2^2/\Lambda\right)
\nonumber\\
&=
Nm(g_{\bias}-g_\eff)L\,\xi_2
+O\!\left(\xi_2^2/\Lambda\right).
\label{e:Fres-homo-landscape}
\end{align}
We can express this in terms of pressure drops.
Define the fixed-$\bar v$ thermal-pressure drop by
\begin{align}
\Delta P_{\bias}
\equiv
p(\bar v,\botT)-p(\bar v,\topT).
\label{e:DeltaP-bias}
\end{align}
Using $\Delta P_{\bias}=-p_T\Xi+O(\ep^2)$ and the hydrostatic force balance, \eqref{e:g-bias} can be written as
\begin{align}
mg_{\bias}L
=
\bar v(\Delta P_{\mech}-\Delta P_{\bias}),
\label{e:g-bias-pressure}
\end{align}
where $\Delta P_{\mech}$ is the mechanical pressure drop defined in \eqref{e:pressure-drops}.
Then, \eqref{e:Fres-homo-landscape} becomes equivalent to
\begin{align}
\calF_{g,\res}
&=
V(\Delta P_{\sat}-\Delta P_{\bias})\,\xi_2
+O\!\left(\xi_2^2/\Lambda\right).
\label{e:Fres-homo-landscape2}
\end{align}
Here the saturation pressure drop $\Delta P_{\sat}$ is defined in \eqref{e:pressure-drops}.

Equation \eqref{e:Fres-homo-landscape} explains why two systems with the same
$g_\eff$ need not have identical local barrier shapes.
The effective gravity organizes the large-scale hierarchy of stationary values,
as summarized in Fig.~\ref{fig:global-hierarchy}, whereas the residual difference
at fixed $g_\eff$ changes the local bias of the landscape near a homogeneous state.
The pressure-drop form \eqref{e:Fres-homo-landscape2} shows that this residual
bias is the mismatch between the saturation-pressure response entering $g_\eff$
and the fixed-$\bar v$ thermal-pressure response entering $g_{\bias}$.
This is the mechanism behind the heat-flow dependence visible in the
one-dimensional sections in Fig.~\ref{fig:vdw_1d_sections_pm5}.

\subsection{Full landscapes at fixed effective gravity}

\begin{figure*}[hbtp]
\centering
\includegraphics[width=0.88\linewidth,trim=0 0 0 0,clip]{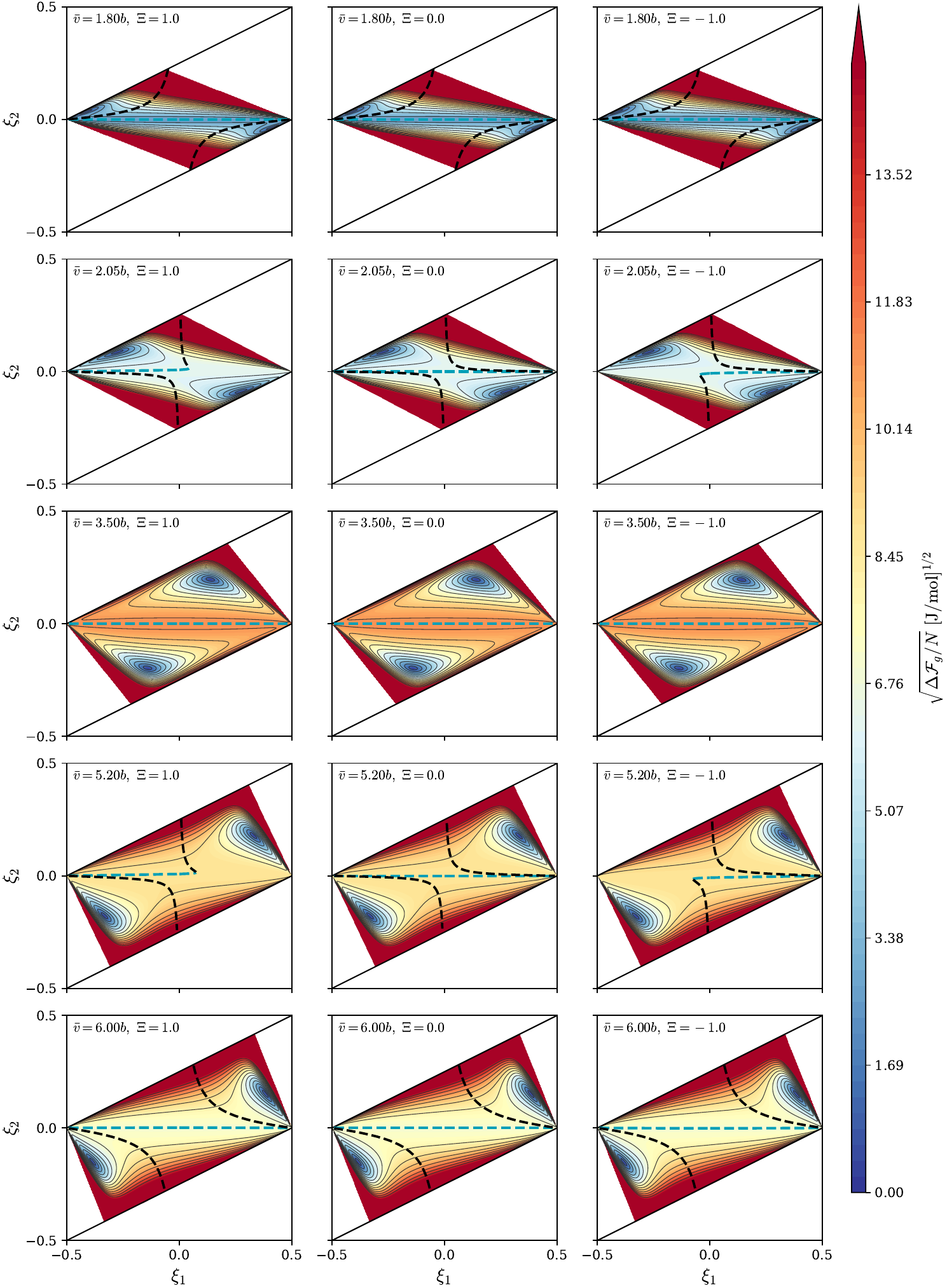}

\caption{Matrix of variational landscapes $\mathcal{F}_g$ in $(\xi_1,\xi_2)$ space for the van der Waals model at fixed $mg_\eff L=0.0$. The rows sample representative values of $\bar v$, and the columns compare $\Xi=1.0$, $0.0$, and $-1.0$. The color map shows $\sqrt{\Delta \mathcal{F}_g/N}$ in molar units after subtracting the minimum value of $\mathcal{F}_g$ in each panel. Dashed cyan and black curves denote the near-homogeneous valley curve $\xi_2^\homo$ and its companion curve $\xi_2^\saddle$ derived from \eqref{e:xi2_solutions}.}
\label{fig:landscape_matrix_5x3_pm1}
\end{figure*}

Figure~\ref{fig:landscape_matrix_5x3_pm1} displays landscapes for $\bar{v}=1.8b$, $2.05b$, $3.5b$, $5.2b$, and $6.0b$ with $\Xi=1.0$, $0.0$, and $-1.0$ under the common constraint $mg_\eff L=0$.
For reference, the spinodal volumes are $v_{\mathrm{spn}}^{\subL}=2.03b$ and $v_{\mathrm{spn}}^{\subG}=5.10b$.
Fixing $g_\eff$ removes the $O(\ep)$ macroscopic free-energy difference between the two separated configurations, so the column dependence displays the local barrier reshaping caused by $\calF_{g,\res}$.
For $\Xi=\pm1.0\,\mathrm{K}$, the corresponding $\ep$ values are below $0.008$ for all five rows.
To locate these rows within the coexistence interval, we use the edge-distance variables introduced in Sec.~\ref{s:hierarchy}: $z'$ measures the distance from the liquid coexistence edge, while $z$ measures the distance from the gas coexistence edge as in \eqref{e:z-def}.
They are used only to classify the rows in Fig.~\ref{fig:landscape_matrix_5x3_pm1}.
Their values are $z'=0.017$ and $0.051$ for $\bar v=1.8b$ and $2.05b$, and $z=0.526$ and $0.419$ for $\bar v=5.2b$ and $6.0b$.
Thus, the first row is a liquid-edge case, the second row is a liquid-side near-spinodal case close to the edge, and the fourth and fifth rows are gas-side cases away from the coexistence edges.

The dashed cyan and black curves are the near-homogeneous valley curve $\xi_2^\homo$ and its companion curve $\xi_2^\saddle$ derived from \eqref{e:xi2_solutions}.
Their positions are determined by the quadratic equation \eqref{e:quadratic_master}.
The strongest heat-flow sensitivity appears near the spinodal rows, especially $\bar v=2.05b$ and $5.2b$.
Where the dashed curves approach or disappear, the discriminant $\Delta(\xi_1)$ in \eqref{e:discriminant_main} approaches or crosses zero.
This is the same near-merging mechanism discussed after \eqref{e:discriminant-Khomo}, not a new global reordering of separated stationary values.

\begin{figure*}[t]
\centering
\includegraphics[width=0.92\linewidth]{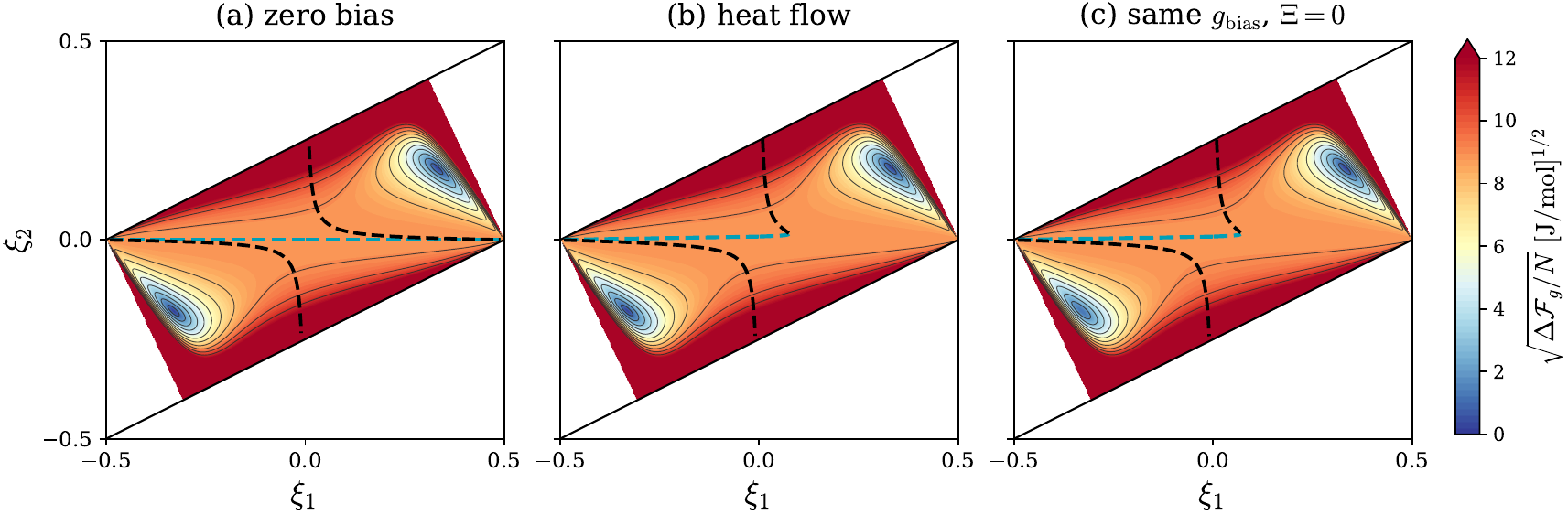}
\caption{
Local-bias comparison at $\bar v=5.2b$ for the van der Waals model.
Panel (a) is the unbiased reference $(mgL,\Xi)=(0,0)$.
Panel (b) is the heat-conducting case with $\Xi=1.0$ and $mg_\eff L=0$, giving $mgL=-14.5\,\mathrm{J/mol}$ and $mg_{\bias}L=-4.2\,\mathrm{J/mol}$.
Panel (c) is a gravity-only comparison case with $\Xi=0$ and the same $mg_{\bias}L=-4.2\,\mathrm{J/mol}$ as in panel (b).
Matching $g_{\bias}$ fixes the constant-bias term of \eqref{e:quadratic_master}, but the heat-conducting and gravity-only landscapes remain different because the full discriminant $\Delta(\xi_1)$ is not the same.
The same color scale is used in all three panels.
}
\label{fig:v52-bias-match}
\end{figure*}

Figure~\ref{fig:v52-bias-match} tests how far the local bias field $g_{\bias}$ summarizes the near-homogeneous deformation at the representative gas-side value $\bar v=5.2b$.
In terms of the gas-side edge-distance variable of Sec.~\ref{s:hierarchy}, this value has $z=0.526$, so the comparison lies outside the edge-layer scaling.
Panel (a) is the reference case $(mgL,\Xi)=(0,0)$.
Panel (b) is the heat-conducting case with $\Xi=1.0$ and $mg_\eff L=0$, while panel (c) is a gravity-only case with $\Xi=0$ chosen so that $mg_{\bias}L$ has the same value as in panel (b).
The heat-conducting case has $\ep=0.0067$, and the gravity-only bias-matched case has $\ep=0.0019$.
Matching $g_{\bias}$ fixes the constant term in \eqref{e:quadratic_master}, but it does not fix the full discriminant $\Delta(\xi_1)$ in \eqref{e:discriminant-Khomo}.
The comparison therefore separates the part captured by the local bias from the remaining stiffness and heat-flow contributions.
The comparison can be read directly from \eqref{e:discriminant-Khomo}.
In panel (a), only the positive stiffness-square term remains.
In panel (b), the second term proportional to $mgL$ is negative because $mgL$ is chosen to keep $mg_\eff L=0$, while the third term proportional to $\Xi$ partially compensates it.
In panel (c), the third term is absent and the negative second term alone lowers the discriminant.
At the representative value $\xi_1=0.45$, the discriminant is positive in panel (a), while it is negative in both panels (b) and (c).
Thus the bias-matched gravity-only case reproduces the disappearance of the near-homogeneous root pair at this point, although the full landscapes are not identical.

\section{Experimental window for detecting effective gravity}
\label{s:experiment}

Ref.~\cite{NSgeff26} proposed a cylindrical H$_2$O cell as an experimental route to the effective-gravity inversion.
Here we use that estimate as a diagnostic tool and compare several working fluids using NIST saturation-property values.
Consider a cylindrical cell of radius $r$ and height $L$ containing $n$ mol of a working fluid, so that $V=\pi r^2L$.
We set $T_1=T(0)$ and $T_2=T(L)$ as in the setup.
For the boundary-temperature choice $T_1>T_2$, the saturation-pressure drop in \eqref{e:pressure-drops} is
\begin{align}
\Delta P_{\sat}= \Ps(T_1)-\Ps(T_2).
\end{align}
Equation~\eqref{e:g-eff-Ps} shows that the effective-gravity inversion is the condition
\begin{align}
g_\eff=0
\quad \Longleftrightarrow \quad
\Delta P_{\mech}=\Delta P_{\sat}.
\end{align}
For the cylindrical cell, the mechanical pressure drop is written in molar variables as
\begin{align}
\Delta P_{\mech}
=
\frac{nMgL}{V}
=
\frac{nMg}{\pi r^2},
\end{align}
where $M$ is the molar mass and $V=\pi r^2L$ has been used.
Thus the radius at which $g_\eff=0$ is determined by
\begin{align}
\frac{nMg}{\pi r_c^2}=\Delta P_{\sat},
\end{align}
which gives the effective-gravity inversion radius
\begin{align}
r_c
=
\left[
\frac{nMg}{\pi\Delta P_{\sat}}
\right]^{1/2}.
\label{e:rc-experiment}
\end{align}
Thus, changing $n$ shifts the inversion radius $r_c$.
For $T_1>T_2$, one has $g_\eff>0$ for $r<r_c$ and $g_\eff<0$ for $r>r_c$.
The thermodynamically favored configuration is liquid below gas for $r<r_c$ and liquid floating for $r>r_c$.

The formula for $r_c$ in \eqref{e:rc-experiment} does not specify the height $L$ of the container.
This does not mean that $L$ is arbitrary. In order for liquid to coexist with gas at a fixed amount $n$,
the system should satisfy the lever rule.
Let $\phi_\subL$ be the liquid volume fraction in $0<\phi_\subL<1$.
For a given $r_c$, we can identify a height $L_c$ realizing $\phi_\subL$ by the lever rule,
\begin{align}
L_c
&=
\frac{nM}
{\pi r_c^2\{\phi_\subL \rho_{\mathrm{m},\subC}^\subL+(1-\phi_\subL)\rho_{\mathrm{m},\subC}^\subG\}}\nm
&=
\frac{\Delta P_{\sat}}
{g\{\phi_\subL \rho_{\mathrm{m},\subC}^\subL+(1-\phi_\subL)\rho_{\mathrm{m},\subC}^\subG\}}.
\label{e:Lc-experiment}
\end{align}
The second line follows by using \eqref{e:rc-experiment}.
$\rho_{\mathrm{m},\subC}^\subL$ and $\rho_{\mathrm{m},\subC}^\subG$ are the saturated mass densities of the liquid and gas phases, evaluated along the saturation curve.
Because the estimates below use small temperature differences, these densities may be evaluated at a representative temperature between $T_2$ and $T_1$ within the present accuracy.
We choose this reference temperature as the midpoint temperature, $T_\mathrm{m}=(T_1+T_2)/2$.

Summarizing the above, choosing the fluid, the amount $n$, and the boundary temperatures $T_1$ and $T_2$ determines $\Delta P_{\sat}$ and hence the critical inversion radius $r_c$.
Specifying the liquid volume fraction $\phi_\subL$ then determines the height $L_c$ for realizing $\phi_\subL$ at the critical radius $r_c$.

The inversion should be observed in a geometry where the interface is neither wall-dominated nor already prone to large deformation.
The capillary length $\ell_{\mathrm{cap}}$ sets the distance over which a wall-induced meniscus deformation relaxes,
 where
\begin{align}
\ell_{\mathrm{cap}}
=
\left[
\frac{\sigma}{(\rho_{\mathrm{m},\subC}^\subL-\rho_{\mathrm{m},\subC}^\subG)g}
\right]^{1/2},
\label{e:cap-length-experiment}
\end{align}
with the surface tension $\sigma$.
If $r_c$ is not larger than $\ell_{\mathrm{cap}}$, wall wetting and meniscus curvature can affect most of the interface.
Conversely, taking $r_c$ much larger can make long-wavelength interface deformation and fluid instability harder to suppress.
We therefore use $r_c/\ell_{\mathrm{cap}}$ as a diagnostic of the capillary margin; very large values may indicate an experimentally unfavorable geometry even when the Rayleigh-number bound is satisfied.
The phase inversion should also be observed in a cell for which buoyancy-driven convection is not triggered.
The relevant measure is the Rayleigh number of each layer.
To evaluate it, we first estimate the temperature drops across the liquid and gas layers by a one-dimensional series-conduction model.
Using the heights of the liquid and gas layers, $h_\subL=\phi_\subL L_c$ and $h_\subG=(1-\phi_\subL)L_c$, the temperature drops in the respective layers are
\begin{align}
\Delta T_\subL
&=
(T_1-T_2)
\frac{h_\subL/\kappa^\subL}{h_\subL/\kappa^\subL+h_\subG/\kappa^\subG},
\qquad
\Delta T_\subG=(T_1-T_2)-\Delta T_\subL,
\label{e:DeltaTl-experiment}
\end{align}
where $\kappa^\subL$ and $\kappa^\subG$ are the thermal conductivities of the liquid and gas phases.
The Rayleigh numbers are then
\begin{align}
Ra_\subL
=
\frac{g\alpha^\subL \Delta T_\subL h_\subL^3}{\nu^\subL\chi^\subL},
\qquad
Ra_\subG
=
\frac{g\alpha^\subG \Delta T_\subG h_\subG^3}{\nu^\subG\chi^\subG},
\label{e:cap-Ra-experiment}
\end{align}
where $\alpha^i$, $\nu^i$, and $\chi^i$ are the thermal expansion coefficient, kinematic viscosity, and thermal diffusivity of phase $i=\subL,\subG$.
A standard no-convection estimate for a horizontal layer is $Ra\lesssim 1.7\times10^3$.
In the numerical search below, we use the more conservative condition $\max(Ra_\subL,Ra_\subG)\leq 5.0\times10^2$.
The conductivities $\kappa^i$ determine how the imposed temperature difference is divided between the two layers, whereas the thermal diffusivities $\chi^i$ enter the Rayleigh numbers.

The design protocol used for Table~\ref{tab:effective-gravity-experiment} is therefore as follows.
We first choose $T_\mathrm{m}$ for each working fluid and evaluate all saturation properties at $T_\mathrm{m}$.
We evaluate the saturation densities and transport coefficients entering \eqref{e:DeltaTl-experiment} and \eqref{e:cap-Ra-experiment} at $T_\mathrm{m}$; $\alpha^\subL$ is obtained from the saturation-density slope, and $\alpha^\subG$ is approximated by $1/T_\mathrm{m}$.
We then choose an experimentally natural amount of material.
For H$_2$O we take $n=0.5\,\mathrm{mol}$.
For CO$_2$, SF$_6$, and He, we choose $n$ so that the whole amount would occupy $100\,\mathrm{mL}$ if it were saturated gas at $T_\mathrm{m}$.
For Table~\ref{tab:effective-gravity-experiment}, we fix this amount $n$ and set $\phi_\subL=0.5$ while varying $\Delta T=T_1-T_2$.
At each trial $\Delta T$, the saturation-pressure drop $\Delta P_{\sat}$ gives $r_c$ through \eqref{e:rc-experiment}, and \eqref{e:Lc-experiment} gives the corresponding cell height $L_c$ required to realize the prescribed $\phi_\subL=0.5$ at $r=r_c$.
Equations~\eqref{e:DeltaTl-experiment} and \eqref{e:cap-Ra-experiment} then give $Ra_\subL$ and $Ra_\subG$.
For each representative temperature, Table~\ref{tab:effective-gravity-experiment} lists the largest $\Delta T$ satisfying $\max(Ra_\subL,Ra_\subG)\leq 5.0\times10^2$ under this $\phi_\subL=0.5$ convention.
We denote this largest allowed temperature difference by $\Delta T_{\max}$.

\begin{figure}[!tbp]
\centering
\includegraphics[width=0.95\linewidth]{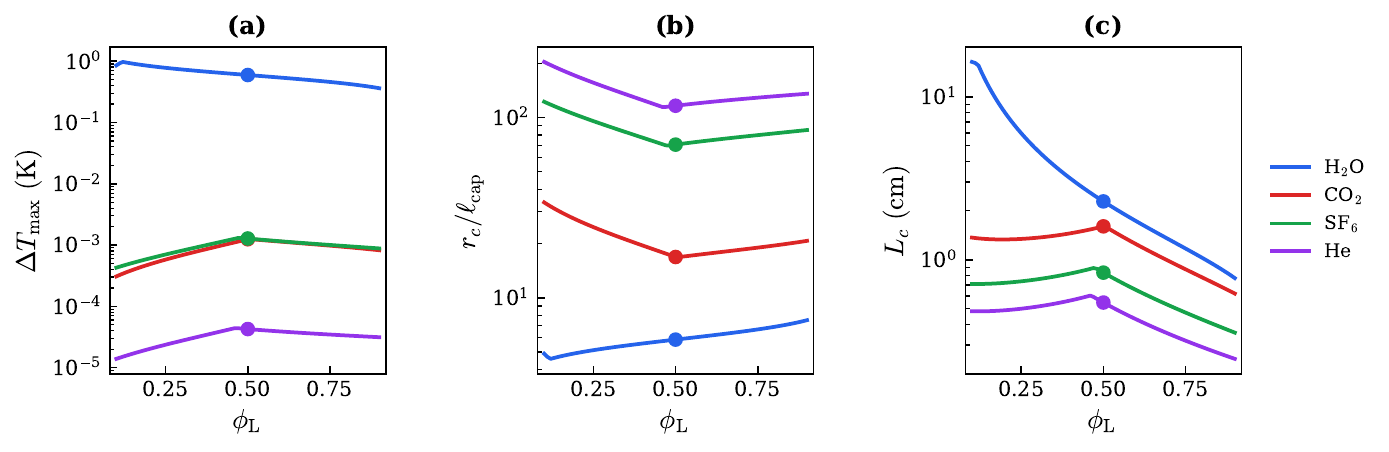}
\caption{Dependence of the candidate estimates on the liquid volume fraction $\phi_\subL$ at the representative temperatures in Table~\ref{tab:effective-gravity-experiment}.  For each fluid, the amount $n$ is fixed to the value used in Table~\ref{tab:effective-gravity-experiment}.  (a) Maximum allowed temperature difference $\Delta T_{\max}$.  (b) Capillary ratio $r_c/\ell_{\mathrm{cap}}$.  (c) Cell height $L_c$ determined by the lever rule at $r=r_c$.  The markers indicate the $\phi_\subL=0.5$ values listed in Table~\ref{tab:effective-gravity-experiment}.}
\label{fig:effective-gravity-experiment}
\end{figure}

These values and Fig.~\ref{fig:effective-gravity-experiment} show the qualitative experimental hierarchy.
In Fig.~\ref{fig:effective-gravity-experiment}, the same construction is repeated as $\phi_\subL$ is varied over the plotted range while keeping the chosen amount $n$ fixed for each fluid, and Table~\ref{tab:effective-gravity-experiment} reports the representative values at $\phi_\subL=0.5$.
The different shapes of the H$_2$O curve and the other curves reflect which layer reaches the Rayleigh bound: for H$_2$O the limiting layer changes from gas to liquid at a smaller $\phi_\subL$, whereas for CO$_2$, SF$_6$, and He this change occurs near the middle of the plotted range.
The resulting kink in $\Delta T_{\max}$ occurs where the active constraint in $\max(Ra_\subL,Ra_\subG)$ switches between the liquid and gas layers.
For CO$_2$, SF$_6$, and He, the saturated gas is sufficiently dense in the accessible coexistence range that $Ra_\subG$ can remain comparable to $Ra_\subL$.
H$_2$O near room temperature gives the widest temperature window: a temperature difference of order $1\,\mathrm{K}$ remains compatible with the conservative non-convection estimate while keeping both $r_c$ and $L_c$ in the centimeter range.
CO$_2$ gives a centimeter-scale radius and height, but the allowed temperature difference is only of order $10^{-3}\,\mathrm{K}$.
SF$_6$ and He also give centimeter-scale radii for the present choice of $n$, with sub-centimeter heights in the representative cells; He in the $4\,\mathrm{K}$ liquid-gas range requires a temperature difference of order $10^{-5}\,\mathrm{K}$.
The He point lies in the normal-liquid He-I regime: it is below the liquid-gas critical temperature, $5.2\,\mathrm{K}$, but above the superfluid transition temperature, $2.17\,\mathrm{K}$.
Thus the main difficulty beyond H$_2$O is not the formal existence of an effective-gravity inversion point, but the combined temperature-control and cell-scale window required to observe it in a conduction-dominated cell.

\begin{table}[!tbp]
\centering
\scriptsize
\setlength{\tabcolsep}{1.0pt}
\begin{tabular}{|c|c|c|c|c|c|c|c|c|}
\hline
fluid & $T_\mathrm{m}$ (K) & $n$ (mol) & $\Delta T_{\max}$ (K) & $r_c$ (cm) & $L_c$ (cm) & $\ell_{\mathrm{cap}}$ (cm) & $Ra_\subL$ & $Ra_\subG$ \\
\hline
H$_2$O & $298$ & $0.500$ & $0.594$ & $1.59$ & $2.28$ & $0.271$ & $5.0\times10^2$ & $1.6\times10^{-1}$ \\
CO$_2$ & $258$ & $0.137$ & $1.24\times10^{-3}$ & $1.50$ & $1.60$ & $0.0891$ & $4.8\times10^2$ & $5.0\times10^2$ \\
SF$_6$ & $293$ & $0.131$ & $1.28\times10^{-3}$ & $3.03$ & $0.833$ & $0.0429$ & $5.0\times10^2$ & $3.3\times10^2$ \\
He & $4.10$ & $0.376$ & $4.26\times10^{-5}$ & $3.51$ & $0.546$ & $0.0302$ & $5.0\times10^2$ & $2.7\times10^2$ \\
\hline
\end{tabular}
\caption{Candidate experimental scales estimated from NIST saturation-property values \cite{NISTWebBook}. The representative state is defined by the midpoint temperature $T_\mathrm{m}$ and $\phi_\subL=0.5$. We set $n=0.5\,\mathrm{mol}$ for H$_2$O and use the $100\,\mathrm{mL}$ saturated-gas equivalent for CO$_2$, SF$_6$, and He. Here $L_c$ is the cell height determined by the lever rule at $r=r_c$. The listed $\Delta T_{\max}$ satisfies $\max(Ra_\subL,Ra_\subG)\leq 5.0\times10^2$.}
\label{tab:effective-gravity-experiment}
\end{table}

The estimates above also separate the laboratory target from the model fluid used in the numerical landscape calculations.
H$_2$O near room temperature appears to be the most accessible first target for observing the effective-gravity inversion, whereas the van der Waals CO$_2$ model used above should be regarded as a thermodynamic model for displaying the landscape structure.

\section{Concluding remarks and discussion} \label{s:conclusion}
In this paper, we extended global thermodynamics to liquid-gas coexistence under gravity and steady heat conduction. Building on the heat-conducting coexistence theory \cite{NS17,NS19,NS22}, the equilibrium weak-gravity formulation \cite{NSgravity25}, and the effective-gravity ordering rule \cite{NSgeff26}, we constructed a variational free-energy function $\calF_g$ for the fixed-$\bT$ description.

The central result is the decomposition
\[
\calF_g=\calF_{g,\eff}+\calF_{g,\res}.
\]
For separated liquid--gas stationary states, the role of $\calF_{g,\eff}$ is direct. Its values reproduce the weak-gravity equilibrium free energy with $mgL$ replaced by $mg_\eff L$.  Since $\calF_{g,\res}=O(\ep^2)$ at the minimizing point of these states, the minimum value of $\calF_g$ is given by $\calF_{g,\eff}$.  Thus the comparison between $(\subL,\subG)$ and $(\subG,\subL)$ is reduced to the equilibrium weak-gravity comparison with the replacement $g\to g_\eff$.  When both phases occupy finite fractions, the minimum switches at $g_\eff=0$, giving the first-order configurational transition with order parameter $X-\xm$ or equivalently $\Psi_g=N(\xm-X)/L$.

This reduction of the minimum value should not be confused with a reduction of the thermodynamics to $mg_\eff L$ alone.  The fundamental relation is obtained by differentiating the minimized $\calF_g$ with respect to the laboratory variables.  The result is
\[
dF_g
=
-(S^\subl+S^\subu)d\bT
-\bP dV
+\bmu_g dN
-N\frac{\xm-X}{L}d(mgL)
-V\frac{\xm-X}{L}\frac{d\Ps}{d\bT}d\Xi .
\]
At fixed $(\bT,V,N)$, the last two terms reduce to $-N(\xm-X)d(mg_\eff L)/L$.  In the full thermodynamic relation, however, $(mgL,\Xi)$ are laboratory variables and their derivatives must be taken separately.  This is where $\calF_{g,\res}$ is essential: it does not set the ordering of separated heterogeneous states, but its derivatives are needed to recover thermodynamic observables such as the spatially averaged pressure and the global chemical potential.

The derivative information that is not read from the minimum value of $\calF_g$ alone also appears in local fields.  The chemical potential profile becomes discontinuous at the interface under heat conduction, and the interface temperature is shifted from the equilibrium saturation temperature, forming a supercooled or superheated layer near the interface.  These local anomalies are effects of the residual contribution $\calF_{g,\res}$.

Beyond the fundamental relation, the fixed-$\bT$ landscape shows how $\calF_{g,\res}$ changes the paths between stationary states.  The Hessian stability of the homogeneous branch remains tied to the ordinary spinodal structure, but the residual contribution changes the barrier shape and the ridge/valley arrangement.  Near a homogeneous state, this deformation appears through the local bias field $g_{\bias}$, and the disappearance of the nearby valley is described by the discriminant $\Delta(\xi_1)$.  The landscape may suggest features to be reproduced in a future dynamical theory of heat-conducting coexistence.  The fixed-$\bT$ landscape is not itself the fluctuation functional for boundary-controlled dynamics at fixed $(T_1,T_2)$ \cite{Touchette09,Bertini15}; instead, it gives the thermodynamic structure to which such dynamics should be compared in stationary states.  We expect that the disappearance of the near-homogeneous valley would appear, in an appropriate dynamical description, as a saddle-node bifurcation or an observable loss of a local barrier.

The estimates in Sec.~\ref{s:experiment} turn the effective-gravity ordering rule into a concrete design problem: the experimentally accessible window is constrained simultaneously by saturation pressure, cell size, capillarity, and convection.  Among the representative fluids examined there, water near room temperature gives the most accessible scale: both the inversion radius and the corresponding cell height are in the centimeter range, while the Rayleigh numbers remain within the conservative bound used in the estimate.  This suggests a realistic route to testing the sign change of $g_\eff$ by suppressing convection and observing whether the liquid layer moves to the gas side.
This configurational inversion is already the primary experimental target.
Residual effects should be regarded as secondary, more delicate signatures: they may appear as an interfacial metastable layer even without a complete inversion, or, in a more demanding measurement, as small changes of the interface position at nearly fixed $g_\eff$.

Numerical studies outside the present global-thermodynamic formulation have already probed neighboring aspects of this problem: gravity modifies liquid-gas phase separation in lattice-Boltzmann and molecular-dynamics simulations \cite{Cristea10,Wagner23,DavisGupta25}, and nonequilibrium molecular dynamics resolves interfacial thermal-resistance layers under heat flow \cite{Muscatello17}.
Within the global-thermodynamic line, the heat-flux control of interfacial metastable states was identified in a Hamiltonian Potts model \cite{KNS23}, and heat-induced liquid hovering was observed in molecular dynamics \cite{YNS24}.
A forthcoming molecular-dynamics study will examine whether the configurational change expected near $g_\eff=0$ is realized \cite{UrunoNakagawa26}.

The landscape described here is based on the assumption of a sharp interface.  The theory does not contain an independent surface-free-energy term, an interfacial width, or a capillary-wave term, unlike classical interfacial theories \cite{CahnHilliard58,RowlinsonWidom82}.  Therefore the barriers discussed here are not classical nucleation barriers with a surface-free-energy cost.  Nevertheless, the residual contribution and the local anomalies suggest that an apparent interfacial free-energy cost may contain both a global nonequilibrium component and a microscopic interfacial component.  A natural next problem is to separate these contributions.

More broadly, phase coexistence can amplify linear-response nonequilibrium effects into qualitative changes of configuration ordering.  The configurational transition, residual barrier deformation, and disappearance of a near-homogeneous valley arise from the degeneracy and metastability already present in coexistence, not from strongly nonlinear driving.  Similar effective-field descriptions may be useful in other coexistence systems where a conservative external field couples differently to the two phases and is balanced by heat-flow-induced or other nonequilibrium shifts of the coexistence condition.  Examples include liquid-liquid coexistence under gravity or centrifugal force, dielectric phases in an electric-field gradient inside a capacitor, magnetic phases in a magnetic-field gradient, and liquid-crystal coexistence under electric or magnetic fields.

Heat-flow-induced configurational bias may also be relevant to film boiling, Leidenfrost states, atmospheric phase coexistence such as cloud layers, and other situations where heat flow, gravity, and phase coexistence are coupled \cite{Quere13,Bourrianne19}.  These phenomena include hydrodynamic and interfacial mechanisms beyond the present theory.  The present framework may nevertheless help isolate the heat-flow-induced configurational bias within such more complex dynamics.

\section*{Acknowledgments}

The authors thank A. Yoshida, S. Uruno, R. Higa, A. Hisada, and F. Kagawa for useful discussions.

\section*{Statements and Declarations}

\noindent\textbf{Funding}
This work was supported by JSPS KAKENHI Grant Numbers JP23K22415, JP25K00923, JP25K22002, JP25H01975, and JP26H00383.

\noindent\textbf{Competing interests}
The authors declare that they have no competing interests.

\noindent\textbf{Data availability}
The parameter values used for the estimates in Sec.~\ref{s:experiment} are given in the manuscript. The saturation-property data used there were taken from the NIST Chemistry WebBook \cite{NISTWebBook}. Additional numerical data generated during the current study are available from the corresponding author on reasonable request.

\end{document}